\documentclass[12pt,a4paper]{iopart}
\pdfoutput=1
\usepackage{iopams,setstack,mathrsfs,amsthm,cite,enumerate,float,graphicx,slashed,ytableau}
\usepackage[utf8]{inputenc}
\usepackage[T1]{fontenc}
\usepackage{lmodern}
\usepackage[colorlinks,linkcolor=blue,citecolor=blue,urlcolor=blue]{hyperref}
\newcommand{\al}{\alpha}
\newcommand{\be}{\beta}
\newcommand{\de}{\delta}

\newcommand{\vep}{\varepsilon}
\newcommand{\ga}{\gamma}
\newcommand{\ka}{\kappa}
\newcommand{\la}{\lambda}

\newcommand{\si}{\sigma}
\renewcommand{\th}{\theta}
\newcommand{\vp}{\varphi}

\newcommand{\ze}{\zeta}
\newcommand{\De}{\Delta}

\newcommand{\La}{\Lambda}
\newcommand{\Si}{\Sigma}

\newcommand{\bk}{\mathbf{k}}

\newcommand{\bn}{\mathbf{n}}
\newcommand{\bp}{\mathbf{p}}

\newcommand{\bx}{\mathbf{x}}
\newcommand{\by}{\mathbf{y}}

\newcommand{\tS}{\widetilde{S}}

\newcommand{\tsi}{\widetilde\si}

\newcommand{\tbx}{\widetilde{\bx}}
\newcommand{\tby}{\widetilde{\by}}

\newcommand{\hbe}{\hat{\be}}

\newcommand{\CC}{{\mathbb C}}
\newcommand{\NN}{{\mathbb N}}
\newcommand{\RR}{{\mathbb R}}

\newcommand{\QQ}{{\mathbb Q}}
\newcommand{\cE}{{\mathcal E}}

\newcommand{\cH}{{\mathcal H}}

\newcommand{\cN}{{\mathcal N}}
\newcommand{\cP}{{\mathcal P}}

\newcommand{\cS}{{\mathcal S}}
\newcommand{\cT}{{\mathcal T}}

\newcommand{\cZ}{{\mathcal Z}}
\newcommand\bbe{\bar\beta}

\newcommand\Hsc{H_{\mathrm{sc}}}
\newcommand\Hsp{H_{\mathrm{spin}}}
\newcommand\Zsc{Z_{\mathrm{sc}}}
\newcommand\Zsp{Z_{\mathrm{spin}}}

\newcommand{\pd}{\partial}

\newcommand{\id}{1\hspace{-.25em}{\rm I}}
\newcommand{\noi}{\noindent}

\newcommand{\ket}[1]{|#1\rangle}

\newcommand{\mss}{\kern 1pt}

\renewcommand{\le}{\leqslant}
\renewcommand{\ge}{\geqslant}
\newcommand{\tends}[1]{\bbuildrel{\hbox to 2em{\rightarrowfill}}_{#1}^{}}

\newcommand{\operatorname}[1]{\mathop{\rm #1}\nolimits}
\newcommand{\sech}{\operatorname{sech}}

\newcommand{\iu}{\mathrm i}
\newcommand{\diff}{\mathrm{d}}

\newcommand{\su}{\mathrm{su}}

\newcommand{\implies}{\Longrightarrow}

\newcommand{\en}{\enspace}
\newcommand{\all}{\forall}
\newcommand{\pdf}[2]{\frac{\partial #1}{\partial #2}}

\newcommand{\Int}[1]{\,\mathop{\!#1}\limits^{\lower1ex\hbox{$\scriptstyle\circ$}}{}}

\newcommand{\vpmax}{\vp_{\mathrm{max}}}

\theoremstyle{remark}
\newtheorem{remark}{Remark}

\let\eqref\eref
\eqnobysec
\newcommand{\bN}{\mathbf N}

\newcommand{\bsv}{\mathbf s}

\def\clap#1{\hbox to 0pt{\hss#1\hss}}

\def\mathclap{\mathpalette\mathclapinternal}

\def\mathclapinternal#1#2{%
  \clap{$\mathsurround=0pt#1{#2}$}}
\begin{document}

\title{The open Haldane--Shastry chain: thermodynamics and criticality}

\author{Federico Finkel and Artemio González-López }

\address{Depto.~de Física Teórica, Facultad de Ciencias Físicas,\\
  Universidad Complutense de Madrid,\\
  Plaza de las Ciencias 1, 28040 Madrid, SPAIN}

\eads{\mailto{ffinkel@ucm.es}, \mailto{artemio@ucm.es}}

\vspace{10pt}
\begin{indented}
\item[]June 6, 2022
\end{indented}
\begin{abstract}
  We study the thermodynamics and criticality of the $\su(m|n)$ Haldane--Shastry chain of $BC_N$
  type with a general chemical potential term. We first derive a complete description of the
  spectrum of this model in terms of $BC_N$-type motifs, from which we deduce a representation for
  the partition function as the trace of a product of site-dependent transfer matrices. In the
  thermodynamic limit, this formula yields a simple expression for the free energy per spin in
  terms of the Perron--Frobenius eigenvalue of the continuum limit of the transfer matrix.
  Evaluating this eigenvalue we obtain closed-form expressions for the thermodynamic functions of
  the chains with $m,n\le2$. Using the motif-based description of the spectrum derived here, we
  study in detail the ground state of these models and their low energy excitations. In this way
  we identify the critical intervals in chemical potential space and compute their corresponding
  Fermi velocities. By contrast with previously studied models of this type, we find in some cases
  two types of low energy excitations with linear energy-quasimomentum relation. Finally, we
  determine the central charge of all the critical phases by analyzing the low-temperature
  behavior of the expression for the free energy per spin.
   
\end{abstract}


\maketitle


\section{Introduction}\label{sec.intro}

The original (trigonometric) Haldane--Shastry (HS) spin chain~\cite{Ha88,Sh88} describes an array
of $N$ spins $1/2$ fixed at equally spaced points on a circle and interacting through a two-body
exchange potential whose strength is inversely proportional to the square of the chord distance.
This model turns out to be both integrable~\cite{BGHP93} and exactly solvable, and is in fact
invariant under the Yangian algebra $Y(\su(2))$ even for a finite number of
sites~\cite{HHTBP92,Hi95npb}. In fact, its energy levels can be fully classified in terms of
finite-dimensional representations of the Yangian labeled by a type of skew Young diagrams, the so
called border strips~\cite{KKN97,NT98,BBS08}, which in part explains the exceptionally high
degeneracies found in its spectrum~\cite{FG15}. The physical properties of the HS chain are also
remarkable, being relevant in such disparate areas as the theory of one-dimensional anyons and
fractional statistics~\cite{Ha91b,Ha93,BS96}, conformal field theory~\cite{Ha91,BBS08,CS10,NCS11},
entanglement in low-dimensional systems~\cite{GSFPA10}, and quantum
chaos~\cite{FG05,BB06,BFGR08epl}.

The thermodynamics of the (spin $1/2$) HS chain was studied by Haldane himself, who found an
indirect expression for the entropy using the spinon description of the spectrum~\cite{Ha91}.
Shortly afterwards, the free energy per spin of the rational
(Polychronakos--Frahm)~\cite{Po93,Fr93} and hyperbolic (Frahm--Inozemtsev)~\cite{FI94} versions of
the HS chain (in the absence of a magnetic field or chemical potential term) was computed in
closed form in the latter two references. In fact, the method used in the latter reference was
ultimately based on a heuristic description of the spectrum in terms of border strips and their
associated motifs~\cite{HHTBP92}. This description was rigorously proved and generalized to the
$\su(m|n)$ supersymmetric case in Ref.~\cite{BBH10}. To be more specific, the latter description
entails the equivalence of the HS chain (and its rational/hyperbolic variants) to a classical
(inhomogeneous) vertex model with couplings proportional to the chains' dispersion function. Using
this equivalence, it is possible to express the chains' partition function as the trace of a
product of $N$ (site-dependent) transfer matrices of order $(m+n)\times(m+n)$. This result, when
extended to allow for a chemical potential term, yields a simple expression for the thermodynamic
free energy per spin in terms of the largest eigenvalue (in module) of the continuum limit of the
transfer matrix. This technique has been successfully applied to study the thermodynamics of the
trigonometric, rational and hyperbolic spin chains of HS type, both in the
non-supersymmetric~\cite{EFG12} and supersymmetric cases~\cite{FGLR18}.

The spin chains of HS type discussed above are naturally related to the $A_{N-1}$ classical root
system. In fact, generalizations of the latter models associated to the other classical (extended)
root systems have also been constructed, most notably for the $BC_N$
system~\cite{BPS95,Ya95,YT96,CS02,FGGRZ03,EFGR05,BFGR08,BFGR09} and its $B_N$~\cite{BFG13} and
$D_N$~\cite{BFG09,BFG11} reductions. Although these models have also been solved, i.e., their
partition function has been computed in closed form, to the best of our knowledge the study of
their thermodynamics has not been addressed. This is basically due to the fact that a motif-based
description of their spectrum has not been derived until very recently, and only for the
rational~\cite{BS20} and trigonometric~\cite{CFGR20} $BC_N$ chains.

The first aim of this work is to analyze the thermodynamics of the $\su(m|n)$ HS chain of $BC_N$
type. This model can be regarded as the open counterpart of the supersymmetric version of the
original ($A_{N-1}$-type) HS chain, in which the spins lie on a half-circle (in general not
uniformly spaced) and interact both among themselves and with their reflections with respect to
the origin. As we shall see, in this case it is possible to extend the description of the spectrum
in terms of $BC_N$-type motifs found in Ref.~\cite{CFGR20} to include a chemical potential term.
Using this description and applying the transfer matrix method we have derived a simple expression
for the thermodynamic free energy per spin formally akin to that of the $A_{N-1}$ case, but with a
different dispersion relation depending on a non-negative parameter. More precisely, the latter
expression involves again the largest eigenvalue of the continuum version of the transfer matrix,
which by the Perron--Frobenius theorem is positive and non-degenerate. This eigenvalue can be
easily computed in closed form for $m,n\le2$, thus obtaining simple expressions for the
thermodynamic functions of the corresponding chains in terms of a definite integral. In general,
these functions behave as those of a two- or three-level system (respectively for $m=n=1$ and
$m n>1$), which is not surprising since the partition function can be expressed as the trace of
$(m+n)\times(m+n)$ transfer matrices~\cite{Mu10}. In particular, for $mn>1$ the specific heat
features a double Schottky peak for certain values of the fermionic chemical potential.

As mentioned above, spin chains of HS type are closely connected to $(1+1)$-dimensional conformal
field theories. Indeed, the spectrum of the $\su(0|n)$ HS chain of $A_{N-1}$ type can be described
at low energies by the $\su(n)$ WZWN CFT at level $1$~\cite{Ha91,HHTBP92,BS96}. This result was
soon extended in Ref.~\cite{Hi95} to the $\su(0|n)$ Polychronakos--Frahm (PF) chain. In
particular, this shows that the $\su(0|n)$ HS and PF chains are both critical. Likewise, in
Ref.~\cite{HB00} it was shown that the specific heat of the $\su(m|n)$ PF chain behaves at low
temperatures as that of a CFT with central charge $c=m-1+n/2$ (for $mn\ne0$), or $c=m-1$ in the
purely bosonic/fermionic $\su(m)$ case. A similar analysis for the $\su(m|n)$ supersymmetric HS
chain was carried out in Ref.~\cite{BBS08}. More specifically, it was shown that the ground state
degeneracy is finite only for the $\su(0|n)$ and $\su(1|n)$ chains, which also feature low energy
excitations with linear energy-momentum relation. Thus these models are both critical. Moreover,
the central charge of the $\su(1|n)$ chain was shown in the latter reference to be $n/2$, in
agreement with the formula quoted above. The low-temperature behavior of the free energy per spin
of $\su(m|n)$ HS chains of type $A_{N-1}$ with the addition of a general chemical potential term
was analyzed in Ref.~\cite{FGLR18} for several low values of $m$ and $n$. This made it possible to
determine the regions in the space of chemical potentials in which the models can be critical, and
to compute the central charge of the related CFT. In particular, it was found in the latter
reference that this central charge can take rational (non-integer or half-integer) values for
suitable values of the chemical potentials. It should be stressed that the results mentioned in
this paragraph only apply to HS chains of $A_{N-1}$ type; in fact, we are not aware of similar
results in the literature for chains of HS type associated to the $BC_N$, $B_N$, or $D_N$ root
systems.

The second main aim of this paper is to study the critical behavior of the $\su(m|n)$
supersymmetric (open) HS chain of $BC_N$-type with a chemical potential term for $m,n\le 2$. To
this end, we take advantage of the motif-based description of the spectrum of this model to
determine its ground state and establish the existence of low-energy excitations with a linear
energy-(quasi)momentum relation, characteristic of a critical system. In this way we determine the
critical regions and the value of the Fermi velocity therein. In particular, we find that in some
cases there are two types of excitations, associated to changes in either end of the non-trivial
part of the ground state bond vector. We then use the explicit formula for the free energy per
spin derived in this paper to determine the central charge of the associated CFT. This provides a
complete description of the criticality properties of the model under study for all values of the
chemical potential.

The paper is organized as follows. In Section~\ref{sec.model} we recall the definition of the
$BC_N$-type $\su(m|n)$ HS chain introduced in Ref.~\cite{CFGR20}, and explain how to construct a
chemical potential term commuting with its Hamiltonian. The partition function of this model is
computed in Section~\ref{sec.PF} using Polychronakos's freezing trick argument~\cite{Po94}. In
Section~\ref{sec.VM} we establish the equivalence of the model under study with a suitable
inhomogeneous vertex model, which yields an explicit description of the spectrum (including the
correct degeneracy of the energy levels) with the help of $BC_N$-type motifs. This description is
used in Section~\ref{sec.FETD} to derive a general expression for the model's thermodynamic free
energy per spin in terms of the Perron--Frobenius eigenvalue of an appropriate transfer matrix. By
evaluating this eigenvalue, we then obtain a simple explicit formula for the free energy in the
case $m,n\le 2$. In Section~\ref{sec.GSlow} we determine the ground state of the latter models
using the motif-based description of the spectrum, and identify the low-energy excitations with
linear energy-momentum relation and their corresponding Fermi velocities. The results of the
previous sections are used in Section~\ref{sec.crit} to analyze the critical behavior of the
chains with $m,n\le 2$ for arbitrary values of the fermionic chemical potential, and in particular
to determine the central charge of the associated CFT in the critical regions by studying the
low-temperature behavior of the free energy per spin. In Section~\ref{sec.thermo} we derive
closed-form expressions for the main thermodynamic functions of the latter chains and discuss
their qualitative behavior. We summarize our main results in Section~\ref{sec.conc}, where we also
point out several possible lines for future research suggested by the present work. Finally, in
the Appendix we present the detailed calculation of the low-temperature asymptotic expansion of
the free energy per spin for the $\su(0|2)$, $\su(1|2)$, $\su(2|1)$ and $\su(2|2)$ chains, which
is used in Section~\ref{sec.crit} to derive their central charge.

\section{The model}\label{sec.model}

The open (i.e., $BC_N$-type) supersymmetric Haldane--Shastry chain consists of an array of $N$
particles, which can be either bosons or fermions, lying on the upper unit half-circle at fixed
angles~$2\theta_i\in(0,\pi)$ (with $1\le i\le N$), where $\theta_i$ is a root of the equation
\begin{equation}\label{roots}
  P^{(\beta_1-1,\beta_2-1)}_N(\cos2\theta)=0.
\end{equation}
Here $\be_1$ and $\be_2$ are two positive parameters, and $P^{(\be_1-1,\be_2-1)}_N$ is a Jacobi
polynomial of degree $N$. We shall respectively denote by $m$ and $n$ the number of bosonic and
fermionic internal degrees of freedom. In order to label these degrees of freedom, we divide the
set $\{1,\dots,m+n\}$ into two non-intersecting subsets $B=\{b_1,\ldots,b_m\}$ and
$F=\{f_1,\ldots,f_n\}$, where $b_1<b_2<\cdots<b_m$ and $f_1<f_2<\cdots<f_n\,$. The single particle
state~$\ket{s}$ (with $s=1,\dots,m+n$) shall then be regarded as bosonic if $s\in B$ or fermionic
if $s\in F$. We shall accordingly define the grading $p:\{1,\dots,m+n\}\to\{0,1\}$ as $p(s)=0$ if
$s\in B$ and $p(s)=1$ if $s\in F$. The Hilbert space of the system is the linear space
$\cS^{(m|n)}=\otimes_{i=1}^N\cS_i^{(m|n)}$, with $\cS_i^{(m|n)}=\CC^{m+n}$, spanned by the basis
vectors
\begin{equation}\label{basis}
  |s_1\cdots s_N\rangle:=|s_1\rangle\otimes\cdots\otimes|s_N\rangle,\qquad 1\le s_i\le m+n.
\end{equation}
Following reference~\cite{CFGR20}, we take the model's Hamiltonian as
\begin{equation}\label{Hchain}
  \fl
  H_0^{(m\vep|n\vep')}=\frac J4\sum_{i<j}\bigg(\frac{1-S_{ij}^{(m|n)}}{\sin^{2} \theta_{ij}^{-} }
  +\frac{1-\widetilde{S}_{ij}^{(m\vep|n\vep')}}{\sin^{2} \theta_{ij}^{+} }\bigg)
  +\frac{J}{8}\sum_i\left(\frac{\be_1}{\sin^{2} \theta_i} + \frac{\be_2}{\cos^2
      \theta_i}\right)\big(1-S_i^{(m\vep|n\vep')} \big),
\end{equation}
where $J\in\RR\setminus\{0\}$, the Latin indices (as in the sequel, unless otherwise stated) run
from $1$ to $N$, $\th_{ij}^\pm:=\th_i\pm\th_j$ and\footnote{We shall usually omit in what follows
  the explicit dependence of $H_0$, $S_{ij}$, $\tS_{ij}$ and $S_i$ on $m,n$ $\vep$, and $\vep'$.}
\begin{equation}
  \widetilde{S}_{ij}^{(m\vep|n\vep')}:=S_i^{(m\vep|n\vep')}S_j^{(m\vep|n\vep')}S_{ij}^{(m|n)}\,.
\end{equation}
The supersymmetric spin permutation operators $S_{ij}^{(m|n)}=S_{ji}^{(m|n)}$ in the latter
formula are defined by
\begin{equation}\label{spin-perm}
  S_{ij}^{(m|n)}|\cdots s_i \cdots s_j\cdots\rangle:=
  (-1)^{\nu(s_i,\dots,s_j)}|\cdots s_j\cdots s_i\cdots\rangle\,,
\end{equation}
where $\nu(s_i,\dots,s_j)=p(s_i)$ if $s_i=s_j$ and is otherwise equal to the number of fermionic
spins $s_k$ in the range $k=i+1,\dots,j-1$. Similarly, the supersymmetric spin reversal operators
$S_i^{(m\vep|n\vep')}$ are defined by
\begin{equation}
  \label{spin-rev}
  S_i^{(m\vep|n\vep')}\ket{\cdots s_i\cdots}:=\la_{\vep\vep'}(s_i)\ket{\cdots \imath(s_i)\cdots}\,,
\end{equation}
where $\vep,\vep'\in\{\pm1\}$ are two fixed signs and
\[
  \la_{\vep\vep'}(s)=\cases{\vep,& \(s\in B\)\\\vep',&\(s\in F\)\,.}
\]
Here $\imath$ is any involution leaving invariant the bosonic and fermionic sectors, i.e.,
$\imath^2=I$, $\imath(B)=B$ and $\imath(F)=F$, with at most one fixed point in each sector. The
existence of fixed points of~$\imath$ obviously depends only on the parity of the integers $m$ and
$n$: indeed, there is a bosonic (respectively fermionic) fixed point if and only if $m$ (resp.
$n$) is odd. We can without loss of generality fix the action of the involution $\imath$ by
setting
\[
  \imath(b_\al):=b_{m+1-\al}\,,\qquad \imath(f_{\be}):=f_{n+1-\be}\,,
\]
where, as in the sequel (unless otherwise stated), the Greek indices are assumed to label the
elements of the sets $B$ and $F$. One can intuitively think of $\imath$ as reversing the spin of a
site, by relabeling the bosonic degrees of freedom by $b_\al-(m+1)/2$ or the fermionic ones by
$f_\be-(n+1)/2$. In view of the above we shall say from now on, with a slight abuse of notation,
that a spin is of type $\pm\al$ if it is either of type $\al$ or of type $\imath(\al)$. Since
$\pm\al=\pm\imath(\al)$, all the different values of $\pm\al$ can be obtained by restricting $\al$
to the set
\[
  B_0\cup F_0:=\left\{b_1,\dots,b_{\lceil m/2\rceil}\right\}\cup\left\{f_1,\dots,f_{\lceil
      n/2\rceil}\right\},
\]
where $\lceil x\rceil$ is the lowest integer greater that or equal to $x$ (for an integer $k$ we
have $\lceil k/2\rceil=(k+\pi(k))/2$, where $\pi(k)\in\{0,1\}$ is the parity of $x$).

\begin{remark}\label{rem1}
  As mentioned in the Introduction, the model~\eqref{Hchain} can be regarded as the \emph{open}
  version of the (supersymmetric) Haldane--Shastry chain. Indeed, the chain sites
  $z_j:=\e^{2\iu\theta_j}$ lie on the upper unit circle, and the spin at $z_j$ interacts not only
  with the remaining spins at $z_k$ (with $k\ne j$) but also with their reflections $\bar z_k$
  with respect to the real axis. The strength of these interactions is inversely proportional to
  the square of the (chord) distance between $z_j$ and the points $z_k$ and $\bar z_k$,
  respectively. Note also that, since the last sum in Eq.~\eqref{Hchain} can be written as
  \[
    \sum_i\bigg(\frac{\be_1-\be_2}{\sin^2\theta_i}+\frac{4\be_2}{\sin^2(2\theta_i)}\bigg)(1-S_i),
  \]
  the Hamiltonian~\eqref{Hchain} is obviously related to the $BC_N$ extended root system with
  positive roots $\theta_{ij}^{\pm}$, $\theta_i$ and $2\theta_i$, with $1\le i<j\le N$.
\end{remark}

We next define the number operators~$\cN_\al$  by
\[
  \cN_\al\ket{s_1\cdots s_N}=N_\al(\bsv)\ket{s_1\cdots s_N}\,,\qquad 1\le\al\le m+n\,,
\]
where
\[
  N_\al(\bsv):=\sum_{i=1}^N\de_{\al,s_i}
\]
is the number of spins of type $\al$ in the basis vector $\ket{s_1\cdots s_N}$. It is clear that
$[S_{ij},\cN_\al]=0$, since $S_{ij}$ preserves the spin content of a basis vector. On the other
hand, the operators $S_i$, and hence $\tS_{ij}$, do \emph{not} commute with $\cN_\al$ unless
$\imath(\al)=\al$, since $S_i$ changes a spin of type $\al$ into one of type $\imath(\al)$. For
this reason, the Hamiltonian $H_0$ in general does \emph{not} commute with the linear combination
\begin{equation}\label{H1}
  H_1:=-\sum_\al\mu_\al\cN_\al\,,
\end{equation}
where $\mu_\al$ is an arbitrary real number. However, it is also clear that $S_i$ preserves the
number $N_\al(\bsv)+N_{\imath(\al)}(\bsv)$ of spins of type $\pm\al$, so that
\[
  [S_i,\cN_\al+\cN_{\imath(\al)}]=[\tS_{ij},\cN_\al+\cN_{\imath(\al)}]=0
\]
and hence
\[
  [H_0,\cN_\al+\cN_{\imath(\al)}]=0\,.
\]
Thus $H_0$ will commute with $H_1$ provided that
\begin{equation}\label{mucond}
  \mu_\al=\mu_{\imath(\al)}\,,\qquad\all\al\in B\cup F\,,
\end{equation}
a condition that we shall assume from now on to hold.

The partition function of the Hamiltonian $H_0$ was computed in closed form in Ref.~\cite{CFGR20}.
In order to study the thermodynamics of the open supersymmetric HS chain, in the next section we
shall evaluate the partition function of the modified Hamiltonian
\begin{equation}
  H:=H_0+H_1
  \label{Hdef}
\end{equation}
for arbitrary values of the parameters $\mu_\al$ satisfying condition~\eqref{mucond} above. The
coefficient $\mu_\al=\mu_{\imath(\al)}$ can then be regarded as the chemical potential for the
spins of type $\pm\al$, since $H$ commutes with $H_0$ and $\cN_{\al}+\cN_{\imath(\al)}$ for all
$\al$.
More precisely, $H_0$ leaves invariant the subspaces~$\cH(\bN)$ with well defined numbers
$N_{\pm\al}$ of particles of all types $\pm\al$, where
$\bN:=\big(N_{\pm\al}\big)_{\al\in B_0\cup F_0}$ and the non-negative integers $N_{\pm\al}$
satisfy the condition $\sum_{\al\in B_0\cup F_0}N_{\pm\al}=N$.
It follows that the
partition function $Z_N$ of the modified Hamiltonian $H$ can be expressed as
\[
  Z_N=\sum_{\sum\limits_{\mathclap{\al\in B_0\cup F_0}}N_{\pm\al}=N}q^{\hphantom{\scriptstyle\al\in1
      }-\sum\limits_{\mathclap{\al\in B_0\cup F_0}}\mu_\al N_{\pm\al}}Z_{0,\bN},
    \qquad q:=\e^{-1/T}\equiv\e^{-\be}\,,
\]
where $Z_{0,\bN}$ denotes the partition function of the restriction of $H_0$ to~$\cH(\bN)$ and we
have taken Boltzmann's constant $k_B$ as $1$. Since $Z_{0,\bN}$ is independent of the
$\mu_{\al}$'s, the thermal average~$n_{\pm\al}:=\langle N_{\pm\al}\rangle/N$ of the density of
particles of type~$\pm\al$ is given by
\[
  n_{\pm\al}=\sum_{\sum\limits_{\mathclap{\al\in B_0\cup F_0}}N_{\pm\al}=N}\frac{N_{\pm\al}}N\,
  q^{\hphantom{\scriptstyle\al\in1 }-\sum\limits_{\mathclap{\al\in B_0\cup F_0}}\mu_\al
    N_{\pm\al}}Z_{0,\bn}=-\frac{\pd f_N}{\pd\mu_\al}\,,
\]
with
\[
  f_N:=-(N\be)^{-1}\log Z_N\,.
\]
In the thermodynamic limit this becomes
\begin{equation}\label{npmal}
  n_{\pm\al}=-\frac{\pd f}{\pd\mu_\al}\,,
\end{equation}
where
\[
  f:=\lim_{N\to\infty}f_N
\]
is the thermodynamic free energy per particle. All the other thermodynamic functions can then be
computed from $f$. For instance, the energy, specific heat and entropy per particle are
respectively given by
\begin{equation}\label{thermofs}
  u=\pdf{(\be f)}{\be}\,,\qquad c_V=-\be^2\pdf{u}{\be}\,,\qquad s=\be^2\pdf{f}{\be}=\be(u-f)\,.
\end{equation}
Likewise, a straightforward calculation shows that the variance
$\De_{\pm\al}:=(\langle N_{\pm\al}^2\rangle-\langle N_{\pm\al}\rangle^2)/N$ of the number of spins
of type~$\pm\al$ is equal to
\begin{equation}\label{varnF}
  \De_{\pm\al}=-\frac1\be\,\frac{\pd^2f}{\pd\mu_\al^2}\,.
\end{equation}

\section{Partition function}\label{sec.PF}

In this section we shall compute the partition function of the modified Hamiltonian~\eqref{Hdef}
in closed from. Since we shall closely follow the method used in Ref.~\cite{CFGR20} to evaluate
the partition function of the Hamiltonian $H_0$ in Eq.~\eqref{Hchain}, we shall only outline the
main steps in the calculation and omit unnecessary details.

The method is based on the connection of the spin chain Hamiltonian~\eqref{Hchain} with the
dynamical one
\begin{eqnarray}
  \fl
  \Hsp=&-\sum_i\pd_{x_i}^2+a\sum_{i\neq
         j}\!\left(\frac{a-S_{ij}}{\sin^2x_{ij}^{-}}+\frac{a-\widetilde{S}_{ij}}{\sin^2x_{ij}^{+}}\right)
         +\sum_i\!\left(\frac{\hbe_1(\hbe_1- S_i)}{\sin^2x_i}+\frac{\hbe_2(\hbe_2-
         S_i)}{\cos^2x_i}\right)\nonumber\\
  \fl
       &-\frac{8a}{J}\sum_\al\mu_\al\cN_\al
         \label{Hsp}
\end{eqnarray}
and its scalar version
\begin{equation}
  \label{Hsc}
  \fl
  \Hsc=-\sum_i\pd_{x_i}^2+a(a-1)\sum_{i\neq j}\!\bigg(\frac{1}{\sin^2x_{ij}^{-}}+\frac{1}{\sin^2x_{ij}^{+}}\bigg)+\sum_i\!\bigg(\frac{\hbe_1(\hbe_2- 1)}{\sin^2x_i}+\frac{\hbe_2(\hbe_2-1)}{\cos^2x_i}\bigg),
\end{equation}
where $a$ is a positive parameter and we have set $\hbe_{1,2}:=\be_{1,2} a$,
$x_{ij}^{\pm}:=x_i\pm x_j$. Indeed, we clearly have
\[
  \Hsp=\Hsc+\frac{8a}{J}H(\bx)=-\sum_i\pd_{x_i}^2+a^2U(\bx)+O(a)\,,
\]
where $H(\bx)$ is obtained from the spin chain Hamiltonian~$H$ replacing the fixed sites
$\theta_i$ by the dynamical variables (coordinates) $x_i$, and
\[
  U(\bx):=\sum_{i\ne j}\left(\frac{1}{\sin^2x_{ij}^{-}}+\frac{1}{\sin^2x_{ij}^{+}}\right)
  +\sum_i\!\left(\frac{\be_1^2}{\sin^2x_i}+\frac{\be_2^2}{\cos^2x_i}\right).
\]
As $a$ tends to infinity the particles oscillate with decreasing amplitude about the coordinates
of the equilibrium position of the scalar potential $U(\bx)$ on the configuration space
\[
  C'=\{\bx:=(x_1,\dots,x_N)\in\RR^N:0<x_1<x_2<\cdots<x_N<\pi/2\}\,.
\]
This equilibrium is unique~\cite{CS02}, and its coordinates coincide with the chain sites
$\theta_i$ \cite{OS02}. Thus in the large $a$ limit the spin degrees of freedom decouple from the
dynamical ones, and are governed by the Hamiltonian $H(\theta_1,\dots,\theta_N)=H$. As a
consequence, when $a\gg 1$ the eigenvalues $E_{ij}$ of $\Hsp$ behave as
\[
  E_{ij}= E_{\mathrm{sc},i}+\frac{8\mss a}J E_j+o(a),
\]
where $E_{\mathrm{sc},i}$ and $E_j$ denote any two eigenvalues of the scalar (dynamical)
Hamiltonian~\eqref{Hsc} and the spin chain Hamiltonian~$H$, respectively. Denoting by $\Zsp$ and
$\Zsc$ the partition functions of $\Hsp$ and $\Hsc$, the partition function $Z_N$ of $H$ can then
be computed from the \emph{exact} expression
\begin{equation}\label{Z}
  Z_N(T)=\lim_{a\to\infty}\frac{\Zsp(8\mss a T/J)}{\Zsc(8\mss a T/J)}\,.
\end{equation}
This method of computing the partition function of the spin chain Hamiltonian~$H$ exploiting its
connection with the dynamical spin Hamiltonian~\eqref{Hsp} goes by the name of ``Polychronakos's
freezing trick'' in the literature.

We shall next compute the large $a$ limit of the partition functions $\Zsc(8aT/J)$ and
$\Zsp(8aT/J)$. To begin with, the $a\to\infty$ limit of the former partition function can be
derived from the result in Ref.~\cite{EFGR05} replacing $q$ by $q^{J}$, namely
\begin{equation}\label{Zsc}
  \lim_{a\to\infty}q^{-JE_0/8a}\Zsc(8aT/J)=\prod_{i=1}^N(1-q^{J\cE(i)})^{-1}\,,
\end{equation}
where the dispersion function~$\cE(i)$ is given by
\begin{equation}\label{dispf}
  \cE(i):=\frac12\,i(2\mss\bar\be+2\mss N-i-1)\,,\qquad \bar\be:=\frac12(\be_1+\be_2)\,,
\end{equation}
and
\[
  E_0=\frac23\,Na^2\big[2N^2+3(2\bbe-1)N+6\bbe(\bbe-1)+1\big]
\]
is the ground state energy of both $\Hsc$ and $\Hsp$. In the spin case, it was shown in
Ref.~\cite{CFGR20} that when $\mu_\al=0$ for all $\al$ the Hamiltonian $\Hsp$ is upper triangular
in a (non-orthonormal) basis with elements
\begin{equation}\label{bpbs}
\ket{\bp,\bsv}:=\La\bigl(\e^{2\iu\mss \bp\cdot\bx}\phi(\bx)\ket{s_1\cdots s_N}\bigr)\,,  
\end{equation}
where $\bsv=(s_1,\dots,s_N)\in (B\cup F)^N$, $\bp=(p_1,\dots,p_N)\in(\NN\cup\{0\})^N$,
$\bx:=(x_1,\dots,x_N)$,
\[
  \phi(\bx):=\prod_{i<j}|\sin x_{ij}^+\sin x_{ij}^-|^a\prod_k|\sin x_k|^{\hbe_1}|\cos x_k|^{\hbe_2}\,,
\]
and $\La$ denotes the total symmetrizer with respect to particle permutations and simultaneous
reversal of a particle's coordinates and spins. In other words, $\La$ is the projector onto states
symmetric under the action of the operators $\Pi_{ij}:=P_{ij}\otimes S_{ij}$ and
$\Pi_i:=P_i\otimes S_i$, where the operators $P_{ij}$ and $P_i$ act on scalar functions as
\[
  \fl
  P_{ij}f(\dots,x_i,\dots,x_j,\dots)=f(\dots,x_j,\dots,x_i,\dots)\,,
  \qquad
  P_if(\dots,x_i,\dots)=f(\dots,-x_i,\dots)\,.
\]
Moreover, for the wave function~\eqref{bpbs} to be a basis the quantum numbers~$\bp$ and $\bsv$
must satisfy the following conditions (or trivial variants thereof):
\begin{enumerate}[(B1)]
\item The integer multiindex $\bp=(p_1,\ldots,p_N)$ is nonnegative and nonincreasing, i.e.,
  $p_i\in\NN\cup\{0\}$ and $p_i\ge p_{i+1}\,$ for all $i$.
\item If $p_i=p_{i+1}$ then $s_i\ge s_{i+1}$ if $s_i\in B$, or $s_i>s_{i+1}$ if $s_i\in F$.
\item If $p_i=0$ then $s_i\in B_\vep\cup F_{\vep'}$, where
  \[
    B_{\vep}:=\{b_1,\ldots,b_{m_\vep}\}\,,\qquad F_{\vep'}:=\{f_1,\ldots,f_{n_{\vep'}}\}\,.
  \]
  with
  \[
    m_\vep:=\frac12\big(m+\vep\pi(m)\big),\qquad n_{\vep'}:=\frac12\big(n+\vep'\pi(n)\big).
  \]
\end{enumerate}
If the wave functions~$\ket{\bp,\bsv}$ are suitably ordered, the
Hamiltonian~$H_{\text{spin},0}:=\Hsp|_{\mu_1=\cdots=\mu_{m+n}=0}$ acts on the resulting basis as
an upper triangular operator with diagonal elements
\begin{equation}\label{ed}
  \fl
  (H_{\text{spin},0})_{\bp\bsv,\bp\bsv}=\sum_i\big(2 p_i+\hbe_1+\hbe_2+2a(N-i)\big)^2=
  4\sum_i\big[p_i+a(\bbe+N-i)\big]^2
\end{equation}
(see~\cite{CFGR20} for details). Since the operator $H_1$ is diagonal in the basis~\eqref{bpbs},
with diagonal elements $-\sum_\al\mu_\al N_\al(\bsv)$, the action of the full Hamiltonian~$\Hsp$
on the latter basis is still upper triangular, with diagonal elements ---i.e., eigenvalues---
\begin{eqnarray}
  E_{\bp\bsv}&=4\sum_i\big[p_i+a(\bbe+N-i)\big]^2-\frac{8a}J\sum_\al\mu_\al N_\al(\bsv)\nonumber\\
             &= E_0+8a\sum_ip_i(\bbe+N-i)-\frac{8a}J\sum_\al\mu_\al N_\al(\bsv)+O(1)\,.
    \label{Ep}
\end{eqnarray}
If we parametrize the multiindex $\bp$ satisfying condition (B1) above as
\begin{equation}\label{ppi}
  \bp=(\underbrace{\pi_1,\dots,\pi_1,}_{k_1}\dots,\underbrace{\pi_r,\dots,\pi_r}_{k_r})\,,
\end{equation}
with $k_i>0$, $k_1+\cdots+k_r=N$ and $\pi_1>\cdots>\pi_r\ge0$, we can write
\[
  \sum_ip_i(\bbe+N-i)=\sum_{j=1}^r\pi_jk_j\bigg[\bbe+N-K_{j-1}-\frac12(k_j+1)\bigg]
  =:\la(\bpi,\bk)\,,
\]
where
\[
  K_j:=\sum_{i=1}^jk_i\,.
\]
We thus have
\[
  \lim_{a\to\infty}q^{-JE_0/8a}\Zsp(8aT/J)=\sum_{\bk\in\cP_N}\sum_{\pi_1>\cdots>\pi_r\ge0}q^{J\la(\bpi,\bk)}
  \sum_{\bsv\in\bp}q^{-\sum_\al\mu_\al N_\al(\bsv)}\,,
\]
where $\cP_N$ denotes the set of partitions of the integer $N$ taking order into account and the
last sum runs over all spin vectors $\bsv=(s_1,\dots,s_N)$ whose components $s_i$ satisfy
conditions (B2)-(B3) for a given multiindex~$\bp$ of the form \eqref{ppi}. We clearly have
\[
  \sum_{\bsv\in\bp}q^{-\sum_\al\mu_\al N_\al(\bsv)}=\prod_{j=1}^{r-1}\si(k_j)\cdot\tsi(k_r,p_r)\,,
\]
where
\begin{eqnarray*}
  \si(k)&=\sum_{s_1\succ\cdots\succ s_k}q^{-\sum_\al\mu_\al N_\al(s_1,\dots,s_k)}\,,\\
  \tsi(k,p_r)&=\cases{\si(k),&\(p_r>0\)\\
    \sum_{\substack{s_1\succ\cdots\succ s_k\cr s_j\in B_\vep\cup F_{\vep'}}}q^{-\sum_\al\mu_\al
      N_\al(s_1,\dots,s_k)},&\(p_r=0\,,\)}
\end{eqnarray*}
and the notation $s\succ t$ indicates that $s\ge t$ if the spin $s$ is bosonic or $s>t$ if it is
fermionic. The sigma functions can be easily expressed in terms of the complete and elementary
symmetric polynomials~\cite{Ma95}
\[
  \fl h_j(x_1,\dots,x_l):=\sum_{i_1\le\cdots\le i_j\le l}x_{i_1}\cdots x_{i_j}\,, \qquad
  e_j(y_1,\dots,y_l):=\sum_{i_1<\cdots<i_j\le l}y_{i_1}\cdots y_{i_j}\,.
\]
(with $e_j(y_1,\dots,y_l):=0$ if $j>l$). Indeed, since
\[
  \sum_{\al}\mu_\al N_\al(s_1,\dots,s_k)=\sum_{i=1}^k\mu_{s_i}\,,
\]
taking into account condition (B2) above we have
\begin{eqnarray}
  \si(k)&=\sum_{j=0}^k\sum_{\al_1\le\cdots\le \al_j}q^{-\mu_{b_{\al_1}}}\cdots
  q^{-\mu_{b_{\al_j}}}
  \sum_{\be_1<\cdots<\be_{k-j}}q^{-\mu_{f_{\be_1}}}\cdots
          q^{-\mu_{f_{\be_{k-j}}}}\nonumber\\
        &=\sum_{j=0}^kh_j(x_1,\dots,x_m)e_{k-j}(y_1,\dots,y_n)=:E_k(\bx,\by)\,,
          \label{sicalc}
\end{eqnarray}
where
\begin{equation}\label{bxy}
  x_\al:=q^{-\mu_{b_\al}}\,,\qquad y_\be:=q^{-\mu_{f_\be}}
\end{equation}
and $E_k(\bx,\by)$ denotes the elementary symmetric function in the variables
$\bx=(x_1,\dots,x_m)$, $\by=(y_1,\dots,y_n)$~\cite{Ma95}. A similar calculation shows that
\[
  \tsi(k,0)=E_k(\tbx,\tby),
\]
where
\begin{equation}\label{tbxy}
  \tbx=(x_1,\dots,x_{m_\vep})\,,\qquad \tby=(y_1,\dots,y_{n_{\vep'}})\,.
\end{equation}
From the equality
\[
  \Si_l:=\sum_{\pi_1>\cdots>\pi_l>0}q^{J\la(\pi_1,\dots,\pi_l,k_1,\dots,k_l)}=\prod_{i=1}^l\frac{q^{J\cE(K_i)}}{1-q^{J\cE(K_i)}}
\]
proved in Ref.~\cite{EFGR05} we then obtain
\begin{eqnarray}
  \fl
  \lim_{a\to\infty}\!q^{-E_0/8a}
  &\Zsp(8aT/J)
  =\sum_{\bk\in\cP_N}\left(\prod_{i=1}^rE_{k_i}(\bx,\by)\cdot\Si_r
  +E_{k_r}(\tbx,\tby)\prod_{i=1}^{r-1}E_{k_i}(\bx,\by)\cdot\Si_{r-1}\right)\nonumber\\
  \fl
  &=\sum_{\bk\in\cP_N}\left(\,\frac{E_{k_r}(\bx,\by)\,q^{J\cE(N)}}{1-q^{J\cE(N)}}+
    E_{k_r}(\tbx,\tby)\right)\prod_{i=1}^{r-1}\frac{E_{k_i}(\bx,\by)\,
    q^{J\cE(K_i)}}{1-q^{J\cE(K_i)}}\,.
    \label{Zsp}
\end{eqnarray}
Using Eqs.~\eqref{Zsc} and~\eqref{Zsp} in the freezing trick formula~\eqref{Z} we obtain the
following closed formula for the partition function of the Hamiltonian~$H$ in Eq.~\eqref{Hdef}:
\begin{eqnarray}
  \fl
  Z_N(T)&=\sum_{\bk\in\cP_N}\left[E_{k_r}(\tbx,\tby)+\Big(E_{k_r}(\bx,\by)
    -E_{k_r}(\tbx,\tby)\Big)q^{J\cE(N)}\right]
        F(q,\bk)\prod_{i=1}^{r-1}E_{k_i}(\bx,\by)\nonumber\\
  \fl
      &=:\cZ_N(q;\bx,\by)\,,
        \label{Zexp}
\end{eqnarray}
where
\begin{equation}\label{Fq}
  F(q,\bk):=\prod_{i=1}^{r-1}q^{J\cE(K_i)}\prod_{j=1}^{N-r}(1-q^{J\cE(K'_j)})
\end{equation}
and $\{K'_1,\ldots,K'_{N-r}\}=\{1,\dots,N\}\setminus\{K_1,\ldots,K_{r}\}$ (with
$K'_1<K'_2<\cdots<K'_{N-r}$). Note, finally, that the vectors~$\bx$, $\by$, $\tbx$, $\tby$
appearing in the latter explicit formula for $Z_N(T)$ are defined in terms of the chemical
potentials~$\mu_\al$ by Eqs.~\eqref{bxy}-\eqref{tbxy}.

\section{Equivalent vertex model}\label{sec.VM}

Following Ref.~\cite{CFGR20}, we shall next construct a vertex model whose partition function
exactly coincides with $Z_N(T)$. The key idea in this respect is the following identity, proved in
the latter reference for \emph{arbitrary} values of the arguments $\bx\in\RR^m$, $\by\in\RR^n$:
\begin{equation}\label{partition}
  \fl
  \cZ_N(q;\bx,\by)=\sum_{\bk\in{\mathcal P}_N}\bigg(S_{\langle
    k_1,\ldots,k_r\rangle,0}(\bx,\by)
  + S_{\langle k_1,\ldots,k_r\rangle,1}(\bx,\by)q^{J\cE(N)} \bigg)
  q^{\sum_{i=1}^{r-1}J\cE(K_i)}.
\end{equation}
Here $S_{\langle k_1,\ldots,k_r\rangle,\ga}(\bx,\by)$ is the $BC_N$-type $\su(m|n)$-supersymmetric
Schur polynomial of type $\ga\in\{0,1\}$, whose definition in terms of the $BC_N$-type motifs
introduced in Ref.~\cite{CFGR20} we shall recall next (see the latter reference for a review of
the notation and relevant results).

\begin{figure}[t]
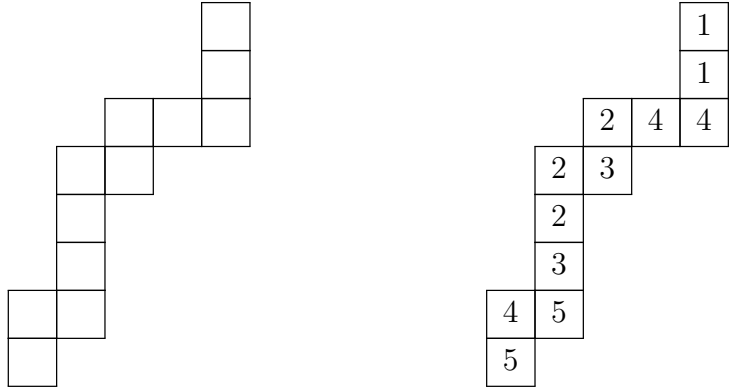

  \hfil
  \ydiagram{4+1,4+1,2+3,1+2,1+1,1+1,2,1}
  \hfil
  \ytableaushort{\none\none\none\none1,\none\none\none\none1,\none\none244,\none23\none\none,
  \none2\none\none\none,\none3\none\none\none,45\none\none\none,5\none\none\none\none}\hfil
  \caption{The border strip $\langle3,1,2,4,2\rangle$ (right) and a (3|2) supersymmetric Young
    tableau of shape $\langle3,1,2,4,2\rangle$ for the choice $B=\{1,2,3\}$, $F=\{4,5\}$
    (left).}
  \label{fig.bs}
\end{figure}

Let us associate to each ordered partition $\bk=(k_1,\dots,k_r)\in\cP_N$ the border strip with
columns of lengths $k_1,\dots,k_r$ (from right to left), which we shall denote by
$\langle k_1,\dots,k_r\rangle$; see, e.g., Fig.~\ref{fig.bs} (left).
An $\su(m|n)$-supersymmetric Young tableau of shape $\langle k_1,\dots,k_r\rangle$ is defined as a
filling of the latter border strip with the integers in $B\cup F=\{1,\dots,m+n\}$ that is:
\begin{enumerate}[(YT1)]
\item Weakly increasing along rows and strictly increasing down columns for integers in $F$.
\item Strictly increasing along rows and weakly increasing down columns for integers in $B$
\end{enumerate}
(see Fig.~\ref{fig.bs} (right) for an example). Such a Young tableau can be identified with a bond
vector $(s_1,\dots,s_N)$, with $s_i\in B\cup F$ for all $i$, by reading the tableau from right to
left and top to bottom. For instance, the tableau in Fig.~\ref{fig.bs} (right) is equivalent to
the bond vector $(1,1,4,4,2,3,2,2,3,5,4,5)$. Let us next order the integers in $B\cup F$ in such a
way that
\begin{equation}\label{order}
  B_\vep\cup F_{\vep'}=\{1,\dots,m_\vep+n_{\vep'}\}\,.
\end{equation}
Two such orderings are, for instance,
\begin{eqnarray}
  \fl
  F&=\{1,\dots,n_{\vep'},n_{\vep'}+m+1,\dots,m+n\}\,,\qquad
     B=\{n_{\vep'}+1,\dots,n_{\vep'}+m\}\label{ord1}\\
  \fl
  B&=\{1,\dots,m_{\vep},m_\vep+n+1,\dots,m+n\}\,,\qquad
     F=\{m_\vep+1,\dots,m_\vep+n\}\,.\label{ord2}
\end{eqnarray}
Let us denote by $\cT_\al(\bk)$ (with $\al\in\{0,1\}$) the set of all $\su(m|n)$-supersymmetric
Young tableau, or equivalently bond vectors $\bsv=(s_1,\dots,s_N)$, of shape
$\langle k_1,\dots,k_r\rangle$ such that
\[
  \cases{s_N\le m_\vep+n_{\vep'}\,,&if \(\al=0\)\\
  s_N>m_\vep+n_{\vep'}\,,&if \(\al=1.\)}
\]
The $BC_N$ type super Schur polynomials $S_{\langle k_1,\ldots,k_r\rangle,\al}(\bx,\by)$ are then
defined for arbitrary~$\bx\in\RR^m$, $\by\in\RR^n$ by
\[
  S_{\langle k_1,\ldots,k_r\rangle,\al}(\bx,\by)=\sum_{\bsv\in\cT_\al(\bk)}x_1^{\nu_1^b(\bsv)}\cdots
  x_m^{\nu_m^b(\bsv)}y_1^{\nu_1^f(\bsv)}\cdots y_n^{\nu_n^f(\bsv)}\,,
\]
where $\nu_\al^b(\bsv)$ (resp.~$\nu_\be^f(\bsv)$) is the number of times the integer $b_\al$
(resp.~$f_\be$) appears in the bond vector $\bsv$ (in other words,
$\nu_\al^b(\bsv)=N_{b_\al}(\bsv)$, $\nu_\be^f(\bsv)=N_{f_\be}(\bsv)$).

Given a border strip~$\langle k_1,\dots,k_r\rangle$, we define its associated \emph{motif} as a
sequence of $N-1$ $0$'s and $1$'s with the $1$'s occupying the positions~$K_1,\dots,K_{r-1}$
(recall that $K_i=\sum_{j=1}^ik_j$). For example, the motif associated to the border strip
$\langle 3,1,2,4,2\rangle$ in Fig.~\ref{fig.bs} is $(0,0,1,1,0,1,0,0,0,1,0)$. There is clearly a
one-to-one correspondence between motifs and border strips, since a motif determines the partial
sums $K_i$, which in turn yield the integers $k_i$ through the relation $k_i=K_{i}-K_{i-1}$
($i=1,\dots,r$, with $K_0;=0$ and $K_r=N$). To rewrite Eq.~\eqref{partition} as the partition
function of an equivalent vertex model, we note that
\begin{equation}\label{Kde}
  \sum_{i=1}^{r-1}\cE(K_i)=\sum_{j=1}^{N-1}\cE(j)\de(s_j,s_{j+1})\,,
\end{equation}
where $\bsv$ is \emph{any} bond vector whose associated tableau has
shape~$\langle k_1,\dots,k_r\rangle$ and the function $\de:(B\cup F)^2\to\{0,1\}$ is defined by
\begin{equation}
  \label{dedefA}
  \de(s,t)=
  \cases{
    0, &\(s<t \text{ or }s=t\in B\),\\
    1, &\(s>t \text{ or }s=t\in F\).
  }
\end{equation}
Indeed, $\de(s_j,s_{j+1})=1$ if and only if $j$ is the last box of one of the columns of the
border strip~$\langle k_1,\dots,k_r\rangle$ except the last one, i.e., if $j=K_1,\dots,K_{r-1}$.
We thus have
\begin{eqnarray}
  \fl
  \sum_{\bk\in{\mathcal P}_N}S_{\langle
  k_1,\ldots,k_r\rangle,0}(\bx,\by)&\,q^{\sum_{i=1}^{r-1}J\cE(K_i)}
  =\sum_{\bk\in{\mathcal P}_N}\sum_{\bsv\in\cT_0(\bk)}x_1^{\nu_1^b(\bsv)}\cdots
  x_m^{\nu_m^b(\bsv)}y_1^{\nu_1^f(\bsv)}\cdots
    y_n^{\nu_n^f(\bsv)}\,q^{\sum_{i=1}^{r-1}J\cE(K_i)}\nonumber\\
  \fl
  &=\sum_{\bk\in{\mathcal P}_N}\sum_{\bsv\in\cT_0(\bk)}x_1^{\nu_1^b(\bsv)}\cdots
  x_m^{\nu_m^b(\bsv)}y_1^{\nu_1^f(\bsv)}\cdots
    y_n^{\nu_n^f(\bsv)}\,q^{J\sum_{j=1}^{N-1}\cE(j)\de(s_j,s_{j+1})}\nonumber\\
  \fl
  &=\sum_{\substack{\bsv\in(B\cup F)^N\cr s_N\le m_\vep+n_{\vep'}}}x_1^{\nu_1^b(\bsv)}\cdots
    x_m^{\nu_m^b(\bsv)}y_1^{\nu_1^f(\bsv)}\cdots
      y_n^{\nu_n^f(\bsv)}\,q^{J\sum_{j=1}^{N-1}\cE(j)\de(s_j,s_{j+1})}\,.
      \label{S0id}
\end{eqnarray}
A similar calculation shows that
\begin{eqnarray}
  \fl
  \sum_{\bk\in{\mathcal P}_N}S_{\langle
  k_1,\ldots,k_r\rangle,1}(\bx,\by)\,&q^{J\cE(N)}
  q^{\sum_{i=1}^{r-1}J\cE(K_i)}\nonumber\\
  \fl
   &=\sum_{\substack{\bsv\in(B\cup F)^N\cr
     s_N>m_\vep+n_{\vep'}}}x_1^{\nu_1^b(\bsv)}\cdots x_m^{\nu_m^b(\bsv)}y_1^{\nu_1^f(\bsv)}\cdots
    y_n^{\nu_n^f(\bsv)}\,q^{J\sum_{j=1}^{N}\cE(j)\de(s_j,s_{j+1})}\,,
      \label{S1id}
\end{eqnarray}
where we have used the fact that $K_r=N$. From Eqs.~\eqref{partition} and
\eqref{S0id}-\eqref{S1id} it then follows that
\begin{eqnarray}
  \fl
  \cZ_N(q;\bx,\by)
  &=\sum_{\substack{\bsv\in(B\cup F)^N\cr s_N\le
    m_\vep+n_{\vep'}}}x_1^{\nu_1^b(\bsv)}\cdots x_m^{\nu_m^b(\bsv)}y_1^{\nu_1^f(\bsv)}\cdots
                        y_n^{\nu_n^f(\bsv)}\,q^{J\sum_{j=1}^{N-1}\cE(j)\de(s_j,s_{j+1})}\nonumber\\
  \fl
  &\hphantom{=\sum}+ \sum_{\substack{\bsv\in(B\cup
    F)^N\cr s_N>m_\vep+n_{\vep'}}}x_1^{\nu_1^b(\bsv)}\cdots x_m^{\nu_m^b(\bsv)}y_1^{\nu_1^f(\bsv)}\cdots
    y_n^{\nu_n^f(\bsv)}\,q^{J\sum_{j=1}^{N}\cE(j)\de(s_j,s_{j+1})}\nonumber\\
  &=\sum_{\bsv\in(B\cup F)^N}x_1^{\nu_1^b(\bsv)}\cdots x_m^{\nu_m^b(\bsv)}y_1^{\nu_1^f(\bsv)}\cdots
    y_n^{\nu_n^f(\bsv)}\,q^{J\sum_{j=1}^{N}\cE(j)\de(s_j,s_{j+1})}\,,
    \label{cZid}
\end{eqnarray}
where we have set
\begin{equation}\label{sNp1}
  s_{N+1}=s^*:=m_\vep+n_{\vep'}+\frac12\,.
\end{equation}
We stress that the identity~\eqref{cZid} is valid for arbitrary values of the variables
$\bx\in\RR^m$, $\by\in\RR^n$. If we now specialize to the particular values in Eq.~\eqref{bxy} we
obtain the following formula for the partition function of the Hamiltonian $H=H_0+H_1$:
\begin{equation}
  \label{Zvertex}
  Z_N(T)=\sum_{\bsv\in(B\cup F)^N}q^{E_V(\bsv)},
\end{equation}
with
\begin{eqnarray}
  E_V(\bsv)&=J\sum_{j=1}^{N}\cE(j)\de(s_j,s_{j+1})-\sum_{\al=1}^m\nu_\al^b(\bsv)\mu_{b_\al}-
  \sum_{\be=1}^n\nu_\be^f(\bsv)\mu_{f_\be}\nonumber\\
  &=\sum_{j=1}^{N}\Big[J\cE(j)\de(s_j,s_{j+1})-\mu_{s_j}\Big]
    \label{Evertex}
\end{eqnarray}
and $s_{N+1}$ given by Eq.~\eqref{sNp1}. The right-hand side of Eq.~\eqref{Zvertex} is the
partition function of a classical vertex model with $N+2$ vertices whose $N+1$ bonds can
respectively take the values $s_1,\dots,s_N\in B\cup F$ and $s_{N+1}= s^*$. The first and last
vertices, which are connected to only one bond, have zero energy, while the $k$-th vertex (with
$k=2,\dots,N+1$) is assigned the energy $J\cE(k-1)\de(s_{k-1},s_k)-\mu_{s_{k-1}}$.
Equations~\eqref{Zvertex}-\eqref{Evertex} are the starting point of the transfer matrix method
used in the following section to evaluate the free energy of the open $\su(m|n)$-supersymmetric
Haldane--Shastry chain in the thermodynamic limit.

\section{Free energy in the thermodynamic limit}\label{sec.FETD}

\subsection{Thermodynamic limit}

In this section we shall obtain a closed-form expression for the thermodynamic limit of the
partition function (per spin) of the open $\su(m|n)$-supersymmetric Haldane--Shastry chain with
the chemical potential term $H_1$. To this end, we shall rely on the equivalence of the latter
model and the vertex model with energies~\eqref{Evertex} established in the previous section. In
order to proceed with the calculation, we first normalize the Hamiltonian $H_0$ so that the
average energy per spin $\langle H\rangle$ of the full Hamiltonian $H=H_0+H_1$ is finite in the
limit $N\to\infty$. We can easily compute $\langle H\rangle$ using Eq.~\eqref{Evertex}, with the
result
\begin{eqnarray*}
  \fl
  \langle H\rangle
  &=(m+n)^{-N}\sum_{\bsv\in( B\cup F)^N}E_V(\bsv)=(m+n)^{-N}\bigg[
    J(m+n)^{N-2}\sum_{j=1}^{N-1}\cE(j)\sum_{s_j,s_{j+1}\in B\cup
    F}\de(s_j,s_{j+1})\\
  \fl
  &\hphantom{=1}{}+
    J(m+n)^{-1}\cE(N)\sum_{s_N\in B\cup F}\de(s_N,s^*)
    -(m+n)^{N-1}\sum_{s_j\in B\cup F}\mu_{s_j}
    \bigg]\\
  \fl
  &=\frac{J}{(m+n)^2}\bigg[\frac12(m+n)(m+n-1)+n\bigg]\sum_{j=1}^{N-1}\cE(j)
    +\frac{J\cE(N)}{m+n}(m+n-m_\vep-n_{\vep'})\\
  \fl
  &\hphantom{=1}{}-\frac{N}{m+n}\sum_{\al}\mu_\al\\
  \fl
  &=\frac J{12}\bigg(1-\frac{m-n}{(m+n)^2}\bigg)N(N-1)(3\bbe+2N-1)
    +\frac{J}2\bigg(1-\frac{m_\vep+n_{\vep'}}{m+n}\bigg)N(2\bbe+N-1)\\
  \fl
  &\hphantom{=1}{}-\frac{N}{m+n}\sum_{\al}\mu_\al\,.
\end{eqnarray*}
Thus for $\langle H\rangle/N$ to tend to a finite limit as $N\to\infty$ we must take
\[
  J=\frac K{N^2}\,,
\]
where $K$ is a real (positive or negative) $N$-independent constant. Setting $x_i:=i/N$ (with
$i=1,\dots,N$) we can therefore write
\[
  J\cE(j)=K\vp(x_j)\,,
\]
with
\begin{equation}\label{vpx}
  \vp(x)=x\bigg(\ga_N-\frac x2\bigg)\,,\qquad \ga_N:=1+\frac{\bbe-1/2}N\,.
\end{equation}
Note that $\lim_{N\to\infty}\ga_N\ge1$ as $\bbe>0$.

In order to apply the transfer matrix method developed in Refs.~\cite{EFG10,FGLR18}, we rewrite
Eq.~\eqref{Evertex} in the equivalent form
\begin{eqnarray*}
  E_V(\bsv)=\sum_{j=1}^{N-1}\bigg[K\vp(x_j)\de(s_j,s_{j+1})&-\frac12\,(\mu_{s_j}+\mu_{s_{j+1}})\bigg]\\
  &+K(\ga_N-1/2)\de(s_N,s^*)-\frac12\,(\mu_{s_1}+\mu_{s_{N}})\,.
\end{eqnarray*}
Introducing the $(m+n)\times(m+n)$ matrices $A(x)$ and $B$ with entries
\begin{equation}\label{ABalbe}
  \fl
  A_{\al\be}(x)=q^{K\vp(x)\de(\al,\be)-\frac12\,(\mu_\al+\mu_\be)}\,,\qquad
  B_{\al\be}=q^{K(\ga_N-1/2)\de(\al,s^*)-\frac12\,(\mu_\al+\mu_\be)}
\end{equation}
and using Eq.~\eqref{Zvertex} we can express the partition function $Z_N(T)$ as
\[
  Z_N(T)=\tr\big[A(x_1)\cdots A(x_{N-1})B\big]\,.
\]
To pass to the thermodynamic limit, we note that all the entries of the matrix $A(x)$ are strictly
positive. By the Perron--Frobenius theorem, the latter matrix has a simple positive eigenvalue
$\la_1(x)$ satisfying
\begin{equation}\label{la1laal}
  \la_1(x)>|\la_\al(x)|\,,\qquad\all\al>1,
\end{equation}
where the $\la_\al(x)$ are the other eigenvalues of $A(x)$. Let $J(x)$ be the Jordan canonical
form of $A(x)$, and let $R(x)$ be the corresponding invertible matrix such that
\[
  A(x)=R(x)J(x)R(x)^{-1}.
\]
The matrix $R(x)$ is of course not unique, but it should be chosen in such a way that it depends
smoothly on the variable $x\in[0,1]$ and so that its first column is an eigenvector of $A(x)$ for
the Perron--Frobenius eigenvalue $\la_1(x)$. Calling $R_i:=R(x_i)$ we have
\[
  \fl Z_N(T)=\tr\Big[J(x_1)(R_1^{-1}R_2)J(x_2)(R_2^{-1}R_3)\cdots
  (R_{N-2}^{-1}R_{N-1})J(x_{N-1})R_{N-1}^{-1}BR_1\Big].
\]
From the smoothness of $R(x)$ it follows that
\[
  R_i^{-1}R_{i+1}=R(x_i)^{-1}R(x_i+1/N)=\id +O(N^{-1}),
\]
and hence
\[
  \log Z_N(T)=\log\tr\Big[J(x_1)\cdots J(x_{N-1})R_{N-1}^{-1}BR_1\Big]+O(1)\,.
\]
If the product $J(x_1)\cdots J(x_{N-1})$ is diagonal we can write
\[
  \tr\Big[J(x_1)\cdots
  J(x_{N-1})R_{N-1}^{-1}BR_1\Big]=\sum_{\al}c_\al\prod_{i=1}^{N-1}\la_\al(x_i),
\]
with $c_{\al}:=(R_{N-1}^{-1}BR_1)_{\al\al}$. When $N\to\infty$ we have
\[
  \frac1N\log\frac{\prod_{i=1}^{N-1}|\la_\al(x_i)|}{\prod_{i=1}^{N-1}\la_1(x_i)}
  =\frac1N\sum_{i=1}^{N-1}\log\bigg(\frac{|\la_\al(x_i)|}{\la_1(x_i)}\bigg)\to
  \int_0^1\log\bigg(\frac{|\la_\al(x)|}{\la_1(x)}\bigg)\,\diff x\,.
\]
Since the integrand is negative everywhere for all $\al>1$ on account of Eq.~\eqref{la1laal}, it
follows that
\[
  \frac{\prod_{i=1}^{N-1}|\la_\al(x_i)|}{\prod_{i=1}^{N-1}\la_1(x_i)}=O(\e^{-CN})\,,
  \qquad\al>1\,,
\]
with $C$ a positive constant. Hence
\begin{eqnarray*}
  \fl
  \frac{\log Z_N(T)}N&=\frac1N\,\sum_{i=1}^{N-1}\log\la_1(x_i)+\frac1N\,\log\left(
    c_1+\sum_{\al>1}c_\al\frac{\prod_{i=1}^{N-1}\la_\al(x_i)}{\prod_{i=1}^{N-1}\la_1(x_i)}
                     \right)+O(1/N)\\
  \fl
  &=\frac1N\,\sum_{i=1}^{N-1}\log\la_1(x_i)+O(1/N)\to \int_0^1\log\la_1(x)\,\diff x\,,
\end{eqnarray*}
provided that
\begin{equation}\label{c1cond}
  \lim_{N\to\infty}c_1=\left(R(1)^{-1}BR(0)\right)_{11}\ne0\,.
\end{equation}
It can be shown that the latter condition is always satisfied (see Remark~\ref{rem.cond} below).
We thus conclude that if the product $J(x_1)\cdots J(x_{N-1})$ is diagonal for sufficiently large
$N$, the free energy per spin in the thermodynamic limit $N\to\infty$ is given by
\begin{equation}\label{ffinal}
  f(T)=-T\lim_{N\to\infty}\frac{\log Z_N(T)}{N}=-T\int_0^1\log\la_1(x)\,\diff x\,,
\end{equation}
where $\la_1(x)$ is the Perron--Frobenius eigenvalue of the matrix $A(x)$ in Eq.~\eqref{ABalbe}
with $\ga_N$ replaced by $\ga:=\lim_{N\to\infty}\ga_N$. Equation~\eqref{ffinal}, from which all
other thermodynamic functions can be easily obtained, is one of the main results of the present
work.

\begin{remark}\label{rem.cond}
  We shall next prove that condition~\eqref{c1cond} is automatically satisfied. Indeed, since by
  construction the numbers $R(x)_{\be1}$ are the components of a Perron--Frobenius eigenvector of
  $A(x)$, by the Perron--Frobenius theorem their sign is independent of $\be$. (In fact, it is
  easy to show that $R(0)_{\be1}$ is proportional to $q^{-\mu_\be/2}$.)
  On the other hand, from the definition of the matrix $R$ it follows that $R^{-1}A=JR^{-1}$
  (where we have omitted, for simplicity, the argument $x$ of the matrices), and thus
  \[
    \sum_\al\big(R^{-1}\big)_{1\al}A_{\al\be}=\sum_\al J_{1\al}\big(R^{-1}\big)_{\al\be}
    =\la_1\big(R^{-1}\big)_{1\be}\,.
  \]
  In other words, the vector with components $\big[R^{-1}(x)\big]_{1\al}$ is an eigenvector of the
  matrix $A(x)^T$ with eigenvalue $\la_1(x)$. Since $A(x)$ and its transpose have the same
  eigenvalues, $\la_1(x)$ is also the Perron--Frobenius eigenvalue of $A(x)^T$. By the
  Perron--Frobenius theorem, all the
  components $\big[R^{-1}(x)\big]_{1\al}$ of its corresponding eigenvector have the same sign.
  Since
  \[
    \left(R(1)^{-1}BR(0)\right)_{11}=\sum_{\al,\be}\big[R^{-1}(1)\big]_{1\al}B_{\al\be}R_{\be1}(0)\,,
  \]
  where $B_{\al\be}>0$ and the sign of $\big[R^{-1}(1)\big]_{1\al}$ and $R_{\be1}(0)$ is
  respectively independent of $\al$ and $\be$, condition~\eqref{c1cond} is trivially satisfied.
\end{remark}

\subsection{General properties of the free energy}\label{sec.genprop}

Let us next discuss several basic properties of the thermodynamic free energy~$f$ that shall be
used to simplify the discussion in the following sections. To begin with, note that $f$ depends
only on the Perron--Frobenius eigenvalue of the matrix $A(x)$ in Eq.~\eqref{ABalbe}. The latter
matrix, in turn, could in principle depend on the parameters $\vep$ and $\vep'$ through the
function $\de$ in Eq.~\eqref{dedefA}, since our analysis requires that the spin degrees of freedom
be ordered so that Eq.~\eqref{order} is satisfied. The value of $f$, however, cannot depend on
this ordering provided that the latter equation holds. In particular, since the
ordering~\eqref{ord1} (resp.~\eqref{ord2}) is independent of $\vep$ (resp.~$\vep'$), $f$ cannot
depend on $\vep$ (resp.~$\vep'$). This shows that, unlike the partition function $Z_N$, the
thermodynamic free energy is independent of the parameters $\vep$ and $\vep'$.

Let us next study the behavior of $f$ under the exchange of bosonic and fermionic degrees of
freedom. To this end, following Refs.~\cite{BBHS07,BFGR09} let us denote by
$U:\cS^{(m|n)}\to\cS^{(n|m)}$ the unitary operator defined by
\[
  U\ket{s_1,\dots,s_N}=(-1)^{\sum_{k=1}^N kp(s_k)}\ket{s_1',\dots,s_N'}\,,
\]
where the prime is defined by (for instance) $b_\al'=f_\al$, $f_\be'=b_\be$. A straightforward
calculation shows that
\begin{eqnarray*}
  \fl
  &US_{ij}^{(m|n)}U^{-1}=-S_{ij}^{(n|m)}\,,\qquad
    US_{i}^{(m\vep|n\vep')}U^{-1}=S_{i}^{(n\vep'|m\vep)}=
    -S_i^{(n,-\vep'|m,-\vep)}\,,\\
  &U\tS_{ij}^{(m\vep|n\vep')}U^{-1}=-\tS_{ij}^{(n\vep'|m\vep)}=-\tS_{ij}^{(n,-\vep'|m,-\vep)}\,,
\end{eqnarray*}
and therefore
\[
  U H_0^{(m\vep|n\vep')}U^{-1}=C-H_0^{(n,-\vep'|m,-\vep)}\,,
\]
where
\[
  C=\frac J2\sum_{i<j}\bigg(\sin^{-2} \theta_{ij}^{-} 
  +\sin^{2} \theta_{ij}^{+} \bigg)
  +\frac{J}{4}\sum_i\left(\frac{\be_1}{\sin^{2} \theta_i} + \frac{\be_2}{\cos^2
      \theta_i}\right)\,.
\]
On the other hand, we obviously have
\[
  U\cN_\al U^{-1}=\cN_{\al'}\,,
\]
and hence (since the prime operator is an involution)
\[
  UH_1U^{-1}=\sum_\al\mu_\al\cN_{\al'}=\sum_\al\mu_{\al'}\cN_{\al}\,.
\]
Combining the previous equations we obtain the relation
\[
  UH^{(m\vep|n\vep')}[\bmu,K]U'=C+H^{(n,-\vep'|m,-\vep)}[\bmu',-K]\,,
\]
where $\bmu=(\mu_1,\dots,\mu_{m+n})$, $\bmu'=(\mu_{1'},\dots,\mu_{(m+n)'})$, and the arguments of
the Hamiltonians denote the values of the parameters (chemical potential and interaction strength)
on which they depend. From the previous identity it follows that the corresponding free energies
per spin are related by
\[
  f^{(m|n)}[\bmu,K]=\lim_{N\to\infty}\frac{C}N+f^{(n|m)}[\bmu',-K]\,.
\]
The constant term can be easily evaluated by noting that $C$ is the maximum energy of the
fermionic Hamiltonian $H_0^{(0|n,-1)}$, and thus
\[
  C=J\sum_{i=1}^N\cE(i)=K\sum_{i=1}^N\vp(x_i)\,,
\]
whence it follows that
\begin{equation}\label{f0def}
  f_0:=\lim_{N\to\infty}\frac{C}N=K\int_0^1\vp(x)\diff x=K\bigg(\frac\ga2-\frac16\bigg)\,.
\end{equation}
(Note that $f_0\ge K/3$, since $\ga\ge1$.) We can thus write
\begin{equation}
  \label{fmnnm}
  f^{(m|n)}[\bmu,K]=f_0+f^{(n|m)}[\bmu',-K]\,.
\end{equation}
From the latter relation it follows that we can restrict ourselves w.l.o.g.\ either to the case
$m\le n$ with arbitrary $K\ne0$, or to the case $K>0$ with arbitrary $m$, $n$. In particular, for
$m=n$ we can always assume that $K>0$.

It is also worth mentioning that in the genuinely supersymmetric case $mn\ne0$ the matrix $A(x)$
always has a zero eigenvalue. Indeed, using (for instance) the ordering~\eqref{ord1} if
$n_{\vep'}<n$ we have $m+n_{\vep'}\in B, m+n_{\vep'}+1\in F$, so that
\[
  \de(\al,m+n_{\vep'})=\de(\al,m+n_{\vep'}+1)=\cases{0,&\(1\le \al\le m+n_{\vep'}\)\\
  1, & \(m+n_{\vep'}<\al\le m+n\),}
\]
and therefore
\[
A_{\al,m+n_{\vep'}+1}(x)=q^{\frac12(\mu_{m+n_{\vep'}}-\mu_{m+n_{\vep'}+1})}A_{\al,m+n_{\vep'}}(x)\,.
\]
A similar argument shows that $A_{1\al}(x)=q^{K\vp(x)+\frac12(\mu_{m+n}-\mu_1)}A_{m+n,\al}(x)$ if
$n_{\vep'}=n$ (i.e., if $n=\vep'=1$). Thus, for fixed $m+n$ the genuinely supersymmetric models are
easier to treat than their non-supersymmetric counterparts.

\subsection{Discussion}


Equation~\eqref{ffinal} in the previous section makes it possible to compute the thermodynamic
free energy in closed form if the Perron--Frobenius eigenvalue $\la_1(x)$ of the transfer matrix
$A(x)$ can be explicitly determined and the product $J(x_1)\cdots J(x_{N-1})$ is diagonal for
sufficiently large $N$. Since, as remarked above, the transfer matrix always has a zero eigenvalue
in the truly supersymmetric case $mn\ne0$, its eigenvalues can be easily found in closed form in
the cases $(m,n)=(1,1), (2,0), (0,2), (2,1)$ and $(1,2)$. Furthermore, it can be readily shown
that in all of these cases $A(x)$ is diagonalizable for $x>0$, so that the method of the previous
section can be applied. It turns out that when $(m,n)=(2,2)$ the zero eigenvalue of the transfer
matrix has algebraic multiplicity $2$, and the product $J(x_1)\cdots J(x_{N-1})$ is diagonal for
$N\ge4$. Thus the Perron--Frobenius eigenvalue $\la_1(x)$ can also be explicitly found in this
case.

We list in Table~\ref{tab.la1mu} the expressions of the Perron--Frobenius eigenvalue $\la_1(x)$
for each of the tractable cases mentioned in the previous paragraph, for which $m,n\le2$. Note
that, as explained in Section~\ref{sec.genprop}, we can take w.l.o.g.\ $K>0$. To simplify the
ensuing formulas, we shall accordingly choose units so that $K=1$. Note also that the condition
$\mu_\al=\mu_{\imath(\al)}$ implies that in the $\su(2|0)$ and $\su(0|2)$ cases we can set
w.l.o.g.\ all the chemical potentials to zero, while in the remaining cases mentioned above only
one chemical potential is needed, which shall always be taken as the one for the fermions.
Moreover, although as explained above the thermodynamic free energy is independent of $\vep$ and
$\vep'$, for definiteness we shall order in all cases the internal degrees of freedom following
the convention~\eqref{ord1} with $\vep'=1$.
\begin{table}[t]
  \centering
  \begin{tabular}{|c|c|c|}
    \cline{2-3}
    \multicolumn{1}{c|}{}
    &$\mu_\al$&$\la_1(x)$\\[1mm]
    \hline
    \vrule width0pt height16pt depth9pt
    $\su(1|1)$& $\mu_1=\mu$, $\mu_2=0$&$1+\e^{-\be(\vp-\mu)}$\\[1mm]
    \hline
    \vrule width0pt height16pt depth9pt
    $\su(0|2)$& $\mu_1=\mu_2=0$& $\e^{-\be\vp/2}+\e^{-\be\vp}$\\[1mm]
    \hline
    \vrule width0pt height16pt depth9pt
    $\su(2|0)$& $\mu_1=\mu_2=0$& $1+\e^{-\be\vp/2}$\\[1mm]
    \hline
    \vrule width0pt height16pt depth9pt
    $\su(1|2)$& $\mu_1=\mu_3=\mu,\mu_2=0$& $\frac12+\e^{-\be(\vp-\mu)}+
                                           \sqrt{\e^{-\be(\vp-\mu)}
                                           +\e^{-\be(\vp-2\mu)}+\frac14}$\\[1mm]
    \hline
    \vrule width0pt height16pt depth9pt
    $\su(2|1)$& $\mu_1=\mu,\mu_2=\mu_3=0$&$1+\frac12\e^{-\be(\vp-\mu)}+
                                           \sqrt{\e^{-\be\vp}+\e^{-\be(\vp-\mu)}
                                           +\frac14\e^{-2\be(\vp-\mu)}}$\\[1mm]
    \hline
    \vrule width0pt height16pt depth9pt
    $\su(2|2)$& $\mu_1=\mu_4=\mu,\mu_2=\mu_3=0$& $\Big(1+\e^{-\be\vp/2}\Big)
                                                 \Big(1+\e^{-\be(\vp/2-\mu)}\Big)$\\[1mm]
    \hline                                                 
  \end{tabular}
  \caption{Chemical potentials and Perron--Frobenius eigenvalue for the (nontrivial) $\su(m|n)$ HS
    chains of $BC_N$ type with $m,n\le2$.}
  \label{tab.la1mu}
\end{table}
We list in Table~\ref{tab.la1mu} the chemical potentials and Perron--Frobenius eigenvalues of the
(nontrivial) $\su(m|n)$ HS chains of $BC_N$ type with $m,n\le 2$, whose thermodynamic free energy
can be computed in closed form through Eq.~\eqref{ffinal}. A cursory inspection of the latter
table evinces the following facts:
\begin{enumerate}[1.]
\item The free energy of the $\su(1|1)$ HS chain of $BC_N$ type is that of a system of free
  fermions with momentum $\pm\pi x$ and dispersion function $\vp(|p|/\pi)$.
  
\item The free energy of the purely bosonic $\su(2|0)$ model is that of a free fermion system with
  dispersion function $\vp(|p|/\pi)/2$ and zero chemical potential. Moreover, the $\su(2|0)$ and
  $\su(0|2)$ models are in fact equivalent, since their thermodynamic free energies differ by a
  constant:
  \begin{equation}\label{f0220}
    f^{(0|2)}=\frac12\int_0^1\vp(x)\diff x+f^{(2|0)}=\frac{f_0}2+f^{(2|0)}\,,
  \end{equation}
  with $f_0$ given by Eq.~\eqref{f0def}. For this reason, in what follows we shall only consider
  the $\su(0|2)$ case.
\item As is well known, the thermodynamic free energy of the inhomogeneous (classical)
  one-dimensional Ising model with Hamiltonian
  \[
    H=\sum_{i=1}^NJ(i/N)\si_i\si_{i+1}\,,\qquad \si_i\in\{\pm1\}\,,\quad\si_{N+1}=\si_1\,,
  \]
  is given by~\cite{Mu10}
  \[
    f(T)=-T\int_0^1\log\Bigl(2\cosh\bigl(\be J(x)\bigr)\Bigr)\diff x\,,
  \]
  where $J(x)$ is the continuum limit of $J(i/N)$. It follows that the $\su(1|1)$ and $\su(0|2)$
  chains are respectively equivalent (up to a trivial additive term) to a one-dimensional
  inhomogeneous Ising model with couplings $J(x)=(\vp(x)-\mu)/2$ and $J(x)=\vp(x)/4$.
\item The $\su(2|1)$ and $\su(1|2)$ free energies are related by
  \[
    f^{(2|1)}(T)=f_0-\mu-f^{(1|2)}(-T)\,,
  \]
  which is in fact a consequence of the general relation~\eqref{fmnnm}.
\item The $\su(2|2)$ free energy
  \begin{equation}\label{fsu22}
    f^{(2|2)}=-T\int_0^1\log\bigl(1+\e^{-\be\vp/2}\bigr)\diff
    x-T\int_0^1\log\bigl(1+\e^{-\be(\vp/2-\mu)}\bigr)\diff x
  \end{equation}
  is the sum of the free energies of two free fermion systems with dispersion function
  $\vp(|p|/\pi)/2$ and chemical potentials $0$ and $\mu$.
\item The free energy density~\eqref{ffinal} of each of the models listed in Table~\ref{tab.la1mu}
  satisfies the condition
  \[
    \lim_{T\to\infty}\frac{f(T)}T=-\log(m+n)\,,
  \]
  which is a direct consequence of the definition of the partition function $Z_N$.
\end{enumerate}

\begin{figure}[t]
  \includegraphics[width=.48\textwidth]{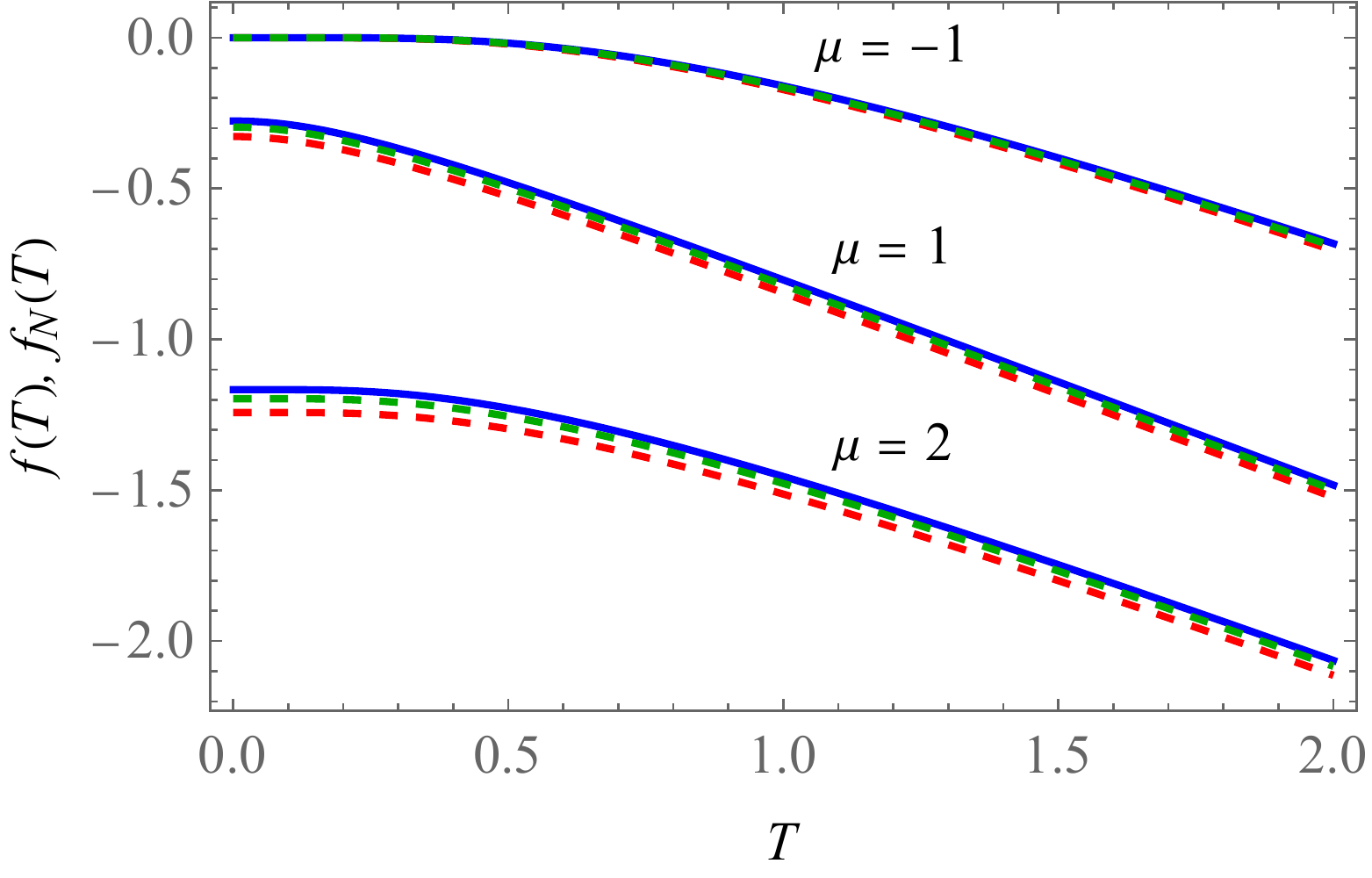}\hfill
  \includegraphics[width=.48\textwidth]{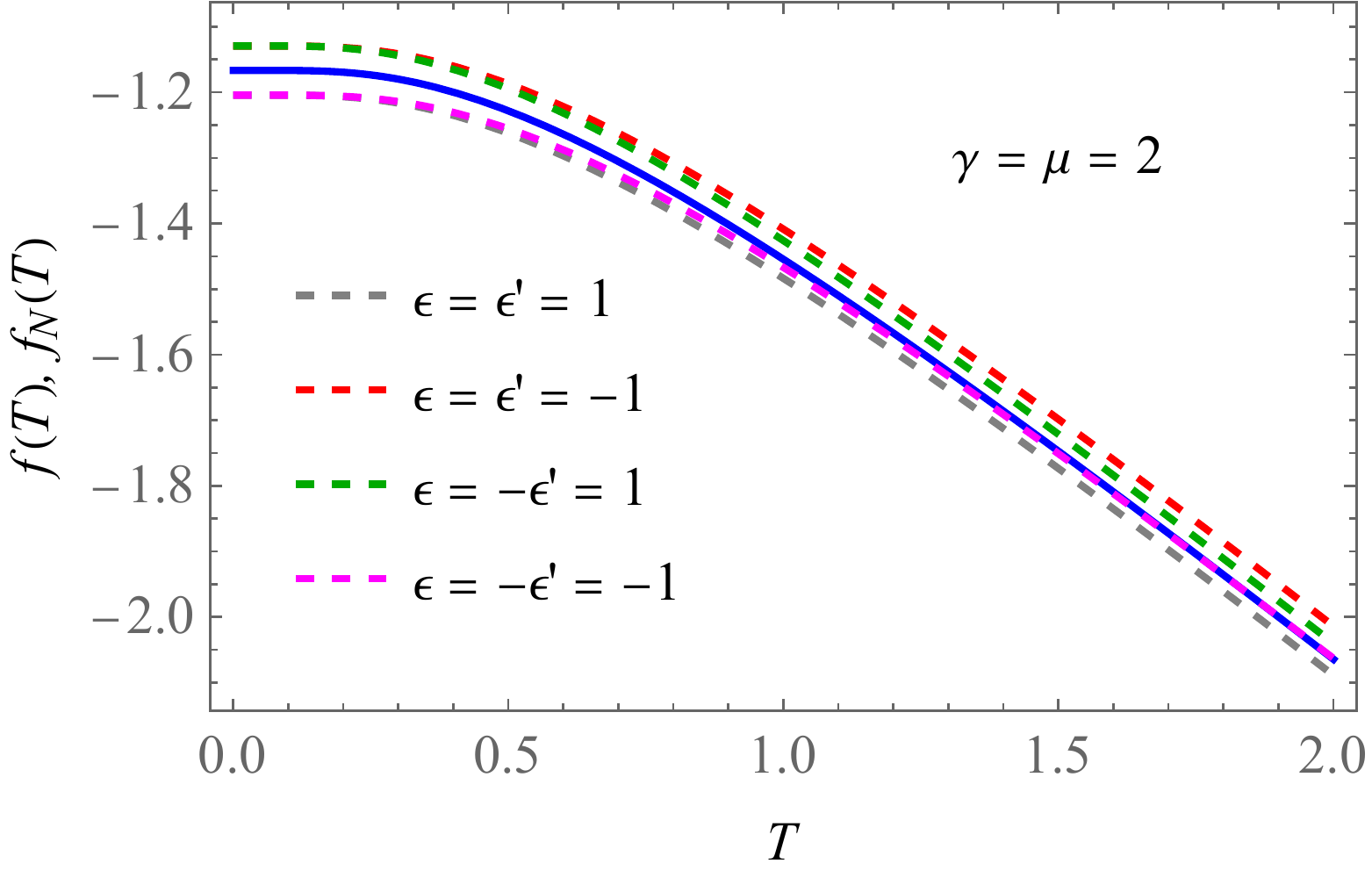}\hfill
  \caption{Left: thermodynamic free energy $f$ of the $\su(1|1)$ chain with $\ga=2$,
    $\vep=\vep'=1$ and $\mu=-1,1,2$ (solid blue lines) compared to their finite counterparts $f_N$
    for $N=10$ and $N=25$ spins (dashed red and green lines, respectively). Right: similar
    comparison for $N=20$ and $\ga=\mu=2$ and all values of $\vep,\vep'\in\{\pm1\}$.}
  \label{fig.ffN}
\end{figure}

\begin{remark}
  The thermodynamic free energy $f$, given by the closed formula Eq.~\eqref{ffinal} for the
  tractable cases listed in Table~\ref{tab.la1mu}, provides a reasonably accurate approximation to
  its counterpart $f_N:=-(T/N)\log Z_N$ even when the number of spins $N$ is relatively low
  ($\sim 10$). To illustrate this fact, we show in Fig.~\ref{fig.ffN} (left) the plots of $f$ and
  $f_N$ for the $\su(1|1)$ chain (with $\vep=\vep'=1$) for $\ga=2$, $N=10,25$ and several values
  of $\mu$. In fact, as remarked above, for large $N$ the dependence of $f_N$ on $\vep$ and
  $\vep'$ is expected to be negligible, since $f$ is independent of the latter parameters. This is
  shown in Fig.~\ref{fig.ffN} (right) for the $\su(1|1)$ chain with $\ga=\mu=2$ and $N=20$ (for
  larger values of $N$ the plots of $f$ and the four types of $f_N$'s become indistinguishable).
\end{remark}

\section{Ground states and low energy excitations}\label{sec.GSlow}

In this section we shall determine the ground state as a function of the (fermionic) chemical
potential $\mu$ for the tractable cases listed in Table~\ref{tab.la1mu}, and compute the zero
temperature fermion and energy densities. We shall also find the infinitesimal energy excitations
above the ground state in the gapless regime, and evaluate their Fermi velocities. These results
will then be applied in the analysis of the critical behavior of the models under study performed
in the next section. The discussion will be based on Eq.~\eqref{Evertex} for the energy spectrum,
which in the present context can be written as
\begin{equation}
  \label{motifvp}
  E(\bsv)=\sum_{i=1}^{N}\vp(x_i)\de(s_i,s_{i+1})-\mu N_f(\bsv)\,,
\end{equation}
where $N_f(\bsv)$ is the number of fermionic components in the bond vector $\bsv$ labeling an
energy eigenstate.
%
%
Although, as explained above, the thermodynamic free energy is independent of the values of the
parameters $\vep$ and $\vep'$, the same is not true for the ground state. It can be shown,
however, that the qualitative features we shall be interested in (namely, the ground state
degeneracy and the nature of the low energy excitations) are essentially the same for all values
of $\vep$ and $\vep'$. For definiteness, we shall restrict ourselves in what follows to the case
$\vep=\vep'=1$ and use the order~\eqref{ord1}.

\subsection{$\su(1|1)$}

In this case $F=\{1\}$, $B=\{2\}$, $m_\vep=n_{\vep'}=1$, so that $s_{N+1}=s_*=5/2$
and $\de(s_N,s_{N+1})=0$. Obviously, if $\mu<0$ the ground state is labeled by the purely
bosonic bond vector\footnote{We conjecture that, just as seems to be the case for the $A_{N-1}$
  Haldane--Shastry chain, the particle content of a state labeled by a bond vector $\bsv$
  coincides with that of the vector $\bsv$ itself.}
\begin{equation}\label{sgb}
  \bsv_g=(2,\dots,2)\,.
\end{equation}
Likewise, if $\mu>\vpmax:=\vp(1)$ the ground state is the purely fermionic one
with bond vector
\begin{equation}\label{sgf}
  \bsv_g=(1,\dots,1)\,.
\end{equation}

Let us next consider the interval $0<\mu<\vpmax$. The state with minimum energy in the subspace
labeled by bond vectors with $k$ fermionic variables (i.e., $k$ $1$'s) has bond vector
\begin{equation}\label{sgcrit}
  \bsv_g=(1,\dots,\underset {\overset\downarrow k}1,2,\dots,2)\,.
\end{equation}
Since the increase in energy caused by replacing the first $2$ by a $1$ in the above bond vector
is $\vp(x_{k})-\mu$, the number of fermions $k_0$ in the ground state is given by\footnote{For the
  sake of simplicity, we shall suppose in the following discussion that
  $\mu\notin\vp(\QQ\cup[0,1])$.}
\begin{equation}\label{k0}
  k_0=\min\{k:\vp(x_{k})>\mu\}=\lceil Nx_0(\mu)\rceil\,,
\end{equation}
where $x_0$ is the inverse function of $\vp:[0,1]\to\RR$:
\begin{equation}
  \label{x0mu}
  x_0(\mu)=\ga-\sqrt{\ga^2-2\mu}\,.
\end{equation}
Note that the condition $0<\mu<\vpmax$ implies that $k_0\in(0,1)$. In the thermodynamic limit
(more precisely, up to terms of order $N^{-1}$) we have
\[
  \mu\simeq\vp(x_{k_0})\en\implies\en n_F(0)=\frac{k_0}N=x_{k_0}=x_0(\mu)\,,
\]
where $n_F(0)$ denotes the fermion density at zero temperature. Note, in particular, that in this
case the ground state is non-degenerate. The zero temperature energy density $u(0)$ is also easily
computed:
\begin{eqnarray}
  \fl
  u(0)&=\lim_{N\to\infty}\frac{E(\bsv_g)}N=\lim_{N\to\infty}\left[\frac1N\sum_{i=1}^{k_0-1}\vp(x_i)
        -\frac{k_0}N\,\mu\right]=\int_{0}^{x_0(\mu)}\big(\vp(x)-\mu\big)\diff x\label{u01}\\
  \fl
      &=\frac13\Big[\ga(\ga^2-3\mu)-(\ga^2-2\mu)^{3/2}\Big]\,,
        \label{u0}
\end{eqnarray}
where we have used the previous equation for $x_{k_0}=k_0/N$ in the thermodynamic limit.
Obviously, for $\mu<0$ we have $u(0)=0$ (since $E(\bsv_g)=0$), and for $\mu>\vpmax$
\begin{equation}\label{u0su11}
  u(0)=\lim_{N\to\infty}\left[\frac1N\sum_{i=1}^{N-1}\vp(x_i)
    -\mu\right]=\int_{0}^{1}\big(\vp(x)-\mu\big)\diff x=f_0-\mu\,.
\end{equation}
Note that the expressions for $u(0)$ in the non-critical regions $\mu<0$ and $\mu>\vpmax$ can be
obtained from Eq.~\eqref{u01} by setting $x_0(\mu)=0$ for $\mu<0$ and $x_0(\mu)=1$ for
$\mu>\vpmax$. With this convention, the fermion density $n_F(0)$ is given by the simple expression
\begin{equation}\label{nF0}
  n_F(0)=x_0(\mu)\,.
\end{equation}
The limiting cases $\mu=0$ and $\mu=\vpmax$ can be dealt with in much the same way. For instance,
for $\mu=0$ there are two ground states, labeled by the bond vectors
\begin{equation}\label{sg11mu0}
  \bsv_g=(2,\dots,2)\,,\qquad \bsv_g'=(1,2,\dots,2)\,,
\end{equation}
with $n_F(0)=u(0)=0$, i.e., the $\mu\to0$ limits of the expressions~\eqref{u0}-\eqref{nF0}.
Likewise, when $\mu=\vpmax$ the ground state is still the purely fermionic one, with $n_F(0)$ and
$u(0)$ given by the $\mu\to1$ limit of the latter expressions. Note also that the expressions for
the zero temperature fermion and energy densities could also have been obtained by taking the
$\be\to\infty$ limit of
\[
  f(T)=-T\int_0^1\log\bigl(1+\e^{-\be(\vp(x)-\mu)}\bigr)\diff x
\]
(since $u(0)=f(0)$) and the corresponding formula for $n_F$:
\[
  n_F(T)=\int_0^1\frac{\diff x}{1+\e^{\be(\vp(x)-\mu)}}\,.
\]

Let us next discuss the existence of excitations of infinitesimal energy (as $N\to\infty$) above
the ground state, i.e., whether the spectrum is gapless or gapped. To begin with, for $\mu<0$ the
spectrum is clearly gapped with gap energy~$|\mu|$, since replacing a boson by a fermion in the
$i$-th position of the lowest energy bond vector~\eqref{sgb} increases the energy by
$\vp(x_{i-1})+|\mu|$. Similarly, if $\mu>\vpmax$ replacing a fermion by a boson in the $i$-th
position of the lowest energy bond vector~\eqref{sgf} raises the energy by $\mu-\vp(x_{i-1})$, and
hence the spectrum is gapped with energy gap $\mu-\vpmax$.

Consider next the interval $0<\mu<\vpmax$. In this case the lowest energy excitations above the
ground state are labeled by bond vectors~$\bsv$ obtained from $\bsv_g$ in equation~\eqref{sgcrit}
replacing a boson by a fermion, or a fermion by a boson, at the $(k_0+l)$-th position of the
latter vector, provided that $|l|\ll N$. In the former case
the increase in energy of these excitations is given by
\[
  \De E=\vp(x_{k_0+l-1})-\mu\simeq\vp(x_{k_0+l-1})-\vp(x_{k_0-1})\simeq\vp'(x_{k_0-1})\De x\,,
\]
with $l>0$ and
\[
\De x=x_{k_0+l-1}-x_{k_0-1}=\frac lN\,.
\]
Similarly, when a fermion is replaced by a boson we have
\[
\De E=\mu-\vp(x_{k_0+l-1})\simeq\vp(x_{k_0-1})-\vp(x_{k_0+l-1})\simeq\vp'(x_{k_0-1})\De x\,,
\]
with $l<0$ and
\[
  \De x=x_{k_0-1}-x_{k_0+l-1}=\frac{|l|}N\,.
\]
Hence in both cases the excitation energy is of order $N^{-1}$, and the spectrum is therefore
gapless. We shall use in what follows the term ``Fermi excitations'' to refer to these
excitations, since they are obtained by modifying the ground state bond vector near the border
(Fermi ``surface'') between its fermionic and bosonic sectors.

In order to determine the Fermi velocity of these excitations, we need to assign them a
(quasi)momentum. In the $A_{N-1}$ case, where momentum is conserved, it is well known that the
energy and momentum of the energy eigenstate labeled by a bond vector $\bsv$ is given by
\[
  E(\bsv)=\sum_{i=1}^{N-1}i(N-i)\de(s_i,s_{i+1})\,,\qquad
  P(\bsv)=2\pi\sum_{i=1}^{N-1}x_i\de(s_i,s_{i+1})\en\bmod{2\pi}
\]
(see, e.g., \cite{HHTBP92,BBS08}). In the $BC_N$ case the situation is of course different, since
momentum is not conserved. However, motivated by the $A_{N-1}$ result and Eq.~\eqref{motifvp}, we
shall use the following modification of the previous formula to define the effective momentum (or
quasimomentum) $P(\bsv)$ of an energy eigenstate with bond vector~$\bsv$:
\begin{equation}\label{Ps}
  P(\bsv)=\pi\sum_{i=1}^{N}x_i\de(s_i,s_{i+1})\en\bmod{2\pi}\,.
\end{equation}
The missing factor of $2$ is justified by the fact that the dispersion relation~\eqref{vpx} of the
$BC_N$-type HS chain is not symmetric about the midpoint $x=1/2$, so that the interval
$0\le x\le 1$ corresponds only to the positive momentum\footnote{From now on, for the sake of
  conciseness we shall use the shorter term ``momentum'' instead of the more correct one
  ``quasimomentum''.} sector. In other words, the true dispersion relation in this case is the
function $\vp(|p|/\pi)$ with $-\pi\le p\le \pi$. With this definition, the momentum of the
infinitesimal excitations described above (added if $l>0$, or removed if $l<0$) is given
\[
  p=\pi x_{k_0+l-1}=\pi x_{k_0-1}+\frac{\pi l}N=p_0\pm\pi\De x\,,
\]
where the plus sign corresponds to the excitations increasing the fermion number and the Fermi
momentum
\[
  p_0:=\pi x_{k_0-1}
\]
is defined as the largest excited momentum $\pi x_i$ in Eq.~\eqref{Ps}. Calling
\[
  \De p=p-p_0=\pm\pi\De x\,,
\]
the velocity of the Fermi excitations, whose momentum is close to the Fermi momentum $p_0$, is
given by
\begin{equation}\label{vF11}
  v:=\lim_{N\to\infty}\bigg|\frac{\De E}{\De p}\bigg|=\lim_{N\to\infty}\vp'(x_{k_0-1})\bigg|\frac{\De
    x}{\De p}\bigg|=\frac{\vp'\bigl(x_0(\mu)\bigr)}\pi=\Big[\pi x_0'(\mu)\Big]^{-1}\,.
\end{equation}
Finally, the limiting cases $\mu=0$ and $\mu=\vpmax$ can be handled similarly. Indeed, if $\mu=0$
the low energy excitations arise when a component of the ground state bond vectors~\eqref{sg11mu0}
near their left end is modified. We shall refer in the sequel to these excitations, whose momentum
$\De p$ is $\Or(N^{-1})$ as $N\to\infty$, as small momentum excitations. For instance, if
\begin{equation}\label{s212}
  \bsv=(2,\dots,2,\underset{\overset\downarrow{l}}1,2,\dots,2)
\end{equation}
we have
\[
  \De p=\pi x_{l-1},\qquad \De E=\vp(x_{l-1})\simeq\vp'(0)x_{l-1}\,.
\]
The spectrum is thus gapless, with Fermi velocity
\begin{equation}\label{vF0}
  v=\big[\pi x'_0(0)\big]^{-1}\,.
\end{equation}
Likewise, if $\mu=\vpmax$ the low energy excitations are obtained by changing a component of the
bond vector~\eqref{sgf} near its right end. If $\ga>1$, proceeding as before we conclude that the
spectrum is again gapless, with Fermi velocity
\begin{equation}\label{vF1}
  v=\big[\pi x'_0(\vpmax)\big]^{-1}\,.
\end{equation}
In other words, both at $\mu=0$ and at $\mu=\vpmax$ (when $\ga>1$) the Fermi velocity is a
continuous function of the chemical potential $\mu$. The situation is quite different if $\ga=1$
and $\mu=\vpmax=\vp(1)$, since in this case $\vp'(1)=\ga-1=0$ implies that
$x_0'(\mu)=x_0'(\vpmax)=1/\vp'(1)=\infty$. The low energy excitations above the ground state,
obtained by replacing a fermion by a boson in the ($N-l+1)$-th component of the fermionic bond
vector $(1,\dots,1)$ (with $1\le l\ll N$), now carry an energy
\[
  \De E=\mu-\vp(x_{N-l})=\vp(1)-\vp\bigl(1-l/N\bigr)\simeq-\frac12\,\vp''(1)\frac{l^2}{N^2}
  =\frac{l^2}{2N^2}\,.
\]
The fact that $\De E$ is not linear in $\De p=-\pi l/N$ suggests that the system is not critical
when $\ga=1$ and $\mu=\vpmax$. This will indeed be confirmed in the next section by studying the
low temperature behavior of the free energy per spin.

\subsection{$\su(0|2)$}

In this case $\mu=0$, $s_*=3/2$ and $\de(s_N,s_{N+1})=0$ for $s_N=1$. It is easy to
verify that the ground state corresponds to the bond vector
\begin{equation}\label{sgodd}
  \bsv_g=(1,2,\dots,1,2,1)
\end{equation}
for odd $N$, or to the vectors
\begin{equation}\label{sgeven}
  \bsv_g=(2,1,\dots,2,1)\,,\qquad\bsv_g'=(1,1,2,1,\dots,2,1)
\end{equation}
for even $N$. In other words, the ground state bond vectors essentially consist of a sequence of
pairs $(1,2)$ or $(2,1)$. In either case, the ground state energy in the thermodynamic limit is
given by
\[
  u(0)=\frac12\int_0^1\vp(x)\diff x=\frac{f_0}2\,,
\]
since only the even (for odd $N$) or odd (for even $N$) components of the ground state bond
vectors give rise to a non-zero energy. Clearly the lowest energy excitations in this case,
obtained by changing a component of either ground state bond vector near its left end, are all
small momentum excitations. For instance, for odd $N$ such an excitation is labeled by the bond
vector
\[
  \bsv=(1,2,\dots,1,2,1,\underset {\overset\downarrow{2l}}1,1,2,\dots,1,2,1)\,,
\]
with
\[
  \De E=\vp(x_{2l-1})\simeq\vp'(0)x_{2l-1}\,.
\]
This is a low momentum excitation with $\De p=\pi x_{2l-1}$, and hence $v$ is given by
Eq.~\eqref{vF0}.

\subsection{$\su(1|2)$}

Now $F=\{1,3\}$, $B=\{2\}$, $m_\vep=n_{\vep'}=1$, and hence $s_{N+1}=s_*=5/2$ and
$\de(s_N,s_{N+1})=0$ for $s_N\le 2$. Obviously, if $\mu<0$ the ground state is the purely bosonic
one labeled by the bond vector~\eqref{sgb}. The energy spectrum is clearly gapped, with energy gap
$|\mu|$.

Consider next the interval $0<\mu<\vpmax/2$. In this case the ground state has bond vector
\begin{equation}\label{sg12}
  \bsv_g=(1,3,\dots,1,3,\underset{\mathclap{\overset{\downarrow}{2k_0+1}}}1,2,\cdots,2)
\end{equation}
with
\begin{equation}\label{k0su12}
  k_0:=\max\{k:\vp(x_{2k})<2\mu\}=\bigg\lfloor\frac{N}{2}x_0(2\mu)\bigg\rfloor\,,
\end{equation}
and Fermi momentum $p_0=\pi x_{2k_0}$. Indeed, adding a pair $(3,1)$ at positions $(2k,2k+1)$ in a
bond vector of the latter form increases the energy by $\vp(x_{2k})-2\mu$. In the thermodynamic
limit we have
\begin{equation}\label{vpx2k}
  \vp(x_{2k_0})\simeq2\mu\en\implies\en x_{2k_0}=\frac{2k_0}N\simeq x_0(2\mu),
\end{equation}
and the zero temperature fermion and energy densities are thus given by
\[
  \fl
  n_F(0)=x_0(2\mu)\,,\qquad u(0)=\int_{0}^{x_0(2\mu)}\bigg(\frac{\vp(x)}2-\mu\bigg)\diff x=
  \frac16\bigg[\ga(\ga^2-6\mu)-(\ga^2-4\mu)^{3/2}\bigg].
\]

The lowest energy excitations about the ground state~\eqref{sg12} are of two types, depending on
whether $\De N_f=0$ or $\pm2$. In the former case we simply replace a $1$ by a $3$ (or vice versa)
near the start of the bond vector~\eqref{sg12}. These small momentum excitations are gapless, with
Fermi velocity
\begin{equation}
  \label{Fv12}
  v_1=\big[\pi x_0'(0)\big]^{-1}\,.
\end{equation}
For instance, if
\[
  \bsv=(1,3,\dots,1,3,1,\underset{\overset\downarrow{2l}}1,1,3,\dots,1,3,1,2,\dots,2)\,,
\]
we have
\[
  \De E=\vp(x_{2l-1})\simeq\vp'(0)x_{2l-1}=\vp'(0)\,\frac{\De p}{\pi}=\frac{\De p}{\pi x'(0)}\,.
\]
The second type of low energy excitations are Fermi excitations obtained by replacing a pair
$(2,2)$ near the Fermi position $2k_0$ of the bond vector~\eqref{sg12} by the pair $(1,3)$ or
$(3,1)$, or alternatively replacing one of these pairs by a pair of bosons. These excitations are
also gapless, with Fermi velocity
\begin{equation}
  \label{Fv122}
  v_2=\left[\pi x_0'(2\mu)\right]^{-1}\,.
\end{equation}
For instance, if
\[
  \bsv=(1,3,\dots,1,3,1,2,\dots,2,\underset{\mathclap{\overset{\downarrow}{2k_0+l}}}3,1,
  2,\dots,2)
\]
with $l\ge2$ then
\[
  \fl
  \De E=\vp(x_{2k_0+l})-2\mu\simeq\vp(x_{2k_0+l})-\vp(x_{2k_0})\simeq\vp'(x_{2k_0})\frac{l}N\simeq
  \vp'\bigl(x_0(2\mu)\bigr)\frac{l}N=\frac{l}{Nx_0'(2\mu)}\,.
\]
Since the momentum of this excitation is given by
\[
  p=\pi x_{2k_0+l}=\pi x_{2k_0}+\frac{l\pi}N=p_0+\frac{l\pi}N
\]
we have $\De p=l\pi/N$, which immediately yields Eq.~\eqref{Fv122}. When $\ga>1$ a similar
analysis is valid in the limiting case $\mu=\vpmax/2$, in which the ground state bond vector is
$(1,3,\dots,1,3,1)$ for odd $N$ and $(1,3,\dots,1,3,1,2)$ for even $N$. On the other hand, for
$\ga=1$ and $\mu=\vpmax/2$ the Fermi excitations, coming from changes in the last components of
the ground state bond vectors, carry an energy proportional to $(\De p)^2$ instead of $\De p$. As
in the $\su(1|1)$ case, this is an indication that only the small momentum excitations are
critical in this case, with Fermi velocity given by Eq.~\eqref{Fv12}. Likewise, when $\mu=0$ the
ground state is twice degenerate, with bond vectors
\[
  \bsv_g=(1,2,\dots,2)\,,\qquad \bsv_g'=(2,2,\cdots,2)\,.
\]
The spectrum is still gapless, but the only low energy excitations are small momentum ones with
Fermi velocity given again by Eq.~\eqref{Fv12}.

Consider next the case $\mu>\vpmax/2$. Now the ground states are labeled by the purely fermionic
bond vectors~\eqref{sgodd} for odd $N$ and~\eqref{sgeven} for even $N$, with $2$ replaced by $3$.
Indeed, if $\vp(x)<2\mu$ for all $x\in[0,1]$ it is always energetically favorable to introduce a
fermionic pair $(1,3)$ or $(3,1)$. Thus in this case $n_F(0)=1$ and $u(0)=f_0/2$. The low energy
excitations above the ground state are obtained by changing a $1$ by a $3$, or vice versa, near
the start of the ground state bond vector(s). Thus the spectrum is again gapless, with only small
momentum excitations whose Fermi velocity is given by Eq.~\eqref{Fv12}.

\subsection{$\su(2|1)$}

\smallskip\noi In this case $F=\{1\}$, $B=\{2,3\}$, $m_\vep=n_{\vep'}=1$, and hence
$s_{N+1}=s_*=5/2$ and $\de(s_N,s_{N+1})=0$ for $s_N\le 2$. Since $\de(3,s_{N+1})=1$, for all
values of the chemical potential $\mu$ the ground state and the zero temperature densities
$n_F(0)$ and $u(0)$ are as in the $\su(1|1)$ case. When $\mu<0$ the only low energy excitations
above the bosonic ground state are small momentum excitations with bond vector
\[
  \bsv=(2,\dots,2,\underset{\overset\downarrow{l}}3,2,\dots,2),\qquad l\ll N\,,
\]
for which
\[
  \De E=\vp(x_l)\simeq\vp'(0)x_l\,,\qquad \De p=\pi x_l\,.
\]
Hence the spectrum is gapless, with Fermi velocity given by Eq.~\eqref{Fv12}. The same is true for
$\mu=0$, the only difference being that in this case there are additional small momentum
excitations with bond vector~\eqref{s212}. For $0<\mu<\vpmax$ the low energy excitations above the
ground state~\eqref{sgcrit}-\eqref{k0} are obtained, as in the $\su(1|1)$ case, replacing a $2$
boson by a fermion or a fermion by either boson near the Fermi position $k_0$ of the ground state
bond vector. Hence the spectrum is again gapless, with Fermi excitations whose velocity is given
by Eq.~\eqref{vF11}. Likewise, again as in the $\su(1|1)$ case, for $\mu=\vpmax$ and $\ga>1$ the
spectrum is still gapless, with Fermi velocity given by Eq.~\eqref{vF0}. On the other hand, for
$\mu=\vpmax$ and $\ga=1$ the low energy excitations have energy proportional to $(\De p)^2$, thus
signaling that although the spectrum is gapless the system is not critical. Finally, for
$\mu>\vpmax$ the spectrum is clearly gapped, with energy gap $\mu-\vpmax$.

\subsection{$\su(2|2)$}

\smallskip\noi Here $F=\{1,4\}$, $B=\{2,3\}$, $m_\vep=n_{\vep'}=1$, $s_*=5/2$ and hence
$\de(s_N,s_{N+1})=0$ for $s_N\le 2$. The ground state bond vectors are as in the $\su(1|2)$ case,
with $3$ replaced by $4$. In other words,
\[
  \bsv_g=\cases{(2,\dots,2)\,,& \(\mu<0\)\\
    (*,2,\dots,2)\,,& \(\mu=0\)\\
    (1,4,\dots,1,4,\underset{\mathclap{\overset{\downarrow}{2k_0+1}}}1,2,\cdots,2)\,,&
    \(0<\mu<\vpmax/2\)\\
    (1,4,\dots,1,4,1,2)\,,& \(\mu=\vpmax/2\) and \(N\) even\\
    (1,4,\dots,1,4,1)\,,& \(\mu\ge\vpmax/2\) and \(N\) odd\\
    (\dagger,1,4,1,\dots,4,1)\,,& \(\mu>\vpmax/2\) and \(N\) even\\
  }
\]
with $*=1,2$, $\dagger=1,4$ and $k_0$ given by Eq.~\eqref{k0su12}. As a consequence, the zero
temperature densities $n_F(0)$ and $u(0)$ are as in the $\su(1|2)$ case. For $\mu\le0$ the low
energy excitations are obtained replacing a $2$ near the left end of the ground state bond
vector(s) by a $3$ (and, if $\mu=0$, by a $1$ or a $4$). Hence the spectrum is gapless, with only
small momentum excitations of Fermi velocity given by Eq.~\eqref{Fv12}. When $0<\mu<\vpmax/2$ we
have similar small momentum excitations obtained by replacing a $4$ fermion by a $1$ fermion, or
vice versa, near the left end of the ground state bond vector(s). The Fermi velocity of these
excitations is again given by Eq.~\eqref{Fv12}. Moreover, in this case we also have Fermi
excitations obtained by adding or removing a pair $(1,4)$ or $(4,1)$ near the Fermi position
$k_0$, with velocity~\eqref{Fv122}. The analysis is very similar for the case $\mu=\vpmax/2$ and
$\ga>1$, with the second Fermi velocity given by Eq.~\eqref{vF1} ---i.e., the $\mu\to\vpmax/2$
limit of $v_2$ in Eq.~\eqref{Fv122}. On the other hand, for $\mu=\vpmax$ and $\ga=1$ the Fermi
excitations, coming from changes in the right end of the ground state bond vector, do not have the
correct energy/momentum relation. Hence in this case there are only small momentum excitations
with Fermi velocity given by Eq.~\eqref{Fv12}. Finally, if $\mu>\vpmax/2$ the only low energy
excitations are again the small momentum excitations with Fermi velocity~\eqref{Fv12}.

For future reference we display below the expression for the zero temperature densities
$n_F(0)$ and $u(0)$ in the tractable cases we have just discussed. To wit, in the $\su(0|2)$ case
we have
\[
  n_F(0)=1,\qquad u(0)=\frac{f_0}2\,,
\]
while in the remaining cases
\begin{equation}\label{nF0full}
  \fl
  n_F(0)=\cases{0\,,& \(\mu\le0\)\\
    x_0(n\mu)=\ga-\sqrt{\ga^2-2n\mu}\,,& \(0\le\mu\le\frac\vpmax n\)\\
    1\,,& \(\mu\ge\frac\vpmax n\)\,,
  }
\end{equation}
and
\[
\fl
  u(0)=\cases{0\,,& \(\mu\le0\)\\
  \int_0^{x_0(n\mu)}\bigg(\frac{\vp(x)}n-\mu\bigg)\diff x=
  \frac1{3n}\left[\ga(\ga^2-3n\mu)-(\ga^2-2n\mu)^{3/2}\right]\,,& \(0\le\mu\le\frac\vpmax n\)\\
  \frac{f_0}n-\mu\,,& \(\mu\ge\frac\vpmax n\)\,,}
\]
where $n$ denotes the number of fermionic degrees of freedom.
\begin{figure}[t]
  \includegraphics[width=.48\textwidth]{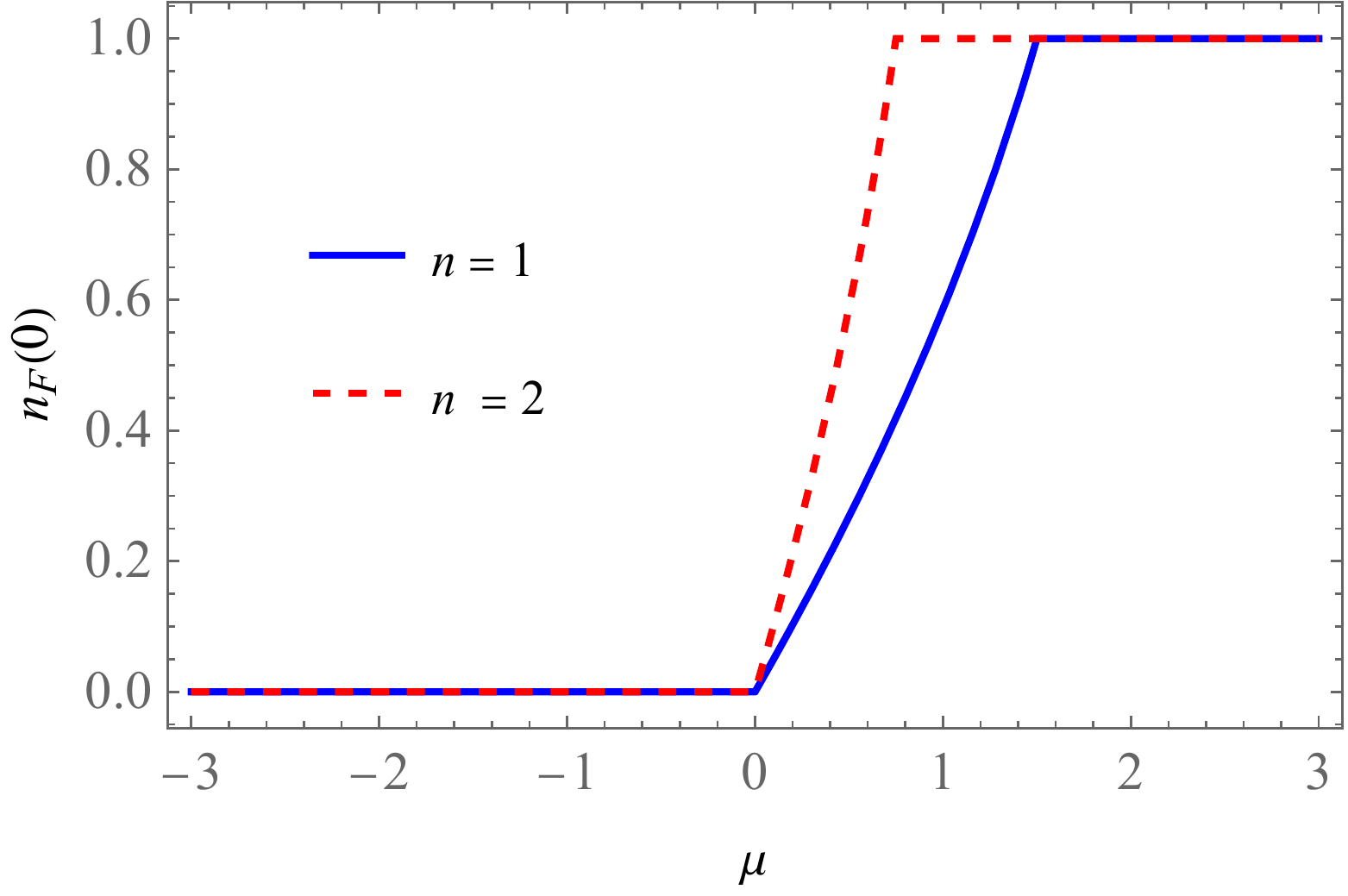}\hfill
  \includegraphics[width=.48\textwidth]{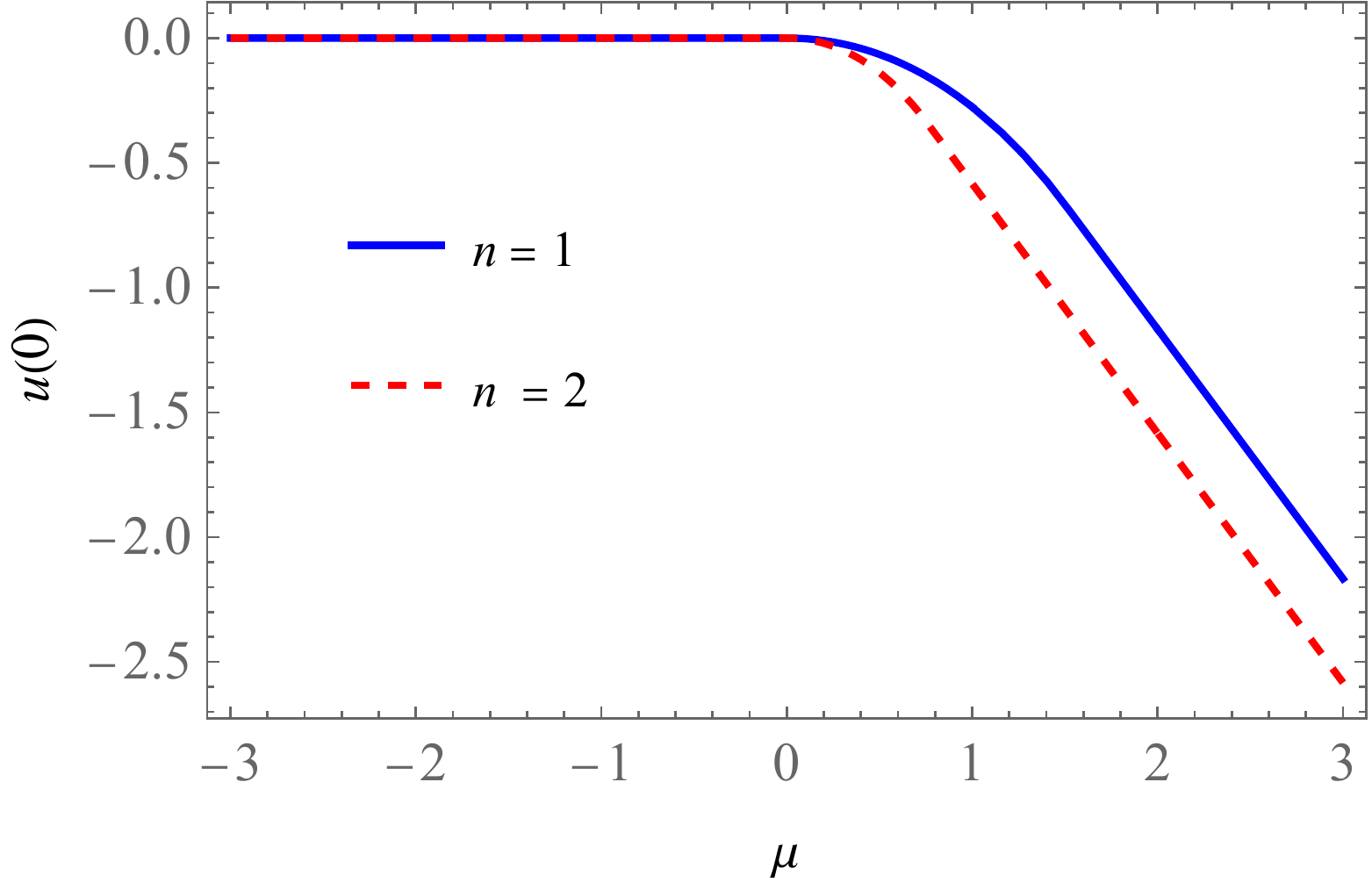}
  \caption{Zero temperature fermion and energy densities for the $\su(m|n)$ chains in
    Table~\ref{tab.la1mu} with $m\ne0$ and $\ga=2$.}
  \label{fig.nFu0}
\end{figure}
\begin{remark}
  %
  %
  From the previous analysis it follows that the ground state degeneracy of the $\su(m|n)$ chains
  with $m,n\le 2$ and $\vep=\vep'=1$ is at most $2$ for all values of the chemical potential.
  (Actually, it is easily checked that the same is true for all other values of
  $\vep,\vep'\in\{\pm1\}$.) By contrast, for $m>1$ and values of the chemical potential inside the
  critical interval, the degeneracy of the type-$A_{N-1}$ HS, PF and FI chains grows with $N$.
  Thus, as pointed out in Ref.~\cite{BBS08}, strictly speaking the latter models cannot be
  critical even if their free energy per spin grows as $T^2$ at low temperatures. The reason for
  this important difference in the ground state degeneracy lies in the last ($N$-th) component of
  $BC_N$-type motifs, which is absent in standard ($A_{N-1}$) motifs. For instance, for the
  $\su(2|0)$ PF chain of $A_{N-1}$ type the ground state bond vectors are of the
  form~\cite{FGLR18}
  \[
    \bsv_k=(\underbrace{1,\cdots,1}_{k},\underbrace{2,\cdots,2}_{N-k}),\qquad k=0,\dots,N\,,
  \]
  with zero energy, and the ground state degeneracy is thus $N+1$. On other hand, in the $BC_N$
  case the energy of $\bsv_k$ is $\vp(1)\de(s_N,s^*)=\vpmax \de(s_N,3/2)$, which vanishes for
  $k=N$ (since in this case $s_N=1<3/2$) but equals $\vpmax>0$ for $0\le k\le N-1$ (since
  $s_N=2>3/2$). Hence in the $BC_N$ case only the bond vector $\bsv_N$ yields the ground state.
 \end{remark}

\section{Critical behavior}\label{sec.crit}

In the last section we showed that the models listed in Table~\ref{tab.la1mu} are critical for
certain values of the fermionic chemical potential $\mu$. In this section we shall compute the
corresponding central charges by studying the low temperature behavior of the free energy per
spin. Indeed, it is well known~\cite{BCN86,Af86} that at low temperatures the free energy per unit
length of a $(1+1)$-dimensional CFT with central charge $c$ behaves as
\begin{equation}\label{fCFT}
f(T)\simeq f(0)-\frac{\pi c T^2}{6 v}
\end{equation}
(in natural units $\hbar=k_ B=1$), where $v$ is the Fermi velocity. The Fermi velocities of the
small momentum and Fermi excitations for all the tractable models were also computed in the
previous section. These velocities are respectively given by
\begin{equation}\label{v12}
  v_1=\frac1{\pi x_0'(0)}\,,\qquad v_2=\frac1{\pi x_0'(n\mu)},
\end{equation}
where $n$ denotes the number of fermionic degrees of freedom. Note, in particular, that the latter
equation for $v_2$ is still valid in the limiting case $\mu=\vpmax/n$.
%
%
Note also that, although Eq.~\eqref{v12} was derived for the case $\vep=\vep'=1$, it is in fact
valid for all other values of the latter parameters.


Consider, for instance, the $\su(1|1)$ chain. For $\mu<0$ we have
\begin{eqnarray*}
  |f(T)|&=T\int_0^1\log\bigl(1+\e^{-\be(\vp(x)+|\mu|)}\bigr)\diff
  x\le T\int_0^1\log\bigl(1+\e^{-\be|\mu|}\bigr)\diff x\\
  &=T\log\bigl(1+\e^{-\be|\mu|}\bigr)=\Or\bigl(T\e^{-\be|\mu|}\bigr).
\end{eqnarray*}
Likewise, if $\mu>\vpmax$ then
\begin{eqnarray*}
  \fl
  f(T)&=-T\int_0^1\log\bigl(1+\e^{\be(\mu-\vp(x))}\bigr)\diff x
        =\int_0^1\big(\vp(x)-\mu)\diff x-T\int_0^1\log\bigl(1+\e^{-\be(\mu-\vp(x))}\bigr)\diff x
  \\
  \fl
      &=f_0-\mu-T\int_0^1\log\bigl(1+\e^{-\be(\mu-\vp(x))}\bigr)\diff x\,.
\end{eqnarray*}
Hence $f(0)=f_0-\mu$ (in accordance with the result of the previous section;
cf.~Eq.~\eqref{u0su11}), and therefore
\begin{eqnarray*}
  |f(T)-f(0)|&=T\int_0^1\log\bigl(1+\e^{-\be(\mu-\vp(x))}\bigr)\diff x\\
           &\le
             T\log\bigl(1+\e^{-\be(\mu-\vpmax)}\bigr)=\Or\bigl(T\e^{-\be(\mu-\vpmax)}\bigr).
\end{eqnarray*}
Thus the model is non-critical for $\mu<0$ and $\mu>\vpmax$, as we had shown in the previous
section by analyzing the low energy excitations above the ground state.

Consider next the range $0<\mu<\vpmax$. To begin with, since $\mu-\vp(x)\ge0$ for
$0\le x\le x_0(\mu)$ and $\mu-\vp(x)\le0$ for $x_0(\mu)\le x\le 1$ we can write
\begin{eqnarray*}
  \fl
  f(T)
        =\int_0^{x_0(\mu)}\big(&\vp(x)-\mu)\diff x\\
        &-T\int_0^{x_0(\mu)}\log\bigl(1+\e^{-\be(\mu-\vp(x))}\bigr)\diff x
          -T\int_{x_0(\mu)}^1\log\bigl(1+\e^{-\be(\vp(x)-\mu)}\bigr)\diff x\,.
\end{eqnarray*}
It follows that
\[
  f(0)=\int_0^{x_0(\mu)}\big(\vp(x)-\mu)\diff x\,,
\]
(in agreement with~Eq.~\eqref{u0}), and hence
\begin{eqnarray*}
\fl
  f(T)-f(0)&=-T\int_0^{x_0(\mu)}\log\bigl(1+\e^{-\be(\mu-\vp(x))}\bigr)\diff x
  -T\int_{x_0(\mu)}^1\log\bigl(1+\e^{-\be(\vp(x)-\mu)}\bigr)\diff x\\
  &=: I_1 + I_2\,.
\end{eqnarray*}
When $\be\to\infty$, the main contribution to each of the integrals $I_k$ comes from the endpoint
$x_0(\mu)$, at which the exponents $\pm\be(\vp(x)-\mu)$ vanish. Performing the change of variable
$\be(\mu-\vp(x))=y$, i.e., $x=x_0(\mu-Ty)$, in the first integral we obtain
\[
  I_1=-T^2\int_0^{\be\mu}x_0'(\mu-Ty)\log\bigl(1+\e^{-y}\bigr)\,\diff y\,.
\]
Since for $\be\to\infty$ the main contribution to the latter integral comes from the endpoint
$y=0$, approximating $x_0'(\mu-Ty)$ by $x_0'(\mu)$ and pushing the upper limit to $+\infty$ we
have
\[
  I_1\simeq-T^2x_0'(\mu)\int_0^\infty\log\bigl(1+\e^{-y}\bigr)\,\diff y
  =-\frac{\pi T^2}{12}\,x_0'(\mu)\,.
\]
Similarly,
\begin{eqnarray*}
  I_2&=-T^2\int_0^{\be(\vpmax-\mu)}x_0'(\mu+Ty)\log\bigl(1+\e^{-y}\bigr)\,\diff y\\
     &\simeq -T^2x_0'(\mu)\int_0^\infty\log\bigl(1+\e^{-y}\bigr)\,\diff y
       =-\frac{\pi T^2}{12}\,x_0'(\mu)\,.
\end{eqnarray*}
Combining both results and using Eq.~\eqref{v12} for the velocity $v_2$ of the Fermi excitations
(with momentum near the Fermi momentum $p_0=\pi x_0(\mu)$) we finally obtain
\[
  f(T)-f(0)\simeq-\frac{\pi T^2}{6}\,x_0'(\mu)=-\frac{\pi T^2}{6v_2}\,.
\]
Thus the central charge in the critical interval $0<\mu<\vpmax$ is $c_2=1$.

A similar analysis can be applied to the limiting point $\mu=0$. Indeed, in this case $f(0)=0$ and
the main contribution to
\[
  f(T)=-T\int_0^1\log\bigl(1+\e^{-\be\vp(x)}\bigr)\diff x
\]
when $\be\to\infty$ comes from the lower limit of the integral. Performing the change of variable
$\be\vp(x)=y$, i.e., $x=x_0(Ty)$, we have
\begin{eqnarray*}
  \fl f(T)&=-T^2\int_0^{\be\vpmax}x_0'(Ty)\log\bigl(1+\e^{-y}\bigr)\,\diff y
            \simeq-T^2x_0'(0)\int_0^{\infty}\log\bigl(1+\e^{-y}\bigr)\,\diff y\\
  \fl
          &=-\frac{\pi
            T^2}{12}\,x_0'(0).
\end{eqnarray*}
From Eq.~\eqref{v12} for the Fermi velocity $v_1$ of small momentum excitations we have
\[
  f(T)\simeq-\frac{\pi T^2}{12v_1}\,,
\]
and hence the central charge is now $c_1=1/2$. Likewise, when $\ga>1$ and $\mu=\vpmax=\ga-1/2$ the
main contribution to the integral
\[
  f(T)-f(0)=-T\int_0^1\log\bigl(1+\e^{-\be(\vpmax-\vp(x))}\bigr)\diff x
\]
only comes from its upper limit (i.e., from Fermi excitations with momentum close to $\pi$).
Performing the analogous change of variable $\be(\vpmax-\vp(x))=y$ we easily obtain
\[
  f(T)\simeq-\frac{\pi^2T^2}{12}\,x_0'(\vpmax)=-\frac{\pi T^2}{12v_2}
\]
so that $c_2=1/2$. On the other hand, when $\ga=1$ and $\mu=\vpmax=1/2$ we have
\[
  \vpmax-\vp(x)=\frac12\,(1-x)^2\,.
\]
Performing the change of variable $\vpmax-\vp(x)=Ty$, i.e., $x=1-\sqrt{2Ty}$, we readily obtain
\begin{eqnarray*}
  \fl
  f(T)-f(0)&=-\frac{T^{3/2}}{\sqrt2}\,\int_0^{\be/2}y^{-1/2}\log\bigl(1+\e^{-y}\bigr)\diff y
        \simeq-\frac{T^{3/2}}{\sqrt2}\,\int_0^{\infty}y^{-1/2}\log\bigl(1+\e^{-y}\bigr)\diff y\\
  \fl
  &=-\frac{\sqrt\pi}2\,(\sqrt2-1)\ze(3/2)T^{3/2}\,,
\end{eqnarray*}
where $\ze(z)=\sum_{n=1}^\infty n^{-z}$ is Riemann's zeta function. The $T^{3/2}$ growth of the
free energy at low temperature shows that the system is not critical in this case, as anticipated
in the previous section from the existence of excitations of energy $\Or(N^{-2})$ above the ground
state energy.

Essentially the same procedure (with a few minor variations) can be applied to the remaining
models listed in Table~\ref{tab.la1mu}. We summarize the results obtained in Table~\ref{tab.crit},
presenting the details of the corresponding calculations in~\ref{app.A}.
\begin{table}[h]
  \centering
  \begin{tabular}{|c|c|c|c|c|c|}
    \cline{2-6}
    \multicolumn{1}{c|}{}
    &$\mu<0$& $\mu=0$ &$0<\mu<\frac{\vpmax\vrule width0pt depth4pt height 6pt}{n}$
    &$\mu=\frac{\vpmax\vrule width0pt depth4pt height 6pt}{n}$
      ($\ga>1$)
    &$\mu>
      \frac{\vpmax\vrule width0pt depth4pt height 6pt}n$\\[1mm]
    \hline\noalign{\vspace*{1pt}}\hline
    $\su(1|1)$& $-$ &$(1/2,-)$& $(-,1)$& $(-,1/2)$& $-$\\[1mm]
    \hline
    $\su(0|2)$
    &\multicolumn{5}{|c|}{$(1,-)$}\\[1mm]
    \hline
    $\su(1|2)$& $-$ &$(1,-)$& $(1,1)$& $(1,3/5)$& $(1,-)$\\[1mm]
    \hline
    $\su(2|1)$& $(1,-)$ &$(3/2,-)$& $(-,2)$& $(-,4/5)$& $-$\\[1mm]
    \hline
    $\su(2|2)$& $(1,-)$ &$(2,-)$& $(1,2)$& $(1,1)$& $(1,-)$\\[1mm]
    \hline
  \end{tabular}
  \caption{Critical behavior of the (nontrivial) $\su(m|n)$ HS chains of $BC_N$ type with
    $m,n\le2$ as a function of the fermionic chemical potential $\mu$. The notation $(c_1,c_2)$ is
    used to indicate that the central charge of the small momentum (resp.\ Fermi) excitations is
    $c_1$ (resp.\ $c_2$), a horizontal dash (``$-$'') denoting that the corresponding excitation
    is not present. A horizontal dash by itself in a cell means that the model is gapped. Note
    that in the case $\mu=\vpmax/n$ and $\ga=1$ all of the models listed above are gapless but
    non-critical, with $f(T)-f(0)\sim T^{3/2}$.}
  \label{tab.crit}
\end{table}
To give an idea of the accuracy of the asymptotic approximations of $f(T)$ obtained in this
section and in the appendix, we show in Fig.~\ref{fig.ffapp} a plot comparing the latter function
to its low temperature approximation $f_{\text{app}}(T)$ for the $\su(1|2)$ chain with $\ga=2$ and
different values of the parameter $\mu$ in each of the four cases in which there is at least a
critical low energy excitation.
\begin{figure}[t]
  \includegraphics[width=.48\textwidth]{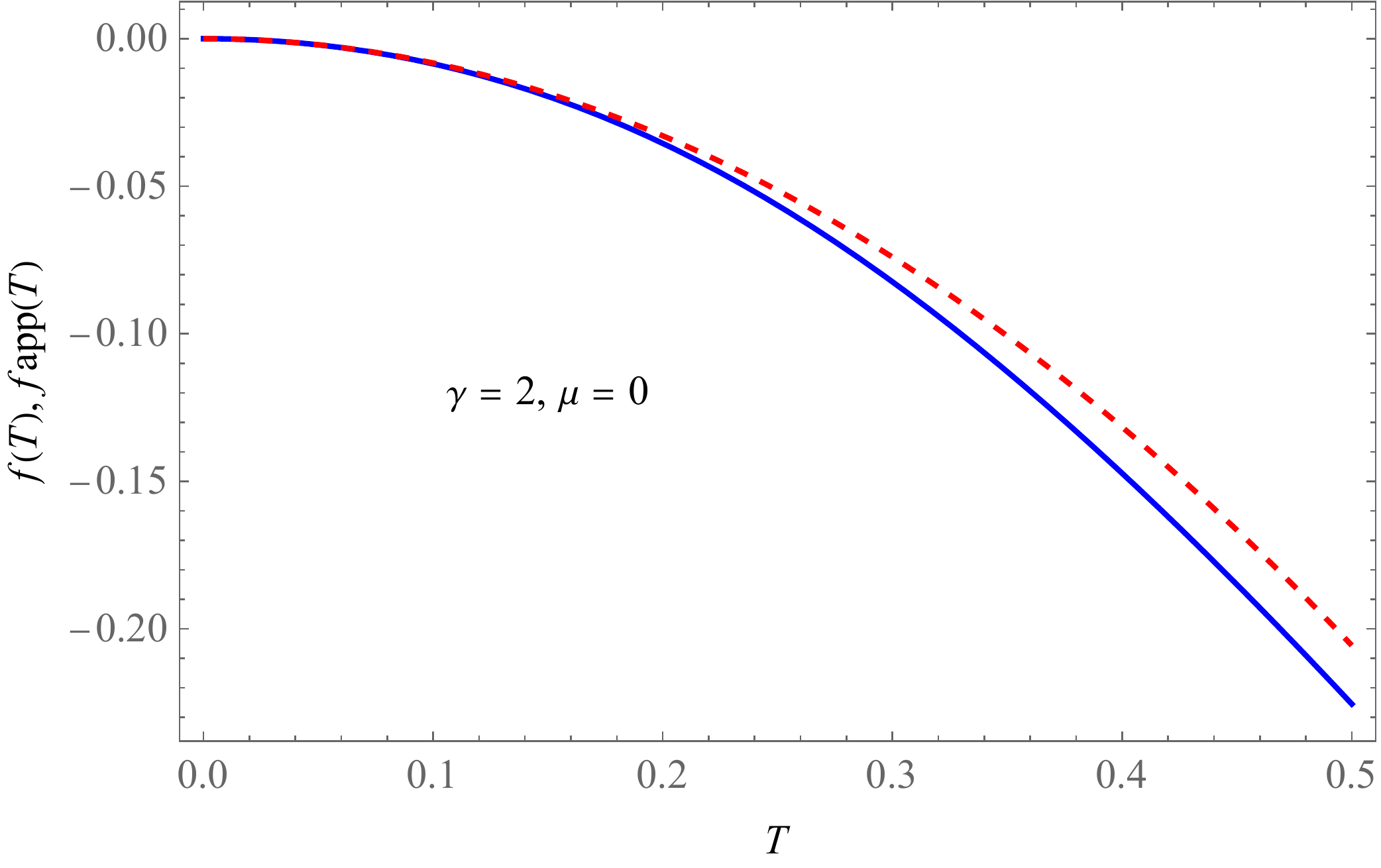}\hfill
  \includegraphics[width=.48\textwidth]{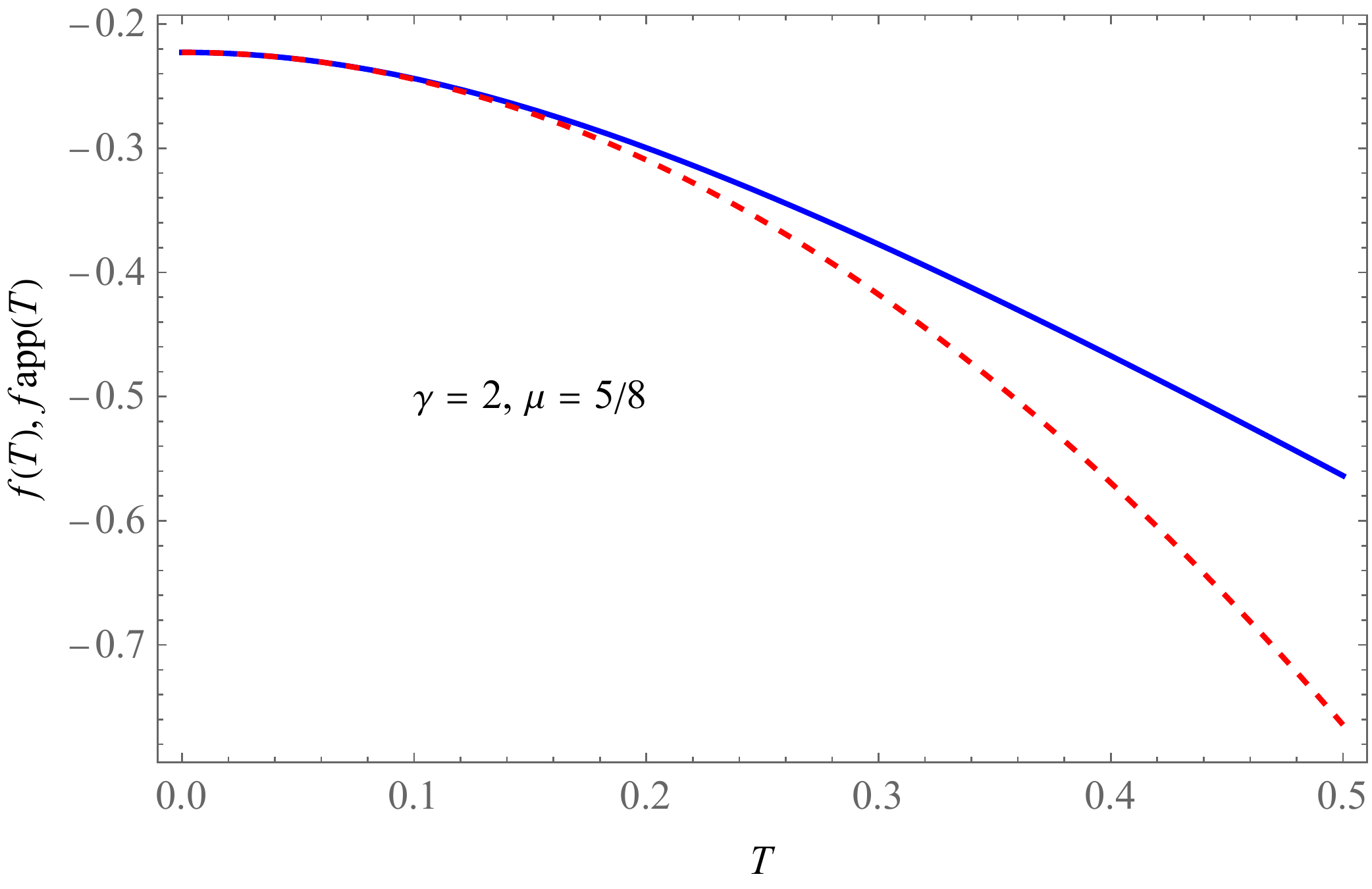}\\[1.5mm]
  \includegraphics[width=.48\textwidth]{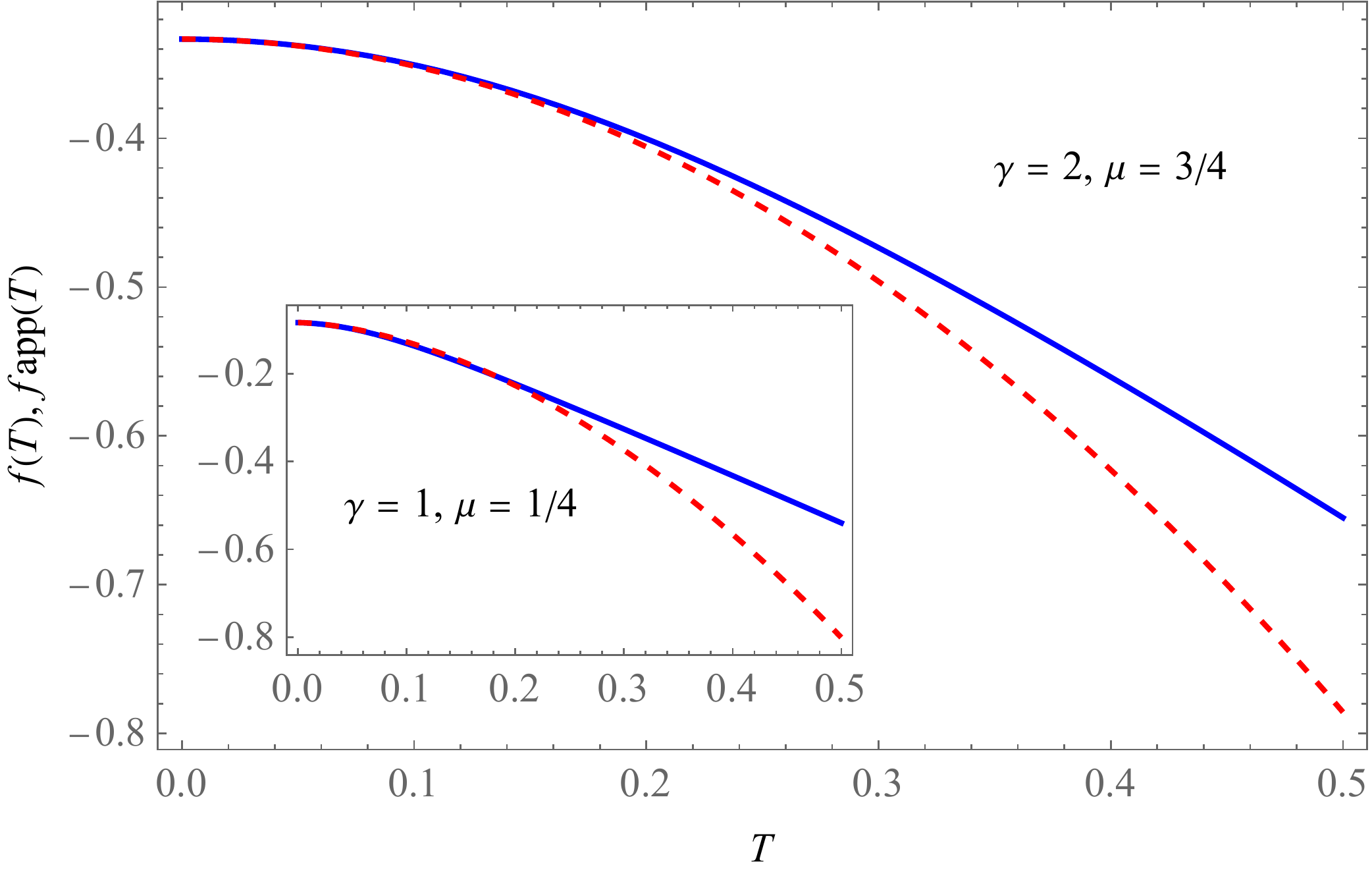}\hfill
  \includegraphics[width=.48\textwidth]{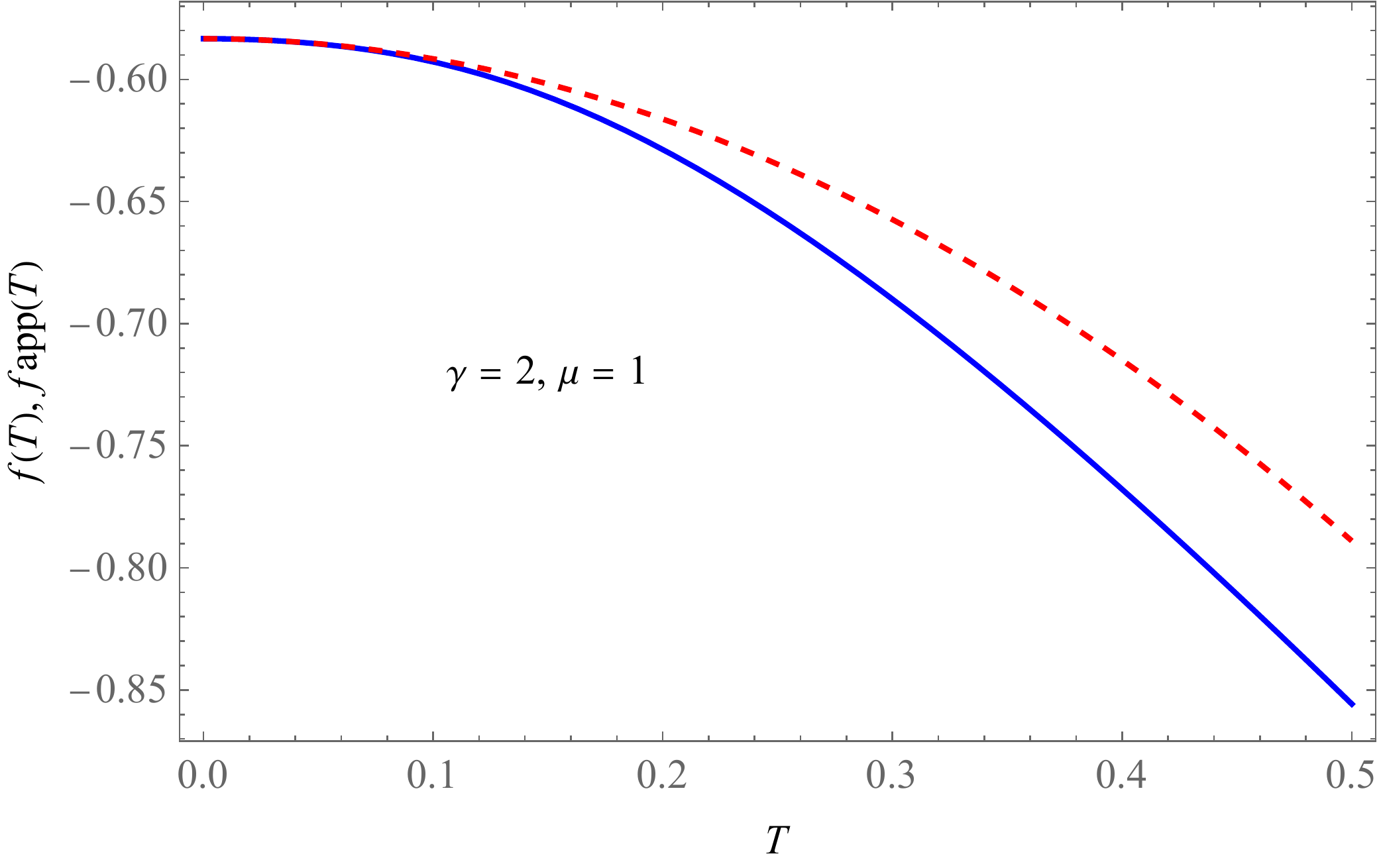}
  \caption{Comparison between the free energy density $f(T)$ (blue line) of the $\su(1|2)$ chain
    and its low temperature approximation $f_{\text{app}}(T)$ (dashed red line) for $\ga=2$ and
    different values of $\mu$. The inset corresponds to the case $\ga=1$, $\mu=1/4$ for which, as
    shown in the appendix, $f(T)-f(0)$ is of order $T^{3/2}$ instead of $T^2$.}
  \label{fig.ffapp}
\end{figure}
\begin{remark}
  It was shown in Ref.~\cite{HB00} that the $\su(m|n)$ Polychronakos--Frahm chain of $A_{N-1}$
  type with zero chemical potential has central charge
  \[
    c(m,n)=\cases{\max(m,n)-1,&\(mn=0\)\\
      m-1+\frac n2\,,& \(mn>0\).}
  \]
  The same formula is actually valid for the $\su(m|n)$ HS chain of $A_{N-1}$ type, as follows
  from the relation between the partition function of these chains uncovered in Ref.~\cite{BBS08}.
  Our results, summarized in Table~\ref{tab.crit}, show that this formula still holds for the open
  counterparts of the HS chain studied in this work.
\end{remark}
\section{Thermodynamics}\label{sec.thermo}

The thermodynamic functions of the models studied in the previous section can be computed in
closed form using Eqs.~\eqref{npmal}-\eqref{varnF} and \eqref{ffinal} together with the explicit
formulas for the Perron--Frobenius eigenvalue $\la_1(x)$ listed in Table~\ref{tab.la1mu}. For
instance, for the $\su(1|1)$ model we easily obtain
\begin{eqnarray}
  \fl
  n_F^{(1|1)}&=\int_0^1\frac{\diff x}{1+\e^{\be(\vp(x)-\mu)}}\,,
       \label{su11nF}\\[1mm]
  \fl
  u^{(1|1)}&=\int_0^1\frac{\vp(x)-\mu}{1+\e^{\be(\vp(x)-\mu)}}\,\diff x\,,
             \label{su11u}\\[1mm]
  \fl
  c_V^{(1|1)}&=\frac{\be^2}4\int_0^1\big(\vp(x)-\mu\big)^2
               \sech^2\biggl(\frac\be2(\vp(x)-\mu)\biggr)\diff
               x\,,
       \label{su11cV}\\
  \fl
  s^{(1|1)}&=\int_0^1\bigg\{\log\biggl[2\cosh\biggl(\frac\be2(\vp(x)-\mu)\biggr)\biggr]
     -\frac\be2(\vp(x)-\mu)
     \tanh\biggl(\frac\be2(\vp(x)-\mu)\biggr)\bigg\}\diff x\,,
     \label{su11s}\\
  \fl
  \De_F^{(1|1)}&=\frac14\int_0^1\sech^2\biggl(\frac\be2(\vp(x)-\mu)\biggr)\diff x\,.
         \label{su11DeF}
\end{eqnarray}
It can be shown that these expressions have the same form as their analogues for the $\su(1|1)$ HS
chain of type $A_{N-1}$~\cite{FGLR18}, the only difference being that in the latter model $\vp(x)$
is proportional to $(x/2)(1-x/2)$. In particular, the $\su(1|1)$ $BC_N$ chain with $\ga=1$ is
thermodynamically equivalent to its (suitably normalized) $A_{N-1}$ counterpart. The thermodynamic
functions of the $\su(0|2)$ chain can also be obtained from Eqs~\eqref{su11nF}-\eqref{su11DeF},
taking $\mu=0$ and replacing $\vp$ by $\vp/2$. Similarly, since the $\su(2|2)$ chain is (in the
thermodynamic limit) to the direct sum of two $\su(0|2)$ chains with chemical potentials $0$ and
$\mu$, its energy, specific heat and entropy (per spin) can also be easily expressed as sums of
the corresponding $\su(1|1)$ functions~\eqref{su11u}-\eqref{su11s} with $\vp$
replaced by $\vp/2$. For instance,
\[
  u^{(2|2)}=\int_0^1\frac{\vp(x)/2}{1+\e^{\be\vp(x)/2}}\,\diff x
  +\int_0^1\frac{\vp(x)/2-\mu}{1+\e^{\be(\vp(x)/2-\mu)}}\,\diff x\,.
\]
On the other hand, for the fermion density and its variance we have the simpler formulas
\[ 
  n_F^{(2|2)}=\int_0^1\frac{\diff x}{1+\e^{\be(\vp(x)/2-\mu)}}\,,\qquad
  \De_F^{(2|2)}=\frac14\int_0^1\sech^2\biggl(\frac\be2\bigg(\frac{\vp(x)}2-\mu\bigg)\biggr)\diff
  x\,.
\]
The expressions of the thermodynamic functions of the $\su(1|2)$ and $\su(2|1)$ chains are more
cumbersome, and will not be presented here, with the only exception of the fermion densities:
\[
  \fl
  n_F^{(1|2)}=1-\int_0^1\frac{\diff x}{\sqrt{1+4\e^{-\be(\vp(x)-\mu)}(1+\e^{\be\mu})}}\,,\quad
  n_F^{(2|1)}=\int_0^1\frac{\diff x}{\sqrt{1+4\e^{\be(\vp(x)-\mu)}(1+\e^{-\be\mu})}}\,.
\]

Note that for critical values of the chemical potential $\mu$ the low temperature behavior of the
thermodynamic functions can be easily determined differentiating the asymptotic expression for
$f(T)$ obtained in the previous section and in the appendix. Indeed, for these values of $\mu$ we
can write
\begin{equation}\label{fasymp}
  f(T)=f(0)-\ka T^2+\mathrm o(T^2)\,,
\end{equation}
with
\begin{equation}\label{kamu}
  \fl
  \ka=\frac\pi6\,\bigg(\frac{c_1}{v_1}+\frac{c_2}{v_2}\bigg)
  =\frac{\pi^2}6\,\big(c_1x_0'(0)+c_2x_0'(n\mu)\big)
  =\frac{\pi^2}6\,\bigg(\frac{c_1}\ga+\frac{c_2}{\sqrt{\ga^2-2n\mu}}\bigg)\,,
\end{equation}
where it is understood that $c_i$ should be taken as $0$ if its corresponding low energy
excitations are not critical. We then have
\[
  u=f(0)+\ka T^2+\mathrm o(T^2)\,,\qquad
  c_V=s=2\ka T+\mathrm o(T)\,.
\]
The behavior of the fermion density $n_F$ and variance $\De_F$ can be computed in the same way
when $\mu$ lies in the interior of one of the critical regions in Table~\ref{tab.crit}. Indeed,
differentiating Eq.~\eqref{fasymp}-\eqref{kamu} with respect to the chemical potential $\mu$ we
obtain
\[
  n_F(T)=-\pdf{}\mu f(0)+\pdf\ka\mu\,T^2+\mathrm o(T^2)\,.
\]
Since, by Eq.~\eqref{kamu},
\[
  \pdf\ka\mu
  =\frac{n\pi^2c_2}{6}\,\big(\ga^2-2n\mu\big)^{-3/2}\,,
\]
it follows that
\[
  n_F(T)-n_F(0)
  =\frac{n\pi^2c_2}{6}\,\big(\ga^2-2n\mu\big)^{-3/2}T^2+\mathrm o(T^2)\,.
\]
Thus when $\mu$ lies in the interior of a critical region the fermion density $n_F(T)$ is always
increasing for sufficiently low $T$. In particular, if $n_F(0)\ge\lim_{T\to\infty}n_F(T)=n/(m+n)$
then the fermion density must have (at least) one local maximum at a positive temperature.
Likewise, differentiating the previous equation with respect to $\mu$ and using
Eq.~\eqref{nF0full} we obtain
\[
  \fl
  \De_F(T)=T\,\pdf{}\mu n_F(T)=\frac{n
    T}{\sqrt{\ga^2-2n\mu}}\,\chi_{[0,\vpmax/n]}+\frac{n^2\pi^2c_2T^3}{2(\ga^2-2n\mu)^{5/2}}+\mathrm
  o(T^3)\,,
\]
where $\chi_S$ denotes the characteristic function of the set $S$.
\begin{figure}[t]
  \includegraphics[width=.5\textwidth]{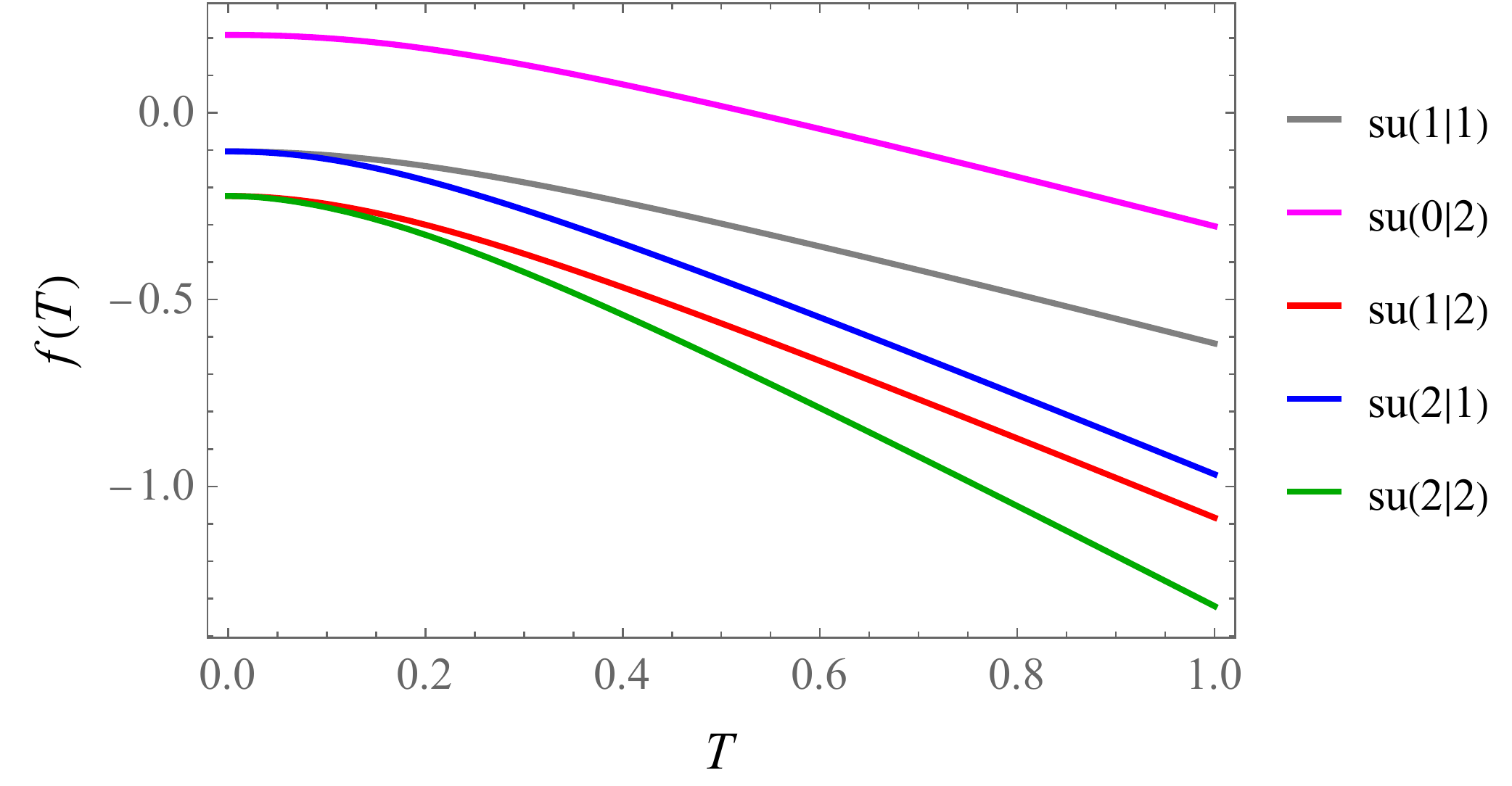}\hfill
  \includegraphics[width=.5\textwidth]{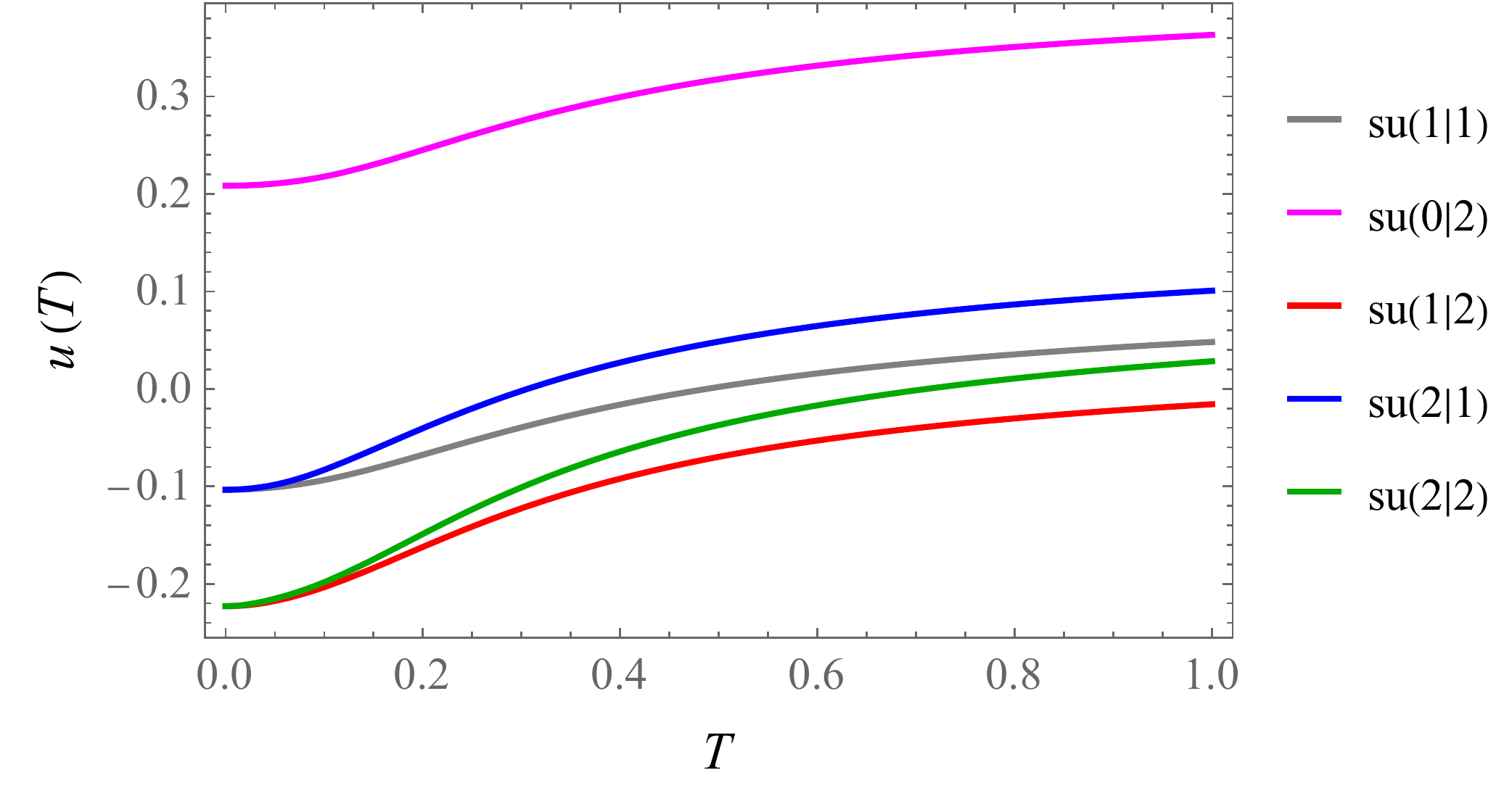}\\[1.5mm]
   \includegraphics[width=.5\textwidth]{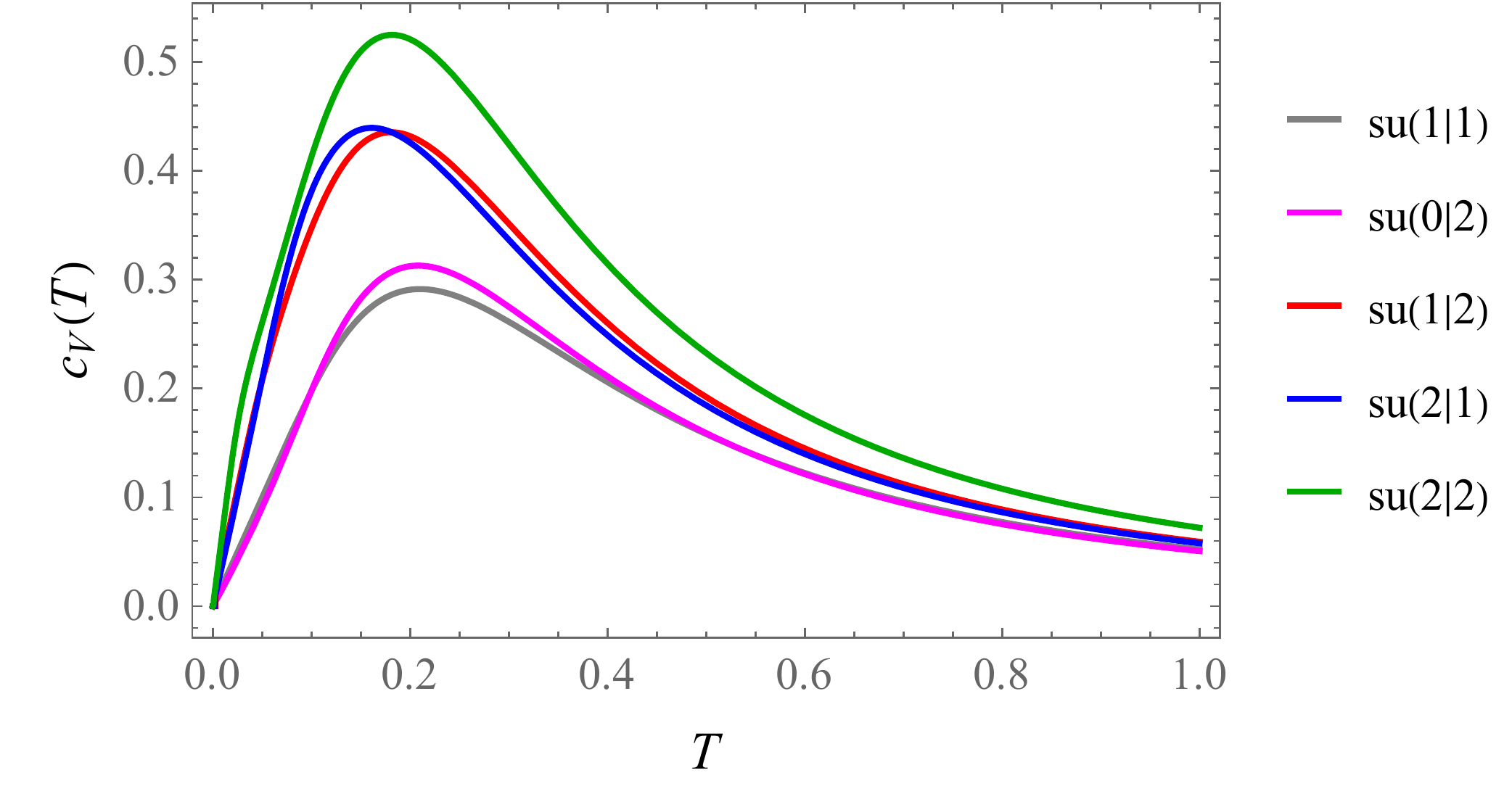}\hfill
   \includegraphics[width=.5\textwidth]{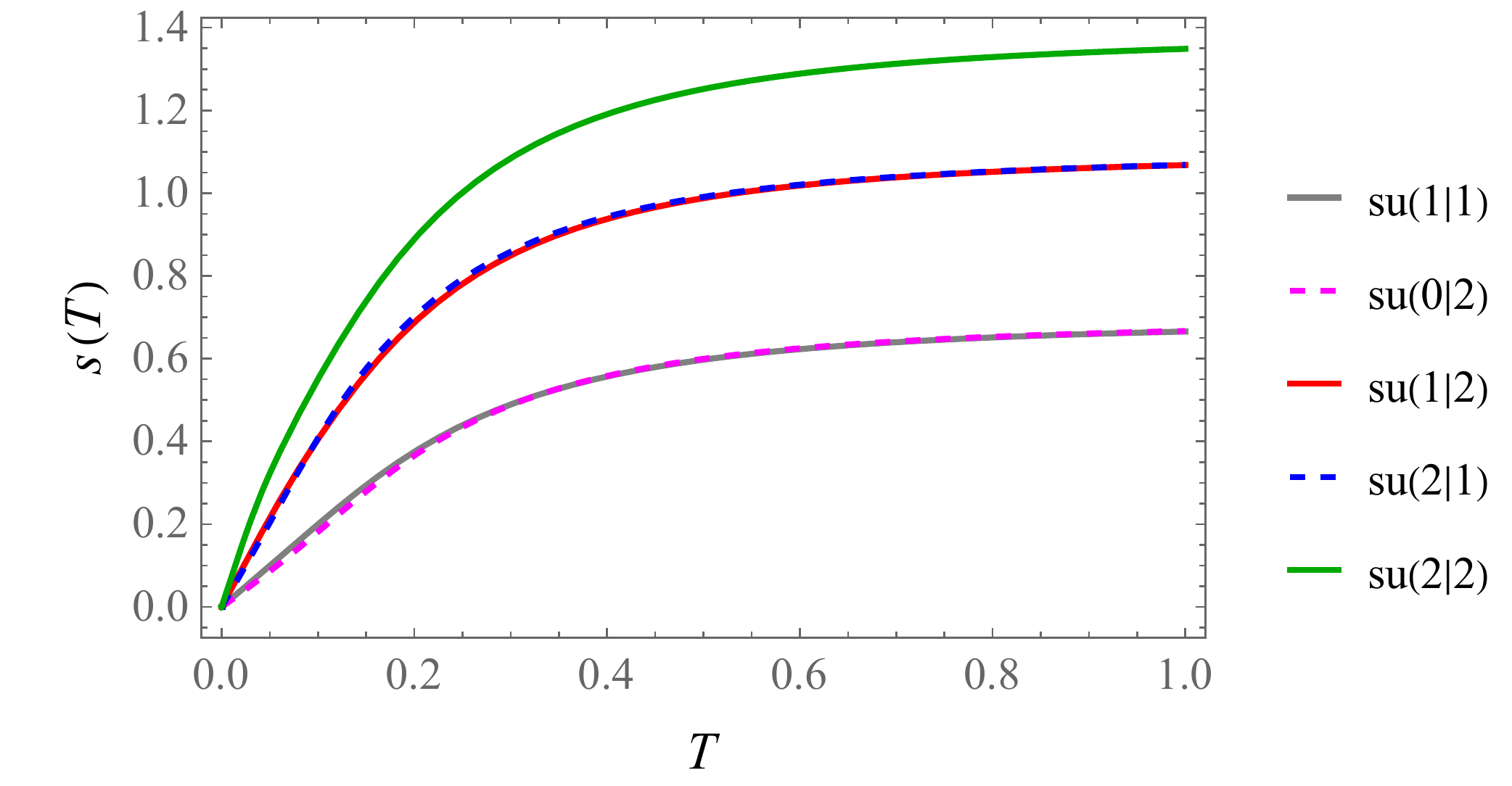}\\[1.5mm]
   \includegraphics[width=.5\textwidth]{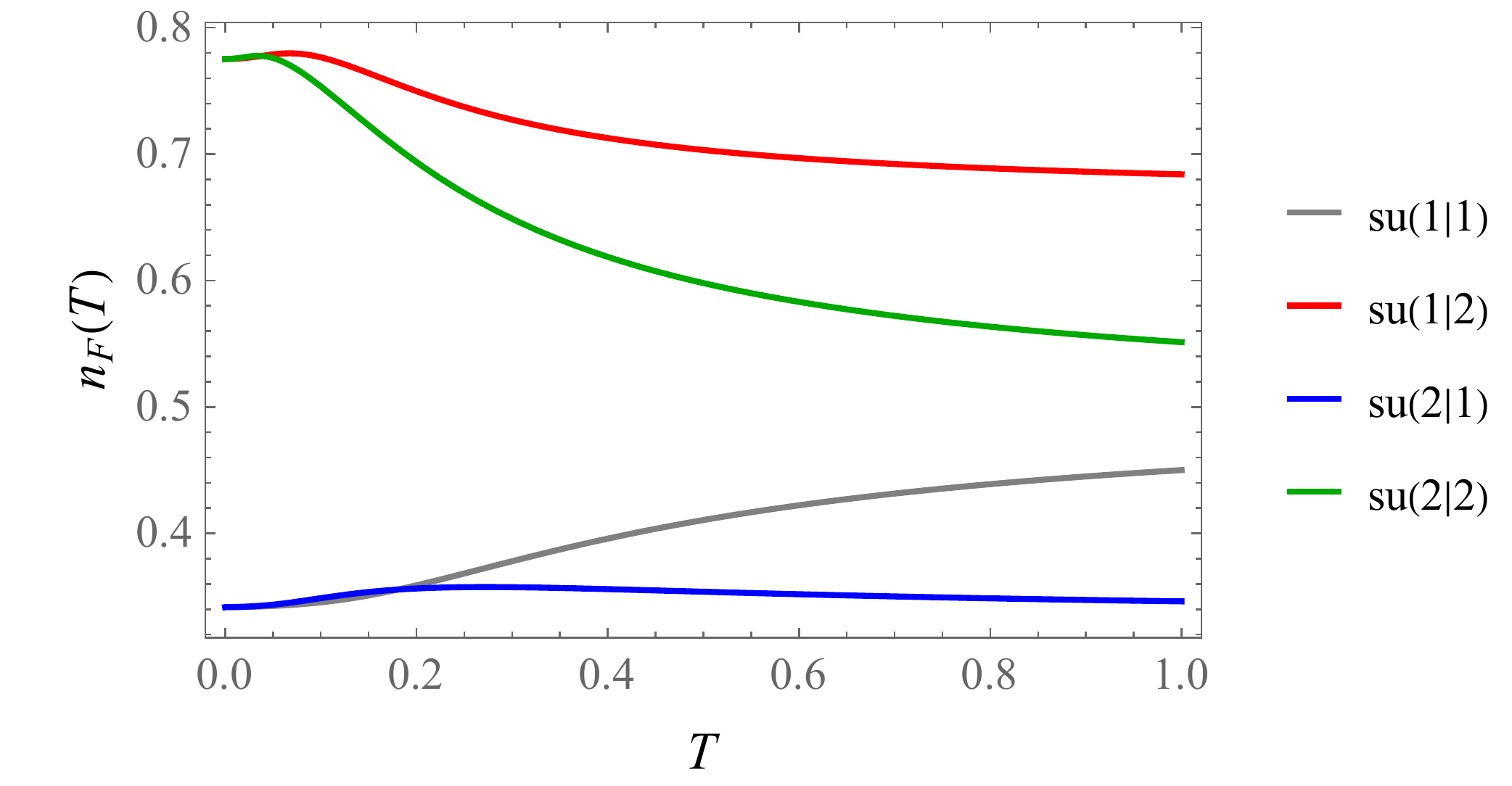}\hfill
   \includegraphics[width=.5\textwidth]{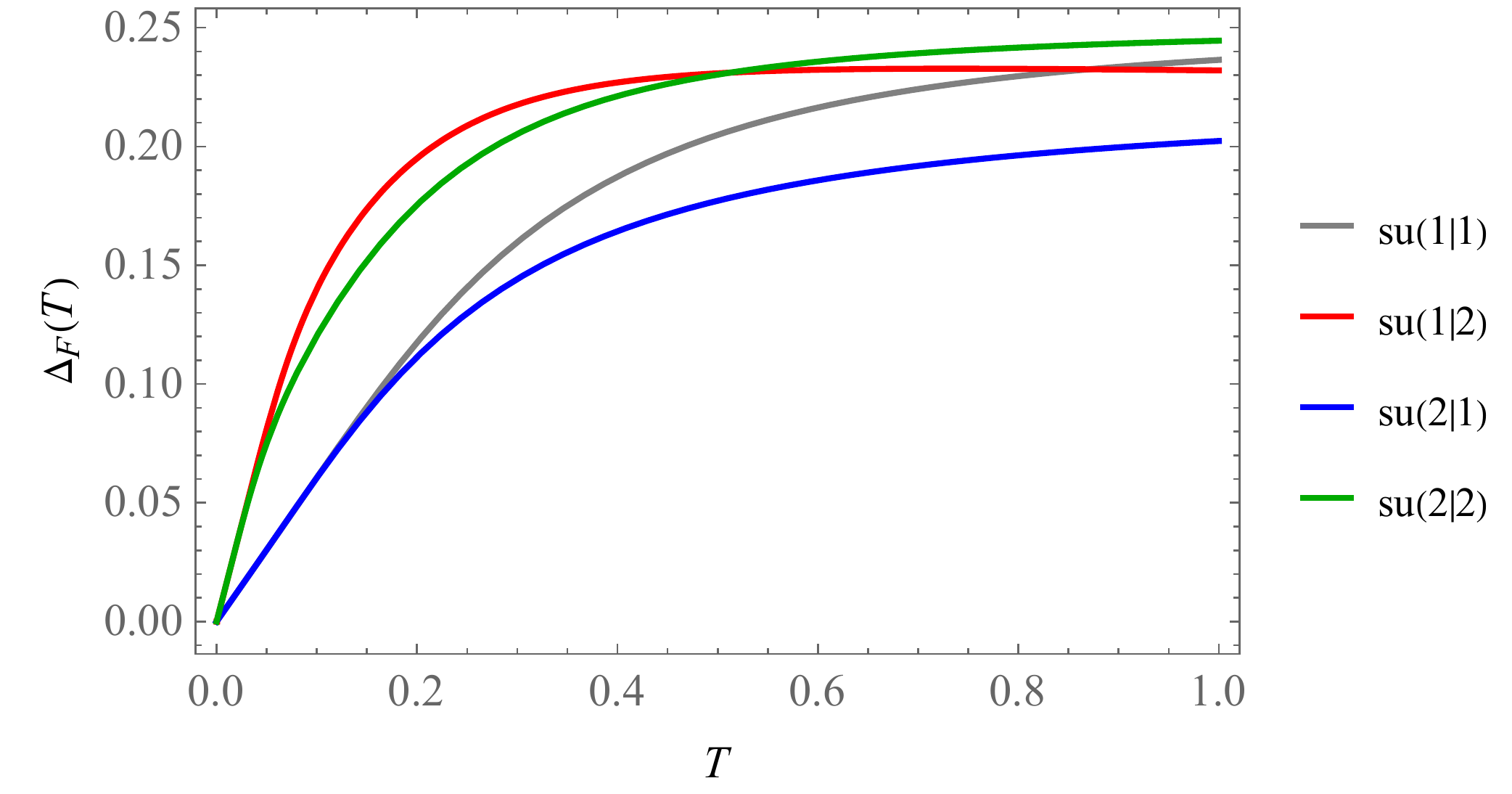}
   \caption{Thermodynamic functions for the (nontrivial) $\su(m|n)$ chains of $BC_N$ type with
     $m,n\le2$ with $\ga=2$ and $\mu=5/8$.}
  \label{fig.thermo}
\end{figure}

\begin{figure}[t]
  \includegraphics[width=.5\textwidth]{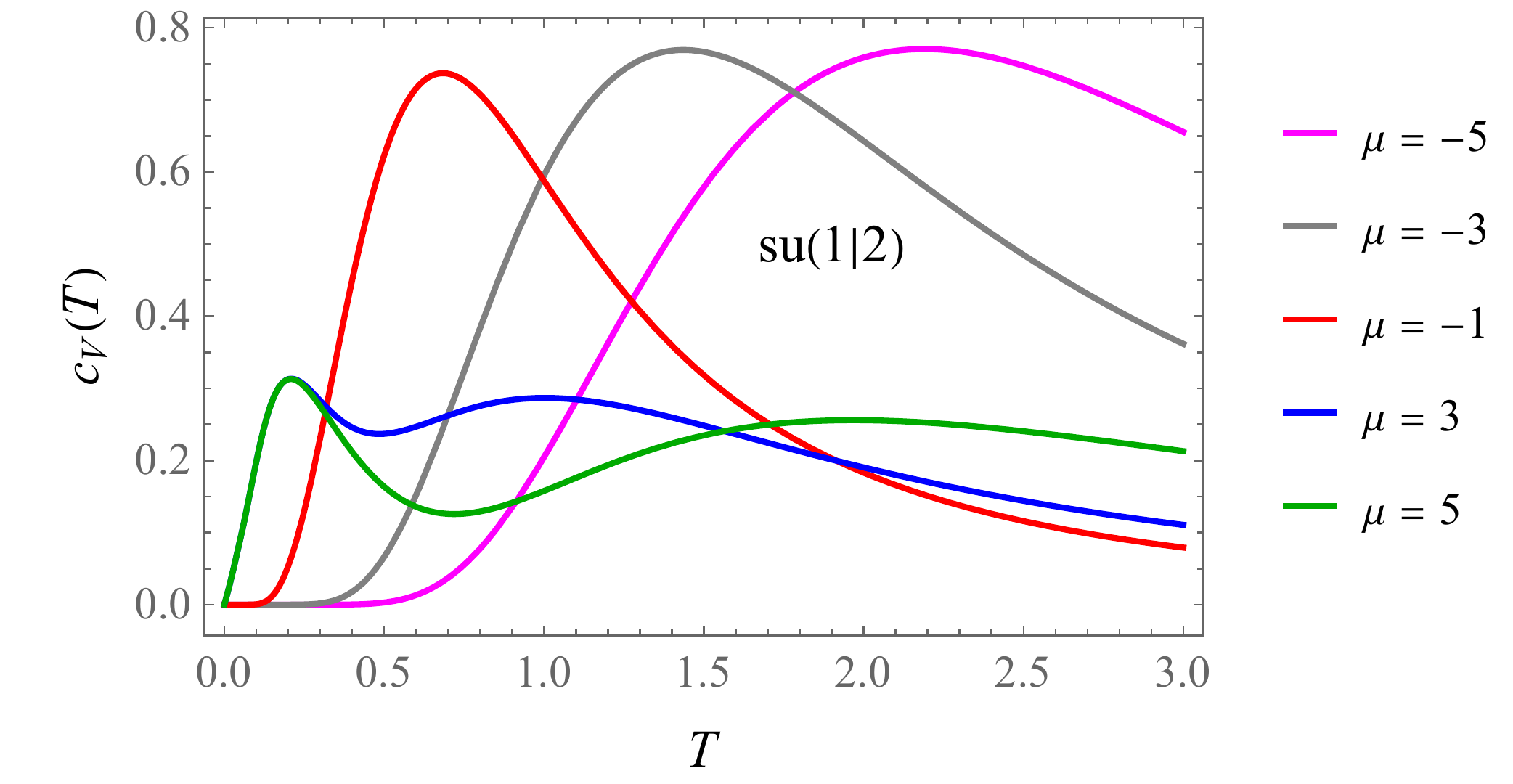}\hfill
  \includegraphics[width=.5\textwidth]{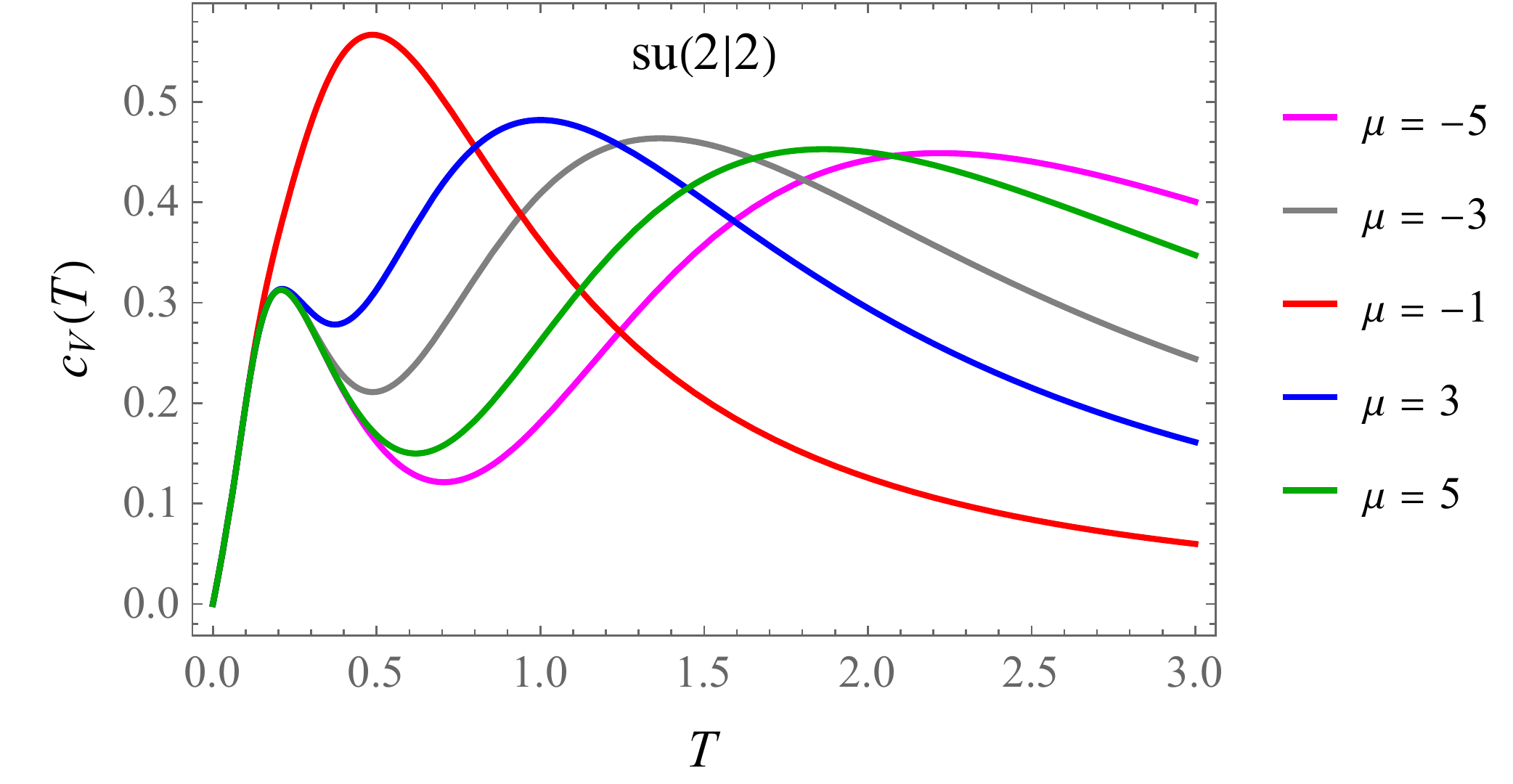}\\
  \includegraphics[width=.5\textwidth]{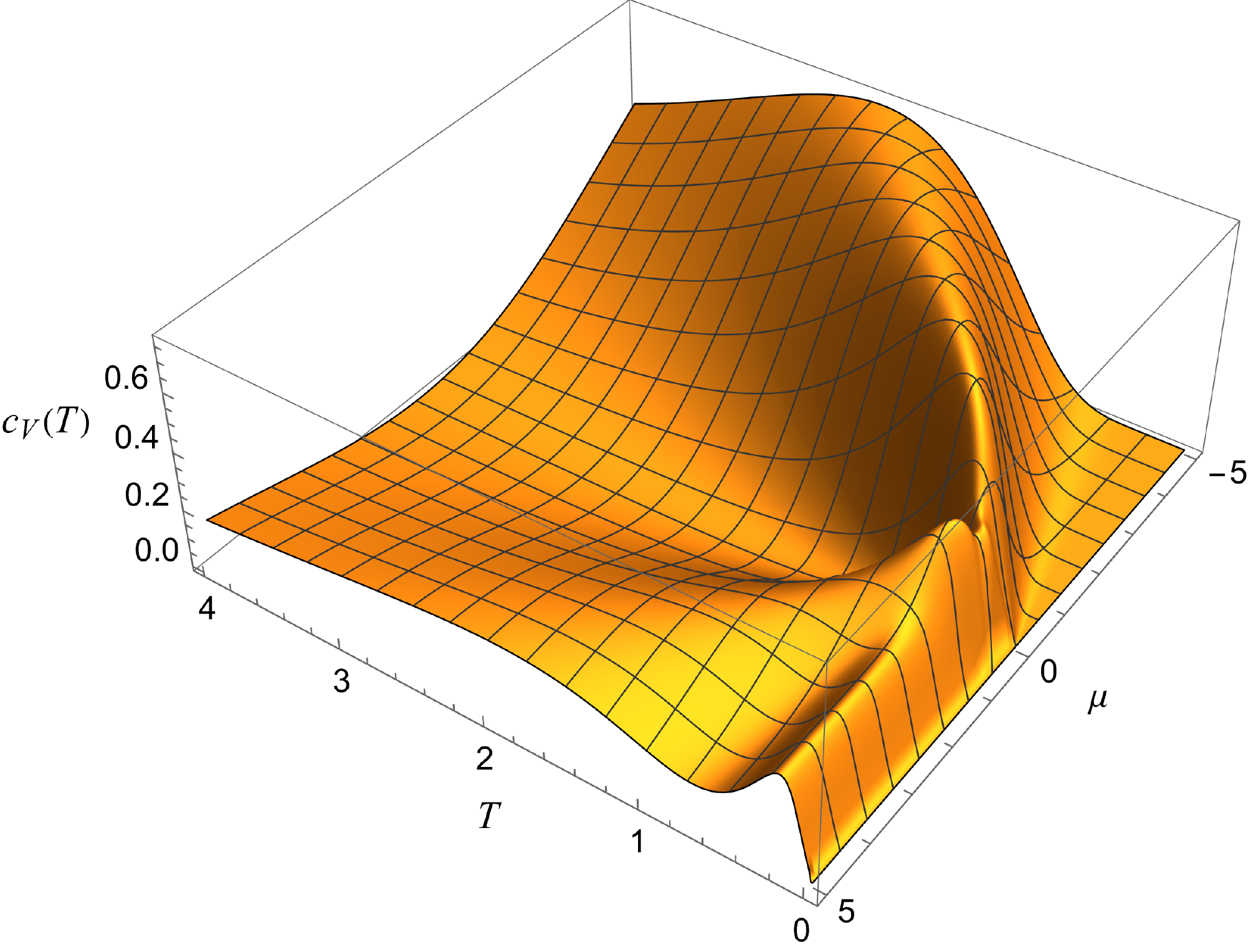}\hfill
  \includegraphics[width=.5\textwidth]{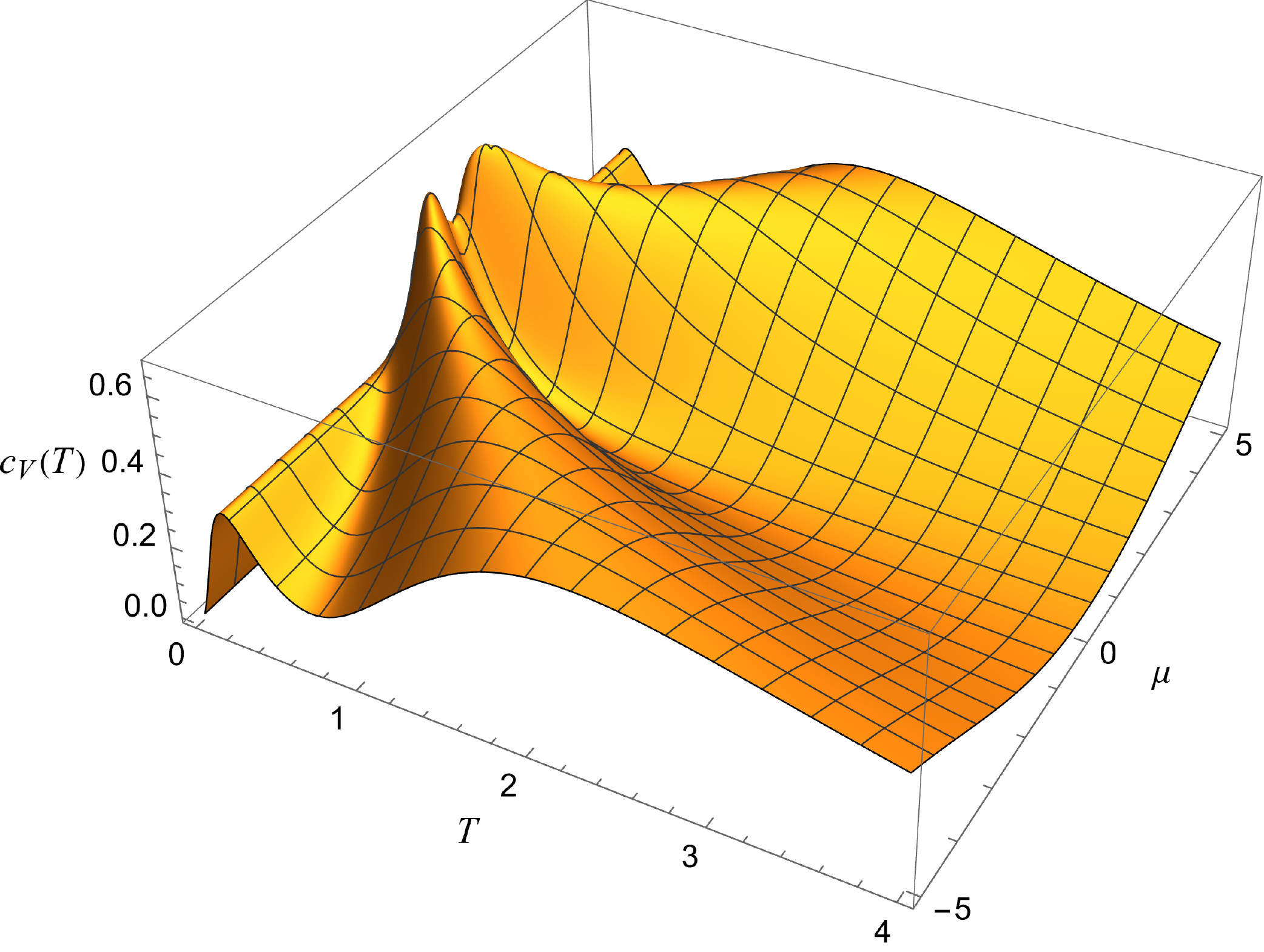}
  \caption{Top: specific heat per spin vs temperature $T$ for the $\su(1|2)$ (left) and $\su(2|2)$
    (right) chains with $\ga=2$ and several values of $\mu$. Bottom: analogous 3D plots for $\mu$
    in the range $[-5,5]$. For the $\su(1|2)$ (resp.~$\su(2|2)$) chain the specific heat per spin
    features a double Schottky peak when $\mu$ (resp.~$|\mu|$) is large enough.}
  \label{fig.peak}
\end{figure}%
As an example, we have presented in Fig.~\ref{fig.thermo} the plots of the thermodynamic functions
discussed above for $\ga=2$ and chemical potential $\mu=5/8$\footnote{For comparison purposes, we
  have subtracted $\mu=5/8$ from the free energy per spin and the energy density of the $\su(0|2)$
  chain.}, which lies inside the critical region $(0,\vpmax/n)=(0,3/(2n))$ both for $n=1$ and
$n=2$. These plots are in agreement with the low-temperature behavior of the latter functions we
have just determined. In general, the behavior of the thermodynamic functions of the HS chains of
$BC_N$ type studied in this paper is similar to that of their $A_{N-1}$
counterparts~\cite{EFG12,FGLR18}. In particular, for $0<\mu<\vpmax/n$ the specific heat features
the Schottky peak characteristic of two-level systems like, e.g., the one-dimensional Ising
model~\cite{Mu10}. On the other hand, when the intervals $(-\infty,0)$ or $(\vpmax/2,\infty)$ are
critical (i.e., respectively for $m=2$, $n\ne0$ and $n=2$, $m\ne0$), our numerical calculations
show that the specific heat develops an additional low-temperature peak for sufficiently large
$|\mu|$ (cf. Fig.~\ref{fig.peak}). Note that this behavior is characteristic of three-level
systems~\cite{SPSL16}, and has been experimentally observed in many low-dimensional quantum
ferrimagnets~\cite{NY02}. As remarked by Mussardo~\cite{Mu10}, this is probably related to the
fact that in the models under study the partition function can be obtained from an
$(m+n)$-dimensional transfer matrix. Note, however, that in our case the situation is less clear
that in the example of the one-dimensional Ising model discussed by the latter author, since (for
$mn\ne0$) the transfer matrix is singular.

\section{Conclusions and outlook}\label{sec.conc}

In this paper we have analyzed the thermodynamics and the critical behavior of the open
($BC_N$-type) supersymmetric Haldane--Shastry chain with a general chemical potential term. As
pointed out in the Introduction, while the thermodynamic and criticality properties of the HS
chain of $A_{N-1}$ type and its rational/hyperbolic counterparts have been studied in great
detail, to the best of our knowledge an analogous study for the HS chain of $BC_N$ type had yet to
be undertaken. The key idea for deriving the thermodynamic functions of the latter chain is to
generalize the description of its spectrum in terms of supersymmetric $BC_N$-type motifs developed
in Ref.~\cite{CFGR20} to allow for a chemical potential term, and to use this expression to
represent the chain's partition function as the trace of a product of suitable site-dependent
transfer matrices. Passing to the thermodynamic limit, we obtain a simple expression for the
thermodynamic free energy per spin in terms of the Perron--Frobenius eigenvalue of the continuum
limit of the transfer matrix. This eigenvalue can be computed in closed form for up to two bosonic
or fermionic degrees of freedom, thus yielding an explicit expression for the thermodynamic
functions in these cases. In general, the behavior of these functions resembles that of their
$A_{N-1}$ counterparts. An interesting new feature we have identified is the existence of a double
Schottky peak (characteristic of certain three-level systems) in the specific heat for certain
values of the chemical potential.

We have also addressed in this paper the study of the criticality properties of the $\su(m|n)$
supersymmetric HS chain of $BC_N$-type for $m,n\le2$. To this end, we have first determined the
ground state of these models for all values of the fermionic chemical potential using the
motif-based description of the spectrum. In particular, we have shown that due to the distinctive
property of the $BC_N$-type motifs the ground state is at most doubly degenerate in all cases, in
contrast with the $A_{N-1}$ case. We have then studied the low energy excitations above the ground
state featuring a linear energy-quasimomentum relation, thus identifying the critical regions in
chemical potential space and the corresponding Fermi velocities. In fact, for certain values of
the chemical potential we have found that there can be two types of critical low-energy
excitations, associated with changes in either end of the ground state bond vector. The second
part of our analysis relies on deriving the low-temperature behavior of the free energy per spin
with the help of the explicit formula found in this paper for $m,n\le2$. More precisely, we have
ascertained the characteristic $T^2$ growth of the free energy in the critical regions, and have
determined the corresponding central charge by computing the $T\to0$ limit of $(f(T)-f(0))/T^2$.
In this way we have completely characterized the critical behavior of the models under study for
all values of the fermionic chemical potential.

The main tools used in this work, namely the motif-based description of the spectrum and the
explicit computation of the free energy per spin via the transfer matrix method, could in
principle be used to analyze the thermodynamics and critical behavior of other chains of
Haldane--Shastry type related to the $BC_N$ root system. To begin with, for the $BC_N$-type
supersymmetric Polychronakos--Frahm chain~\cite{YT96,BFGR08} a description of the spectrum in
terms of branched motifs has been recently found for zero chemical potential~\cite{BS20}. This
description, when suitably generalized to allow for a chemical potential term, could be the basis
for a study similar to the present one. Another model for which there is already a motif-based
description of the spectrum (again for zero chemical potential)~\cite{BFG16} is the
(non-supersymmetric) $BC_N$ Simons--Altshuler chain~\cite{SA94,BPS95}, recently studied in
Ref.~\cite{TS15} in connection with matrix product states from boundary CFTs. The situation is
less clear for HS-type spin chains associated to the $B_N$ and $D_N$ root
systems~\cite{BFG09,BFG11,BFG13}, where it is not yet known how to express the spectrum in terms
of suitable motifs. On the other hand, our method can still be applied to HS-like chains not
directly related to any root system when there is a motif-based description of the spectrum. This
is the case, for instance, for the $\su(m)$ chain recently introduced in Ref.~\cite{BFG20} when
$m$ is even.

\ack This work was partially supported by grants PGC2018-094898-B-I00 from Spain's
Mi\-nis\-te\-rio de Ciencia, Innovaci\'on y Universidades and~G/6400100/3000 from Universidad
Complutense de Madrid.

\appendix

\section{Computation of the central charges for the $\su(0|2)$, $\su(1|2)$, $\su(2|1)$ and
  $\su(2|2)$ chains}\label{app.A}

In this appendix we will provide the details of the evaluation of the central charges of the
$\su(0|2)$, $\su(1|2)$, $\su(2|1)$ and $\su(2|2)$ chains, that were omitted from
Section~\ref{sec.crit} for the sake of conciseness.

\medskip\noi 1.\en$\su(0|2)$

\smallskip\noi
From Eq.~\eqref{f0220} we have
\[
  f(T)=f_0-T\int_0^1\log\bigl(1+\e^{-\be\vp(x)/2}\bigr)\diff x\,.
\]
Thus $f(0)=f_0$, and performing the change of variable $\be\vp(x)/2=y$ we obtain
\begin{eqnarray*}
  f(T)-f(0)&=-2T^2\int_0^{\be\vpmax/2}x_0'(2Ty)\log\bigl(1+\e^{-y}\bigr)\diff y\\
           &\simeq
             -2T^2x_0'(0)\int_0^\infty\log\bigl(1+\e^{-y}\bigr)\diff y
             =-\frac{\pi^2T^2}{6}\,x_0'(0)=-\frac{\pi T^2}{6v_1}\,,
\end{eqnarray*}
where (as in what follows) we have used Eq.~\eqref{v12} for the Fermi velocity of the small
momentum excitations. Hence (as we knew from the analysis of Section~\ref{sec.GSlow}) this model
possesses only small momentum excitations, with central charge $c_1=1$.

\medskip\noi
2.\en $\su(1|2)$

\smallskip\noi
In the half-line $\mu<0$ we have
\[
  f(T)=-T\int_0^1\log\Bigl(1+\Or\bigl(\e^{-\be|\mu|}\bigl)\Bigl)\diff x=\Or(T\e^{-\be|\mu|}),
\]
and thus the system is gapped. Likewise, for $\mu=0$
\begin{eqnarray*}
  f(T)&=-T\int_0^1\log\left(\frac12+\e^{-\be\vp(x)}+\sqrt{\frac14+2\e^{-\be\vp(x)}}\,\right)\diff x\\
      &=-T^2\int_0^{\be\vpmax}x_0'(Ty)\log
        \left(\frac12+\e^{-y}+\sqrt{\frac14+2\e^{-y}}\,\right)\diff y\\
      &\simeq -T^2x_0'(0)\int_0^{\infty}
        \log\left(\frac12+\e^{-y}+\sqrt{\frac14+2\e^{-y}}\,\right)\diff y\
        =-\frac{\pi T^2}{6v_1}\,,
\end{eqnarray*}
Hence in this case there are only small momentum excitations, whose central charge is again
$c_1=1$.

Consider next the interval $0<\mu<\vpmax/2$. We can then write
\[
  \fl
f(T)=\frac12\int_0^{x_0(2\mu)}\big(\vp(x)-2\mu\big)\diff
x-T\int_0^{x_0(2\mu)}\log\widehat\la_1(x)\diff x
-T\int_{x_0(2\mu)}^1\log\la_1(x)\diff x\,,
\]
with
\[
  \widehat\la_1:=\e^{\be(\vp/2-\mu)}\la_1
  =\e^{-\be\vp/2}+\frac12\,\e^{-\be(\mu-\vp/2)}+\sqrt{1+\e^{-\be\mu}+\frac14\,\e^{-\be(2\mu-\vp)}}\,.
\]
Since the exponents in the expressions for $\widehat\la_1(x)$ and $\la_1(x)$ are non-positive
respectively in the intervals $[0,x_0(2\mu)]$ and $[x_0(2\mu),1]$ we have
\begin{equation}\label{f0su12}
  f(0)=\frac12\int_0^{x_0(2\mu)}\big(\vp(x)-2\mu\big)\diff x
\end{equation}
(in agreement with the result of the previous section), and therefore
\begin{equation}\label{fTf012}
  \fl
  f(T)-f(0)=T\int_0^{x_0(2\mu)}\log\widehat\la_1(x)\diff x
-T\int_{x_0(2\mu)}^1\log\la_1(x)\diff x=:I_1+I_2\,.
\end{equation}
The integral $I_2$ is dealt with in much the same fashion as similar integrals in the previous
cases. More precisely, since the main contribution to the latter integral comes from its lower
limit, where several of its exponents vanish, performing the change of variable $\be(\vp/2-\mu)=y$,
or equivalently $x=x_0(2(\mu+Ty))$, we easily obtain
\begin{eqnarray*}
  \fl
  I_2&=-2T^2\int_0^{\be(\vpmax-2\mu)} x_0'\bigl(2(\mu+Ty)\bigr)\log\!\left(\frac12+\e^{-(y+\be\mu)}
       +\sqrt{\frac14+\e^{-y}+\e^{-(y+\be\mu)}}\,\right)\diff y\\
  \fl
     &\simeq-2T^2x_0'(2\mu)\int_0^\infty\log\!\left(\frac12
       +\sqrt{\frac14+\e^{-y}}\,\right)\diff y=-\frac{\pi^2T^2}{15}\,x_0'(2\mu)\,.
\end{eqnarray*}
On the other hand, as the main contribution to the integral $I_1$ comes from both integration
limits, it is convenient to split this integral as
\[
  I_1=-T\int_0^{x_0(\mu)}\log\widehat\la_1(x)\diff x
  -T\int_{x_0(\mu)}^{x_0(2\mu)}\log\widehat\la_1(x)\diff x:=I_{11}+I_{12}\,.
\]
Since the main contribution to $I_{11}$ (resp.\ $I_{12}$) comes only from its lower (resp.\
higher) limit we readily obtain
\begin{eqnarray}
  \fl
  I_{11}&=-2T^2\int_0^{\be\mu/2}x_0'(2Ty)\log\!\left(\e^{-y}+\e^{-(\be\mu-y)}
          +\sqrt{1+\e^{-\be\mu}+\frac14\e^{-2(\be\mu-y)}}\,\right)\diff y\nonumber\\
  \fl
        &\simeq-2T^2x_0'(0)\int_0^{\infty}\log\bigl(1+\e^{-y}\bigr)\diff y
          =-\frac{\pi^2T^2}{6}\,x_0'(0)\,,
  \label{I11}\\
  \fl
  I_{12}&=-2T^2\int_0^{\be\mu/2}x_0'\bigl(2(\mu-Ty)\bigr)\log\!\left(\frac12\e^{-y}+\e^{-(\be\mu-y)}
          +\sqrt{1+\e^{-\be\mu}+\frac14\e^{-2y}}\,\right)\diff y\nonumber\\
  \fl
        &\simeq-2T^2x_0'(2\mu)\int_0^{\infty}\log\!\left(\frac12\e^{-y}
          +\sqrt{1+\frac14\e^{-2y}}\,\right)\diff y
          =-\frac{\pi^2 T^2}{10}\,x_0'(2\mu)\,.
          \label{I12}
\end{eqnarray}
Putting everything together and using the expressions~\eqref{v12} for the Fermi velocities of the
small momentum and Fermi excitations we finally obtain:
\[
  f(T)-f(0)\simeq-\frac{\pi T^2}{6v_1}-\frac{\pi T^2}{6v_2}\,.
\]
We thus see that both types of excitations have central charge $c_1=c_2=1$.

In the limiting case $\mu=\vpmax/2$, i.e., $x_0(2\mu)=1$, the integral $I_2$ in Eq.~\eqref{fTf012}
reduces to zero, while the upper limit in the integrals $I_{11}$ and $I_{12}$ becomes $1$. If
$\ga>1$ then $x_0'(2\mu)=x_0'(\vpmax)=1/\vp'(1)$ is finite, and hence Eqs.~\eqref{I11}-\eqref{I12}
remain valid. We thus obtain
\[
  f(T)-f(0)\simeq-\frac{\pi^2T^2}{6}x_0'(0)-\frac{\pi T^2}{10}\,x_0'(1)= -\frac{\pi
    T^2}{6v_1}-\frac35\,\frac{\pi T^2}{6v_2}\,,
\]
where $f(0)=f_0/2-\mu=(f_0-\vpmax)/2$ (cf.~Eq.~\eqref{f0su12}). Hence in this case the central
charges of the small momentum and Fermi excitations are respectively $c_1=1$ and $c_2=3/5$. On the
other hand, when $\ga=1$ and $\mu=\vpmax/2=1/4$ Eq.~\eqref{I11} still holds, while Eq.~\eqref{I12}
should be replaced by
\begin{eqnarray*}
  \fl
  I_{12}&=-T^{3/2}\int_0^{\be/8}y^{-1/2}\log\!\left(\frac12\e^{-y}+\e^{-(\be/4-y)}
          +\sqrt{1+\e^{-\be/4}+\frac14\e^{-2y}}\,\right)\diff y\\
  \fl
        &\simeq-T^{3/2}\int_0^{\infty}y^{-1/2}\log\!\left(\frac12\e^{-y}
          +\sqrt{1+\frac14\e^{-2y}}\,\right)\diff y=-0.866562\cdots T^{3/2}\,.
\end{eqnarray*}
Thus in this case the Fermi excitations (involving changes in the last components of the ground
state bond vector) are not critical, as anticipated in the previous section, whereas the small
momentum excitations have central charge $c_1=1$.

Finally, when $\mu>\vpmax/2$ proceeding as above we have
\begin{eqnarray*}
  f(T)&=-T\int_0^1\log\left(\e^{-\be(\vp(x)/2-\mu)}\widehat\la_1(x)\right)\diff x
            =\frac{f_0}2-\mu-T\int_0^1\log\widehat\la_1(x)\,\diff x\\
  &=f(0)-T\int_0^1\log\widehat\la_1(x)\,\diff x\,.
\end{eqnarray*}
The main contribution to the last integral comes only from its lower limit, i.e., from small
momentum excitations, so that setting $\be\vp(x)=2y$ we obtain
\begin{eqnarray*}
  \fl
  f(T)&-f(0)\\
  \fl
  &=-2T^2\int_0^{\be\vpmax/2}x_0'(2Ty)\log\!\left(\e^{-y}+\frac12\,\e^{-(\be\mu-y)}
             +\sqrt{1+\e^{-\be\mu}+\frac14\,\e^{-2(\be\mu-y)}}\,\right)\!\diff
             y\\
  \fl
           &\simeq-2T^2x_0'(0)\int_0^\infty\log\bigl(1+\e^{-y}\bigr)\diff y
       =-\frac{\pi^2T^2}{6}\,x_0'(0)=-\frac{\pi T}{6v_1}\,.
\end{eqnarray*}
Hence there in this case there are only small momentum excitations, with central charge $c_1=1$.

\medskip\noi
3.\en $\su(2|1)$

\smallskip\noi
To begin with, when $\mu<0$ setting $\be\vp(x)=2\mu$ in the integral for $f(T)$ we obtain
\begin{eqnarray*}
   \fl
  &f(T)-f(0)\\
  \fl
  &=-2T^2\int_0^{\be\vpmax/2}x_0'(2Ty)\log\!\left(1+\frac12\,\e^{-(\be|\mu|+2y)}
             +\sqrt{\e^{-2y}+\e^{-(\be|\mu|+2y)}+\frac14\,\e^{-2(\be|\mu|+2y)}}\,\right)\!\diff
             y\\
  \fl
           &\simeq-2T^2x_0'(0)\int_0^\infty\log\bigl(1+\e^{-y}\bigr)\,\diff y
       =-\frac{\pi^2T^2}{6}\,x_0'(0)=-\frac{\pi T^2}{6v_1}\,,
\end{eqnarray*}
Hence in this case only the small momentum excitations are present, with central charge $c_1=1$.
When $\mu=0$, a similar calculation shows that
\begin{eqnarray*}
 f(T)-f(0)
  &=-T^2\int_0^{\be\vpmax}x_0'(Ty)\log\!\left(1+\frac12\,\e^{-y}
    +\sqrt{2\e^{-y}+\frac14\,\e^{-2y}}\,\right)\!\diff
    y\\
  &\simeq-T^2x_0'(0)\int_0^\infty\log\!\left(1+\frac12\,\e^{-y}
    +\sqrt{2\e^{-y}+\frac14\,\e^{-2y}}\,\right)\diff y\\
    &=-\frac{\pi^2T^2}{4}\,x_0'(0)=-\frac{\pi T^2}{4v}\,,
\end{eqnarray*}
so that there are only small momentum excitations with central charge $c_1=3/2$.

Consider next the interval $0<\mu<\vpmax$. We then have
\[
  \fl
f(T)=\int_0^{x_0(\mu)}\big(\vp(x)-\mu\big)\diff
x-T\int_0^{x_0(\mu)}\log\widehat\la_1(x)\diff x
-T\int_{x_0(\mu)}^1\log\la_1(x)\diff x\,,
\]
where now
\[
  \widehat\la_1:=\e^{\be(\vp-\mu)}\la_1
  =\frac12+\e^{-\be(\mu-\vp)}+\sqrt{\frac14+\e^{-\be(\mu-\vp)}+\e^{-\be(2\mu-\vp)}}\,.
\]
Since $\widehat\la_1(x)\to0$ as $\be\to\infty$ when $0<x<x_0(\mu)$, whereas $\la_1(x)\to0$ as
$\be\to\infty$ when $x_0(\mu)<x<1$, we conclude that
\begin{equation}\label{f0su21}
  f(0)=\int_0^{x_0(\mu)}\big(\vp(x)-\mu\big)\diff x\,,
\end{equation}
and hence
\begin{equation}\label{fTf021}
  \fl
  f(T)-f(0)=T\int_0^{x_0(\mu)}\log\widehat\la_1(x)\diff x
-T\int_{x_0(\mu)}^1\log\la_1(x)\diff x=:I_1+I_2\,.
\end{equation}
Proceeding as above and using Eq.~\eqref{vF11} for the velocity of the Fermi excitations (with
momentum near $p_0=\pi x_0(\mu)$) we have
\begin{eqnarray*}
  \fl
  I_{1}&=-T^2\int_0^{\be\mu}x_0'(\mu-Ty)\log\!\left(\frac12+\e^{-y}
          +\sqrt{\frac14+\e^{-y}+\e^{-(\be\mu+y)}}\,\right)\diff y\nonumber\\
  \fl
        &\simeq-T^2x_0'(\mu)\int_0^{\infty}\log\!\left(\frac12+\e^{-y}
          +\sqrt{\frac14+\e^{-y}}\,\right)\diff y
          =-\frac{2\pi^2 T^2}{15}\,x_0'(\mu)=-\frac{2\pi T^2}{15v_2}\,,
  \\
  \fl
  I_{2}&=-T^2\int_0^{\be(\vpmax-\mu)}x_0'(\mu+Ty)\log\!\left(1+\frac12\e^{-y}
          +\sqrt{\frac14\e^{-2y}+\e^{-y}+\e^{-(\be\mu+y)}}\,\right)\diff y\nonumber\\
  \fl
        &\simeq-T^2x_0'(\mu)\int_0^{\infty}\log\!\left(1+\frac12\e^{-y}
          +\sqrt{\frac14\e^{-2y}+\e^{-y}}\,\right)\diff y
          =-\frac{\pi^2 T^2}{5}\,x_0'(\mu)=-\frac{\pi T^2}{5v_2}\,.
\end{eqnarray*}
Adding up the asymptotic approximations for the integrals $I_{1,2}$ we obtain
\[
  f(T)-f(0)=-\frac{\pi T^2}{3v_2}\,,
\]
which implies that in this case there are only Fermi excitations having central charge $c_2=2$.
When $\mu=\vpmax$ and $\ga>1$ the contribution from the integral $I_2$ vanishes, while that of the
integral $I_1$ remains unchanged (with $\mu=\vpmax$) since $x_0'(\vpmax)$ is finite. We thus have
\[
  f(T)-f(0)\simeq-\frac{2\pi T^2}{15v_2}\,,
\]
so that the central charge of the Fermi excitations is now $c_2=4/5$. On the other hand, when
$\ga=1$ and $\mu=\vpmax=1/2$ performing the change of variable
$\be(\mu-\vp)=\be(1/2-\vp)=\be/2(1-x)^2=y$, or equivalently $x=1-\sqrt{2Ty}$, in the integral
$I_1$ we obtain
\begin{eqnarray*}
  \fl
  f(T)-f(0)&=-\frac{T^{3/2}}{\sqrt 2}\int_0^{\be/2}y^{-1/2}\log\!\left(\frac12+\e^{-y}
          +\sqrt{\frac14+\e^{-y}+\e^{-(\be/2+y)}}\,\right)\diff y\nonumber\\
  \fl
        &\simeq-\frac{T^{3/2}}{\sqrt 2}\int_0^{\infty}y^{-1/2}\log\!\left(\frac12+\e^{-y}
          +\sqrt{\frac14+\e^{-y}}\,\right)\diff y
          =-1.44084\cdots T^{3/2}\,.
\end{eqnarray*}
Thus the system is gapless but not critical in this case, confirming again the results of the
previous section based on the existence of excitations with $\De E$ proportional to $(\De p)^2$.
Finally, if $\mu>\vpmax$ we have
\begin{eqnarray*}
  \fl
  f(T)&=\int_0^1\big(\vp(x)-\mu\big)\diff x-T\int_0^1\log\widehat\la_1(x)\,\diff x
        =f_0-\mu-T\int_0^1\log\Bigl(1+\Or\bigl(\e^{-\be(\mu-\vpmax)}\bigr)\Bigr)\diff x\\
  \fl
  &=f_0-\mu-\Or\bigl(T\e^{-\be(\mu-\vpmax)}\bigr)\,.
\end{eqnarray*}
Hence the system is now gapped, with energy gap $\mu-\vpmax$.

\medskip\noi
4.\en $\su(2|2)$

\smallskip As shown in the $\su(0|2)$ case, at low temperatures the first integral in
Eq.~\eqref{fsu22} behaves as
\[
  -T\int_0^1\log\bigl(1+\e^{-\be\vp(x)/2}\bigr)\,\diff x\simeq-\frac{\pi^2T^2}{6}\,x_0'(0)\,.
\]
The behavior of the second integral in Eq.~\eqref{fsu22} can also be inferred from the analysis in
Section~\ref{sec.crit}. More precisely, for $\mu<0$ we obtain
\[
  I_2:=-T\int_0^1\log\bigl(1+\e^{-\be(\vp(x)/2-\mu)}\bigr)\diff x=\Or(T\e^{-\be|\mu|})\,,
\]
while for $\mu>\vpmax/2$
\[
  I_2-\frac{f_0}2+\mu=-T\int_0^1\log\bigl(1+\e^{-\be(\mu-\vp(x)/2)}\bigr)\diff
  x=\Or(T\e^{-\be(\mu-\vpmax/2)})\,.
\]
On the other hand, in the interval $0<\mu<\vpmax/2$ we have
\begin{eqnarray*}
  \fl
  I_2-\frac12\int_0^{x_0(2\mu)}\big(\vp(x)-2\mu\big)\diff x=
       &-T\int_0^{x_0(2\mu)}\log\bigl(1+\e^{-\be(\mu-\vp(x)/2)}\bigr)\diff x\\
  \fl
     &-T\int_{x_0(2\mu)}^1\log\bigl(1+\e^{-\be(\vp(x)/2-\mu)}\bigr)\diff x
       \simeq -\frac{\pi^2T^2}{3}\,x_0'(2\mu)\,,
\end{eqnarray*}
where each integral contributes half of the latter value. In the limiting case $\mu=0$ the
integral $I_2$ reduces to the first integral dealt with above, while for $\mu=\vpmax/2$ we have
\[
  \fl
  I_2-\frac{f_0}2+\mu=-T\int_0^1\log\bigl(1+\e^{-\be(\mu-\vp(x)/2)}\bigr)\diff x
  \simeq\cases{-\frac{\pi T^2}{6}\,x_0'(\vpmax),&\(\ga>1\)\\
    -\frac{\sqrt\pi}2(2-\sqrt2)\ze(3/2)T^{3/2},&\(\ga=1\).
  }
\]
Putting all of the above together, and taking into account Eqs.~\eqref{Fv12}-\eqref{Fv122} for the
Fermi velocities of the small momentum and Fermi excitations in this case, we conclude that
\[
  \fl f(T)-f(0)\simeq\cases{
    -\frac{\pi T^2}{6v_1},& \(\mu<0\)\\
    -\frac{\pi T^2}{3v_1},& \(\mu=0\)\\
    -\frac{\pi T^2}{6v_1}-\frac{\pi T^2}{3v_2},& \(0<\mu<\vpmax/2\)\\
    -\frac{\pi T^2}{6v_1}-\frac{\sqrt\pi}2(2-\sqrt2)\ze(3/2)T^{3/2},
    & \(\mu=\vpmax/2\) and \(\ga=1\)\\
    -\frac{\pi T^2}{6v_1}-\frac{\pi T^2}{6v_2},& \(\mu=\vpmax/2\) and \(\ga>1\)\\
    -\frac{\pi T^2}{6v_1},& \(\mu>\vpmax/2\) }
\]
Thus the central charge of the small momentum excitations is $1$ for $\mu\ne0$ and to $2$ for
$\mu=0$, while the Fermi excitations have central charge $c_2=2$ for $0<\mu<\vpmax/2$ and $c_2=1$
for $\mu=\vpmax/2$ and $\ga>1$.

\section*{References}


\end{document}